\pdfoutput=1
\documentclass[usenatbib]{mn2e}
\usepackage{hyperref}
\usepackage{amsfonts,amsmath,amssymb}
\usepackage{txfonts}
\usepackage{graphicx}
\usepackage{bm}
\usepackage{ctable}
\usepackage{url}
\usepackage{breakurl}
\usepackage{fixltx2e} 

\newcommand{\mnras}{Mon. Not. R. Astron. Soc.}

\newcommand{\pop}{Phys. Plasmas}
\newcommand{\pr}{Phys. Rev.}
\newcommand{\apjl}{Astrophys. J. Lett.}

\newcommand{\jfm}{J. Fluid Mach.}
\newcommand{\pfluid}{Phys. Fluids}
\newcommand{\apj}{Astrophys. J.}

\newcommand{\jplp}{J. Plasma Phys.}
\newcommand{\nature}{Nature}
\newcommand{\science}{Science}
\newcommand{\sciencea}{Science Adv.}
\newcommand{\prl}{Phys. Rev. Lett.}
\newcommand{\astronastro}{Astron. Astro.}

\newcommand{\arfm}{Ann. Rev. Fluid Mech.}

\newcommand{\araa}{Ann. Rev. Astron. Astro.}
\newcommand{\arpes}{Ann. Rev. Planet. Earth Sci.}

\newcommand{\astross}{Astrophys. Space. Sci.}
\newcommand{\dansssr}{Dok. Akad. Nauk SSSR}
\newcommand{\icarus}{Icarus}
\newcommand{\aap}{Astron. Astro.}
\newcommand{\pasp}{Pub. Astron. Soc. Pacific}

 \numberwithin{equation}{section}

\newcommand{\plotonesize}[2]{\centering \leavevmode \includegraphics[width={#2\columnwidth}]{#1}}

\newcommand{\acknowledgments}{\begin{small}\section*{Acknowledgments}\end{small}}
\newcommand\altaffilmark[1]{$^{#1}$}
\newcommand\altaffiltext[1]{$^{#1}$}
\newcommand{\driftvel}{{\bf w}_{s}}
\newcommand{\driftvelmag}{{w}_{s}}
\newcommand{\driftvelhat}{\hat{{\bf w}}_{s}}

\newcommand{\driftvelx}{\mathrm{w}_{s,x}}
\newcommand{\driftvely}{\mathrm{w}_{s,y}}
\newcommand{\driftvelz}{\mathrm{w}_{s,z}}
\newcommand{\driftvelj}{\mathrm{w}_{s,j}}
\newcommand{\BV}{Brunt-V\"ais\"al\"a} 

\newcommand{\paperone}{SH17}
\newcommand{\papertwo}{HS17}

\newcommand{\SI}{Settling Instability}
\newcommand{\si}{settling instability}

\newcommand{\YG}{YG}

\title[The RDI in protoplanetary disks]{Resonant Drag Instabilities in protoplanetary disks: the streaming instability and new, faster-growing instabilities\vspace{-0.5cm}}

\author[Squire \&\ Hopkins]{
\parbox[t]{\textwidth}{ 
	Jonathan Squire\altaffilmark{1,2} \&\ Philip F. Hopkins\altaffilmark{1}
} 
\vspace*{6pt} \\
\altaffiltext{1}{Theoretical Astrophysics, Mailcode 350-17, California Institute of Technology, Pasadena, CA 91125, USA} \\
\altaffiltext{2}{Walter Burke Institute for Theoretical Physics, Pasadena, CA 91125, USA\vspace{-0.3cm}}
}

\date{Submitted to MNRAS, November, 2017\vspace{-0.6cm}}
\begin{document}
\maketitle


\begin{abstract}
We identify and study a number of new, rapidly growing instabilities of dust grains in protoplanetary disks, which may be important for planetesimal formation. The study is based on the recognition that dust-gas mixtures are generically unstable to a Resonant Drag Instability (RDI), whenever the gas, absent dust, supports undamped linear modes. 
We show that the ``streaming instability'' is an RDI associated with epicyclic oscillations; this provides simple interpretations for its mechanisms and accurate analytic expressions for its growth rates and fastest-growing wavelengths. We extend this analysis to more general dust streaming motions and other waves, including buoyancy and magnetohydrodynamic oscillations, finding various new instabilities. Most importantly, we identify the disk ``\si,''  which occurs as dust settles vertically into the midplane of a rotating disk. For small grains, this instability grows many orders of magnitude faster than the standard streaming instability, with a growth rate that is independent of grain size. Growth timescales for realistic dust-to-gas ratios are comparable to the disk orbital period, and the characteristic wavelengths are more than an order of magnitude larger than the streaming instability (allowing the instability to concentrate larger masses).   This suggests that in the process of settling, dust will band into rings then filaments or clumps, potentially seeding dust traps, high-metallicity regions that   in turn seed the streaming instability, or even overdensities that  coagulate or directly collapse to planetesimals.
\end{abstract}

\begin{keywords}
protoplanetary disks -- planets and satellites: formation -- hydrodynamics -- instabilities -- accretion, accretion disks
\vspace{-1.0cm}
\end{keywords}

\vspace{-1.1cm}

\section{Introduction}

Explaining the mechanisms of  planetesimal formation---how the micron-sized grains that populate a primordial 
disk are able to coagulate and grow into $\mathrm{km}$-sized planetesimals \citep{1973ApJ...183.1051G,Chiang:2010ew,2014prpl.conf..547J}---is a fundamental problem of 
modern astrophysics. While very small particles can stick together upon colliding, once grains reach approximately millimeter scale or larger in diameter,  not only do they  rapidly fall into the central star, they  are also more  likely 
to  bounce  or shatter in a collision \citep{2008ARA&A..46...21B,2008A&A...480..859B,2010A&A...513A..57Z,2015A&A...574A..83K}. This leads 
one to question how the wide variety of observed exoplanets apparently form so readily \citep{2012Natur.481..167C,2016PASP..128j2001B}. 
One promising solution to this conundrum has emerged in recent years, based on the idea that the dusty gas mixture is unstable to the ``streaming instability'' \citep{2005ApJ...620..459Y,2007ApJ...662..613Y}.
In the course of its nonlinear evolution, the streaming instability acts to concentrate dust 
into pockets and filaments with densities that can be  hundreds of times larger than the background values \citep{2007ApJ...662..627J,2010ApJ...722L.220B,Bai:2010gm,2014ApJ...792...86Y}. With such high densities 
and 
reduced relative velocities, grains may then coagulate due to self gravity, forming the seeds
around which  planetesimals can grow \citep{2007Natur.448.1022J,2012A&A...537A.125J,2016ApJ...822...55S,2041-8205-847-2-L12}. 

However, while this broad picture has garnered some support, there are a variety of aspects that remain unclear. Simulation work has shown that this mechanism depends critically on the dust-to-gas ratio, or metallicity, which we term $\mu$, and that
there may be a critical metallicity below which the concentration is not sufficiently strong to allow gravitational collapse to take over
(see, e.g., \citealp{Johansen:2009ih,Bai:2010gm,2010ApJ...722L.220B,2012A&A...537A.125J,2014ApJ...792...86Y,2041-8205-828-1-L2,2017A&A...597A..69S,0004-637X-839-1-16}). Further, this critical metallicity appears to increase for smaller grains \citep{2015A&A...579A..43C,2016arXiv161107014Y} and it is unclear whether it is feasible to form a sufficiently 
 large population of moderate-sized grains such that the scenario described in the previous paragraph takes place \citep{2014A&A...572A..78D}. 
There have also been a wide variety of other grain concentration mechanisms proposed or observed in simulations---e.g., concentration in background structures (e.g.\ ``traps'') or via externally-driven turbulence \citep{barge:1995.vortex.trap.idea,bracco:1999.keplerian.largescale.grain.density.sims,Johansen:2009jf,2013ApJ...776...48H,Pan:2013bm,2016LPI....47.2661C,dittrich:2013.grain.clustering.mri.disk.sims,zhu:2014.non.ideal.mhd.vortex.traps,hopkins:2014.pebble.pile.formation} or other instabilities \citep{2000Icar..148..537G,2016MNRAS.456.3079H,2016MNRAS.457L..54L,2017arXiv170802945L}---and questions remain regarding the role of these 
mechanisms and/or how they interact with structures produced by the streaming instability.
On the 
 more esoteric side, the detailed  theoretical underpinnings for the critical metallicity remain poorly understood, as 
 do  aspects of the linear streaming instability itself  \citep{2005ApJ...620..459Y,Jacquet:2011cy,2013MNRAS.434.1460K,0004-637X-817-2-140}.


This paper serves two purposes. The first is to give a straightforward interpretation and analytic derivation of the properties of the streaming instability. The second is to introduce several new instabilities of streaming dust, which likely concentrate small grains much more efficiently than the standard Youdin \& Goodman (YG) streaming instability and may play an important role in the planetesimal  formation process.
Our analysis is based on understanding that the streaming instability is a type of \emph{Resonant Drag Instability} (RDI). As introduced in \citet{Squire:2017rdi} (hereafter \paperone), in a dust-gas mixture 
where the dust streams  through the gas with some relative velocity $\driftvel$, an RDI occurs
\emph{whenever the projection of $\driftvel$ along some direction $\hat{\bm{k}}$ matches 
the phase velocity of a wave in the gas}. Equivalently, we can write the resonant condition as $\driftvel\cdot\bm{k} = \omega_{\mathcal{F}}(\bm{k})$, where $\omega_{\mathcal{F}}(\bm{k})$ is the frequency of some natural response in the gas (absent dust), and $\bm{k}=k\hat{\bm{k}}$ is the mode's wavenumber. In the frame of the dust, such a gas wave is stationary, or resonant, and is thus very easily destabilized by the mutual drag interaction between the two phases. In fact, as shown in \paperone, 
when an unstable RDI exists---i.e., when there is a gas wave that resonates with the dust---it  always grows faster than any other drag-induced instabilities of the system at low metallicity.  This idea allows us to identify the YG streaming instability as an RDI (the ``epicyclic RDI''), where the gas wave is an epicyclic oscillation with frequency $\omega_{\mathrm{epi}}=\hat{\bm{k}}\cdot \bm{\Omega}$ (here $\bm{\Omega}$ is the local disk angular rotation velocity). This implies that the resonance, and thus the fastest-growing
modes, occur when $\hat{\bm{k}}\cdot \driftvel = \omega_{\mathrm{epi}}/k$. As another example, 
examined in detail in \citet{Hopkins:2017rdi} (hereafter \papertwo), the resonance with sound waves
of frequency $\omega_{\mathrm{sound}}=k\, c_{s}$ causes an RDI (the ``acoustic RDI'')
at the resonant mode angle $\hat{\bm{k}}\cdot \driftvel =c_{s}$.  We shall see that the analysis 
of the streaming instability within this formalism provides a simple interpretation for the 
mechanism of the instability, as well as straightforward analytical calculation of the fastest-growing 
modes and their growth rates at low-to-moderate dust metallicity ($\mu\lesssim 1$). 

The basic idea of the RDI---that an instability occurs whenever the dust streaming 
is resonant with a fluid wave---suggests that we should consider \emph{other} fluid waves of relevance in disks. Such analyses---including more general  epicyclic resonances, resonance with \BV\ oscillations,
the acoustic resonance, and resonances with ideal and nonideal magnetohydrodynamic (MHD) waves---form the bulk of this work.
Our most important result is that the addition of a vertical settling drift of grains towards the midplane of the
disk dramatically modifies the  streaming instability. We term this the disk ``\si.'' Unlike the YG streaming instability, the maximum growth rate of the disk \si\ at low metallicity does not decrease with grain size, and can be much faster than the time required for
 grains to settle into the midplane.  For plausible disk parameters, the growth timescales can be comparable to, or even shorter than, the disk dynamical time ($\Omega^{-1}$). 
 In fact, in the absence of viscosity,
we find that the growth rate $\Im(\omega)$ of this instability is formally 
infinite, scaling as $\Im(\omega)\sim k^{1/3}$ as $k\rightarrow \infty$ for a particular ``double-resonant'' mode angle, which occurs for any grain size. 
Moreover, the largest unstable wavelengths with significant growth rates are much larger (by one to two orders of magnitude) than the YG streaming instability.

We show these new, fast-growing modes are robust to the addition of gas and dust stratification and gas compressibility. Their existence suggests that in the process of settling towards the disk midplane, small grains may 
clump significantly and will band into radial annuli, essentially segregating into dense dust rings {\em during the process of vertical settling}. This could modify important properties of the dust-gas mixture (e.g., the opacity), enhance
coagulation rates of grains,  act as high-metallicity seeds that improve the planetesimal-formation efficiency of the YG streaming instability in the disk midplane, or even (depending on the nonlinear behavior) cause the direct fragmentation into self-gravitating clumps. Although such processes are necessarily transient---occurring before the dust settles into the disk midplane---for 
smaller grains, the growth time is orders of magnitude shorter than the settling time, suggesting it will evolve well into its nonlinear
stages before the dust stops drifting in the vertical direction.

In addition to this resonance with gas epicycles (the YG streaming instability and the disk \si),
we also study the resonance of dust with inertia-gravity, or \BV\ waves. Although we find that this
 ``\BV\ RDI'' is less important for disks than the epicyclic resonance, it does have relevance in some regimes. Further, the instability is quite generic, occurring whenever grains settle through a stratified gas atmosphere, and forms a likely explanation for observations of clumping in previous numerical experiments \citep{Lambrechts:2016fg}. Finally, we consider RDIs arising from the interaction of dust 
 with ideal and nonideal MHD waves; however, although such instabilities  may be of interest in well ionized regions of disks (e.g., in magneto-centrifugal winds), near the midplane of a cool
 protoplanetary disk they are strongly damped by nonideal effects (Ohmic and ambipolar diffusion).

\subsection{Organization of this work}

We organize the remainder of this work as follows. As a preliminary, in \S\ref{sub: toy model}, we outline a simple, heuristic model
for the operation of RDIs. While the model is simplified by construction, we hope that, by introducing this early on, the reader can gain some intuitive understanding of RDI physics before tackling the more formal calculations later in the work.  To provide 
a quick reference for the remainder of the work, \S\ref{sec:overview.of.modes} then briefly outlines the different instabilities that will be studied and their basic properties.
\S\ref{sec: disk model} is devoted to laying out the details of the disk model we use: the gas and dust equations, 
the drag law governing the interaction between the two phases, the relative drift velocity $\driftvel$,
and  the local and  linear approximations that will be used throughout this work.
In \S\ref{sec: resonance}, we have a short section focused on the algorithm we use to find resonant drag instabilities, which involves computing the wavenumber where the streaming dust resonates with 
a fluid wave and using a simple formula (Eq.~\eqref{eq: dust eval pert}) to compute the growth 
rate of the RDI. 

The next three sections, \S\S\ref{sec: resonance epicyclic}--\ref{sec: MHD and others},  are devoted to studying the different RDIs mentioned above:
the streaming instability and its cousin the disk ``\si'' (from epicyclic oscillations) in \S\ref{sec: resonance epicyclic}, the \BV\ RDI (\S\ref{sub: BV RDI}) and epicyclic-\BV\ RDI (\S\ref{sub: BV epi RDI}) that occur in regions with a stratified background equilibrium, and various other RDIs from sound and MHD waves (\S\ref{sec: MHD and others}).
These sections, which derive analytic expressions for the growth rates of all relevant instabilities, are necessarily somewhat technical. For this reason, following a discussion of  neglected physical effects (\S\ref{sec: other effects}), in \S\ref{sec: astrophysics} we give  an overview of these results and  a discussion
of the astrophysical relevance of each RDI. We have designed \S\ref{sec: astrophysics}
to be accessible without detailed reference to \S\S\ref{sec: resonance epicyclic}--\ref{sec: MHD and others}, 
and a busy reader more interested in astrophysics
should consider focusing on \S\ref{sub: toy model}, \S\ref{sec:overview.of.modes}, \S\ref{sub:How to find an instability}, and \S\ref{sec: astrophysics}, which are relatively short and cover the key ideas of this work without diving into detailed mathematical derivations. 

In App.~\ref{app: high mu streaming}, we cover the important case of the streaming instability 
at high metallicity ($\mu>1$), which is key for grain dynamics in the midplane region.
This is a distinct instability from the low-$\mu$ streaming instability and is not an RDI. We
give simple expressions for its growth rate and fastest-growing wavenumbers (to our knowledge, these have
not appeared in previous works), as well as  discussing its physical mechanism.

\begin{figure}
\begin{center}
\includegraphics[width=1.0\columnwidth]{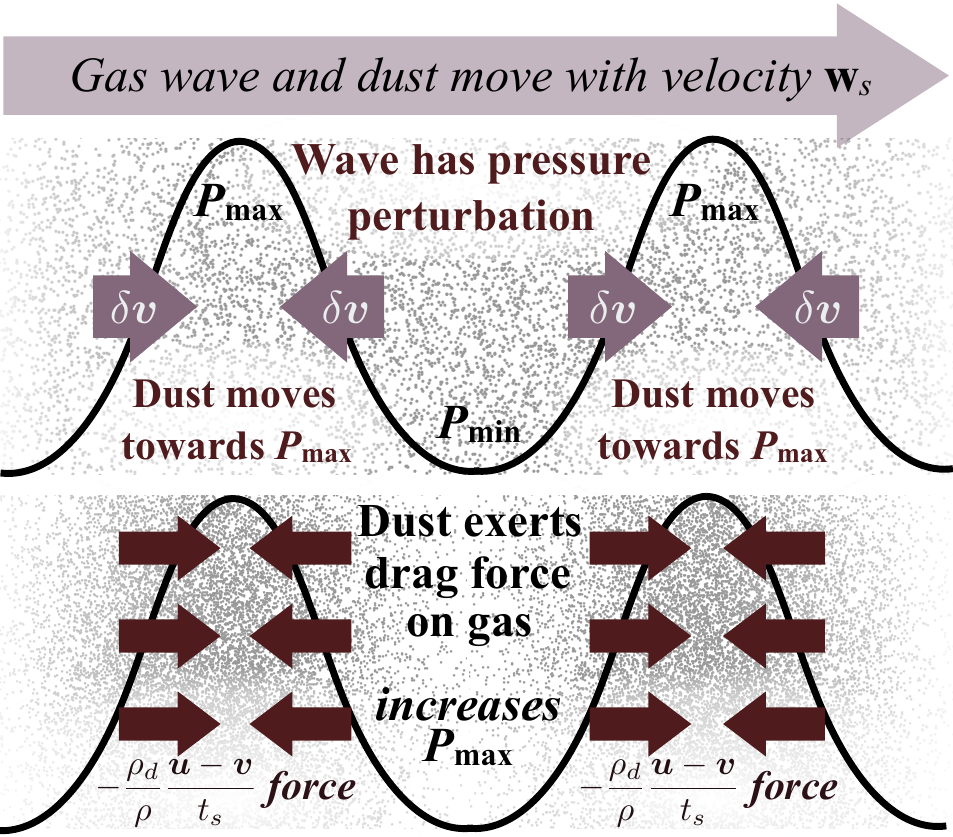}
\caption{Heuristic model for the operation of resonant drag instabilities, including the 
streaming instability, the disk settling instability, and most other RDIs studied in this work (see \S\ref{sec:overview.of.modes} for a
brief overview). 
We consider a generic gas wave that carries an associated pressure perturbation (black, sinusoidal lines). This wave propagates to the right with a phase velocity that matches the dust drift velocity $\driftvel$, \emph{viz.},
the two are \emph{resonant} (note that we neglect misalignment between $\driftvelhat$ and $\hat{\bm{k}}$ here, because this does not modify the general arguments).  In the moving frame where the dust and wave are stationary, 
the dust is attracted to the pressure maxima ($P_{\mathrm{max}}$; see e.g., \citealt{2014MNRAS.440.2136L}), moving towards these with velocity $\delta \bm{v}$ 
(upper panel). This movement then exerts a drag force on the gas (the dust ``backreaction'') of the form $-(\rho_{d}/\rho)(\delta \bm{v}-\bm{u})/t_{s}$, which acts to increase the pressure in the high-pressure regions (lower panel). This 
thus increases the amplitude of the gas wave, causing further accumulation of dust and resulting in exponential instability. }
\label{fig: toy model}
\end{center}
\end{figure}

Finally, we note that in most  figures (excepting Figs.~\ref{fig: toy model}--\ref{fig: 2D epicyclic from YG}) thick colored lines show 
``exact'' results from numerical solutions of the dispersion relation, while 
black or gray crosses and dashed lines illustrate our analytic approximations using the formalism of \S\ref{sec: resonance}.

\subsection{A simple, heuristic model for  Resonant Drag Instabilities}\label{sub: toy model}

Before diving into detailed mathematical calculations, it is useful to give a simple, heuristic
model that describes the  physics of the resonant drag instability. This model applies to the streaming
instability, as well as the other, new instabilities described throughout this work (see \S\ref{sec:overview.of.modes} for an overview). Although the model does not capture 
the full details of the RDI in all cases, we do believe it describes its key elements. It is thus helpful 
for gaining a basic intuitive understanding for why the RDI works, as well as the  properties of the 
wave  and dust-gas interaction that promote instability. 
We give two possible ways the RDI can operate, the first relying on a pressure perturbation in the 
gas wave (see Fig.~\ref{fig: toy model}), the second relying on the dependence of the gas drag on dust parameters. Both 
models apply only at the resonance wavelength, when the dust drift velocity matches the phase velocity of the wave, because they require the  wave to be stationary in the frame of the dust.

In its simplest form, the model is described in Fig.~\ref{fig: toy model}.   We assume that the gas wave (frequency $\omega_{\mathcal{F}}$)
contains a pressure perturbation, and  propagates to the right at the same phase velocity as the streaming dust (velocity $\driftvel$). This assumption, that the two are resonant  (i.e., $\bm{k}\cdot\driftvel = \omega_{\mathcal{F}}$), is by construction: 
we have chosen the wavenumber $\bm{k}$ such that this is the case (see discussion above).
In the frame of the dust, the gas pressure perturbation is effectively constant in time, and the dust is attracted 
towards pressure maxima (this attraction can be formally 
justified in the limit of short stopping times, when the dust quickly reaches its terminal velocity; see, e.g., \citealt{2014MNRAS.440.2136L,2017arXiv170802945L}). 
As it moves towards the pressure maxima, the dust exerts a backreaction force on the gas, which acts in the opposite
direction to the pressure gradient. It thus acts to compress the gas further, increasing the pressure maxima, and thus the 
attraction of the dust towards the pressure maxima. The process runs away as an exponentially growing instability.
We thus expect instability whenever the gas wave contains a pressure perturbation. 
Asymmetric epicyclic oscillations  fulfill this requirement and lead to the streaming instability and disk ``settling instability.'' 

While the gas pressure response is the most common mechanism that causes  RDI, 
a similar effect can occur when the dust drag depends on gas parameters that are perturbed by the wave.
 This is particularly  relevant for waves that perturb gas density and velocity more strongly than the pressure (e.g., inertial gravity 
 oscillations or shear-Alfv\'en waves), but  provides minor modifications to other RDIs also. 
 Consider, for concreteness, a case where the gas wave
involves a density perturbation but no pressure perturbation, and the dust drag time (stopping time) $t_{s}$ depends 
on density also. Dust will naturally accumulate in regions of small $t_{s}$, because this is where it is most tightly coupled 
to the gas. Again, this dust, moving towards such regions, exerts a force on the gas, which 
can further perturb the gas  density in the wave (depending on the details of the gas wave response). 
If this perturbation acts to collect more dust---i.e., if the force from the dust increases the gas density and 
if the  stopping time decreases at higher density, or vice versa---then the effect will 
increase the high density regions, resulting in instability. 
It is also possible that the opposite occurs, in which case the effect will be stabilizing.  
Because differently sized grains have different drag laws (e.g., Epstein drag for small grains, or Stokes drag 
for larger grains; see \S\ref{sub: stopping time} below), whether this mechanism is stabilizing or destabilizing can depend
on details of the drag regime (unlike the gas-pressure mechanism of the previous paragraph).
Similar effects are also possible from the velocity dependence of the dust drag, but we do not
go into detail here (e.g., this is responsible for the RDI with neutral dust and Alfv\'en waves; see \citealt{Hopkins:2018rdi}).

Finally, it is worth clarifying  that, unsurprisingly, the toy models  laid out in the previous paragraphs are oversimplified. In reality, because of the 
time lag between the gas and dust responses, and time lags in the gas response to an applied force, there will be a phase offset between the dust 
and the gas pressure \citep{2000Icar..148..537G,2017arXiv170802945L}, which is not accounted for in the above discussion. However, 
the model does explain the importance of pressure perturbations in RDIs, as 
well as the stabilizing 
or destabilizing influence of the dust drag law and its dependence on gas parameters. It is thus a useful toy model to keep in mind as we wade into more detailed calculations.

\section{Overview of the Instabilities Studied in This Paper}
\label{sec:overview.of.modes}

As discussed above, the RDI is not a single instability but a broad family of instabilities, each associated with a resonance with a particular fluid wave. In this paper we will demonstrate the existence of, and calculate characteristics of, a  range of different RDIs of potential relevance in protoplanetary disks and planetesimal formation. To guide the reader,  here we collect a brief overview of the  distinct instabilities that will be studied and the name that we will use to refer to each.

\begin{itemize}
	\item{\bf The ``\YG\ Streaming Instability'' (Epicyclic RDI)} (\S\ref{subsub: epicyclic horizontal streaming}): We will show that the usual streaming instability,  introduced by \citet{2005ApJ...620..459Y}, is an RDI when the system is gas dominated ($\mu<1$). It  arises from a resonance with  epicyclic oscillations of the gas and occurs when the dust streams  in the midplane of the disk (i.e., the radial and azimuthal directions). 	
	\item{\bf The Disk ``\SI''} (Vertical-Epicyclic or Vertical-Stratified-Epicyclic RDI; \S\ref{subsub: epicyclic vertical streaming}): This is a new instability, which again arises from an RDI resonance with the epicyclic frequency, but when the dust is streaming vertically, \emph{viz.,} when it is settling towards the disk midplane. We will show the growth rates and fastest-growing wavelengths of the \si\ are orders-of-magnitude larger than the YG streaming instability for small grains.
	\item{\bf The ``High-$\mu$ Streaming Instability''} (App.~\ref{app: high mu streaming}): When $\mu>1$ (for horizontal streaming in the midplane), a new mode becomes unstable with faster growth rates than the midplane-epicyclic RDI, albeit at shorter wavelengths. While this is commonly also called the streaming instability, and was also  studied in \citet{2005ApJ...620..459Y} and subsequent works, we show it is a different instability (i.e., not an RDI) that is destabilized only if $\mu>1$.	
	\item{\bf The \BV\ RDI} (\S\ref{sub: BV RDI}): This is another new instability which arises from an RDI resonance with \BV\ oscillations, or gravity waves. This instability cannot occur in isolation in a disk with a standard stratification profile because the rotation  modifies \BV\ oscillations.  However, it may be important in other systems, since it  occurs generically when dust settles through a stratified atmosphere.  
	\item{\bf The Acoustic RDI} (\S\ref{sub: resonance hydro}): This is the RDI studied in \papertwo, which arises from  the resonance with sound waves in compressible gas. While we  briefly discuss its properties, we find its growth rates are uninteresting for the highly subsonic drift usually expected in protoplanetary disks.
	\item{\bf The Magnetosonic RDI} (\S\ref{sub: resonance MHD}): This RDI, introduced in \paperone\ and studied in detail in \citet{Hopkins:2018rdi}, arises from the resonance with magnetosonic waves. In ideal MHD it has growth rates that increase without bound at high $k$; however we show that the strong non-ideal MHD effects in the midplane of protoplanetary disks are usually expected to suppress the magnetosonic RDI's growth rate to values well below those of the \si. The magnetosonic RDI could nonetheless be relevant 
in well-ionized regions far above the midplane, for instance, in outflows and winds.
\end{itemize}

We emphasize that these instabilities are not in any way mutually exclusive. In fact, one can (and we do, for most cases) consider the full system including vertical, radial, and azimuthal drift velocities, vertical and radial stratification, epicyclic forces (centrifugal and Coriolis forces in the rotating frame), gas compressibility (acoustic waves), and non-ideal MHD (magnetic fields including the Hall effect, Ohmic resistivity, and ambipolar diffusion). In this general case {\em all} of the RDIs described above are present,  with different RDIs dominating at different wavenumbers and in different limits. The most important component of this joint analysis is in \S\ref{sub: BV epi RDI}, where we study the joint epicyclic--\BV\ RDI, finding that the buoyancy and compressibility  do not significantly modify the interesting properties of the \si.
For this reason, we will also refer to the epicyclic--\BV\ RDI as the ``\si'' in the discussion of \S\ref{sec: astrophysics}.


\section{Disk model}\label{sec: disk model}

In this section, we describe the basic disk model we use throughout to calculate RDI growth rates and properties. 
This includes the gas and dust equations, the equilibrium, and the relative
streaming velocity between the gas and the dust that arises due to the gas pressure support. 
A summary of important variables and their definitions is given in Table~\ref{tab:}.

We consider a fluid whose density $\rho$, bulk velocity $\bm{u}$, and pressure $P$, satisfy
\begin{gather}
\partial_{t}\,\rho + \nabla\cdot (\rho \bm{u})=0, \label{eq: gas NL rho}\\
\partial_{t}\bm{u} + \bm{u}\cdot \nabla\bm{u} = -\frac{\nabla P}{\rho}-\frac{\rho_{d}}{\rho}\frac{\bm{u}-\bm{v}}{t_{s}} + \bm{g},\label{eq: gas NL u}\\
\partial_{t}P + \bm{u}\cdot \nabla P + \gamma_{\mathrm{gas}} P\, \nabla\cdot \bm{u}=0.\label{eq: gas NL P}
\end{gather}
Here $\bm{g}$ is an external gravitational acceleration, $\gamma_{\mathrm{gas}}$ is the ratio of specific heats (we neglect heat fluxes, cooling etc. for simplicity),  and $\bm{v}$ and $\rho_{d}$ are the bulk velocity and (continuum) density of the dust. As a reasonable approximation for the linear regime \citep{doi:10.1146/annurev.fl.02.010170.002145,doi:10.1146/annurev.fl.15.010183.001401,2007ApJ...662..613Y,Jacquet:2011cy}  we take the dust to  be a pressureless fluid, which  satisfies,
\begin{gather}
\partial_{t}\,\rho_{d} + \nabla\cdot (\rho_{d}\bm{v}),\label{eq: dust NL rho}\\
\partial_{t} \bm{v}+ \bm{v}\cdot \nabla \bm{v} = - \frac{\bm{v}-\bm{u}}{t_{s}} + \bm{F}_{d},\label{eq: dust NL v}
\end{gather}
where $\bm{F}_{d}$ represents arbitrary additional external forces on the dust.
In equations \eqref{eq: gas NL rho}--\eqref{eq: dust NL v} the dust and gas 
are coupled by the  drag law, $\bm{F}_{\mathrm{drag}}\propto (\bm{v}-\bm{u})/t_{s}$, determined by the ``stopping time'' $t_{s}$. This can 
be a general function of fluid parameters ($\rho$, $P$) and relative drift speed ($|\bm{u}-\bm{v}|$) and is described in detail below (\S\ref{sub: stopping time}).  Equations~\eqref{eq: gas NL rho}--\eqref{eq: gas NL P} of course neglect 
many complexities of disk thermodynamics, which can cause other instabilities or oscillation modes (e.g., \citealt{Papaloizou:1985vf,1988ApJ...329..739R,2013PhRvL.111h4501M,2013MNRAS.435.2610N,0004-637X-788-1-21,2015MNRAS.450...21B}). Because the RDI formalism only requires information about the eigenmodes 
of the fluid and dust separately (see \S\ref{sec: resonance}), such effects, or more complex dust physics, could likely be included 
in future work if so desired. 
We have also neglected the influence of magnetic fields at this stage in the discussion; 
this will be addressed (along with nonideal magnetic effects) in \S\ref{sec: MHD and others}.

\begin{table}
\begin{center}
 \begin{tabular}{||c c c ||} 
 \hline
 Symbol &  Description and/or Definition & See \\ [0.5ex] 
 \hline\hline
   $r$, $\phi$, $z$  &  Radial, azimuthal, vertical: global coordinates & \S\ref{sub: local approximation} \\ 
 \hline
    $x$, $y$, $z$  & Radial, azimuthal, vertical: local coordinates  & \S\ref{sub: local approximation} \\
 \hline
      $f_{0}$  & Equilibrium value of variable $f$ & \S\ref{subsub:linearized} \\ 
 \hline
      $\delta f$  & Perturbation (linearization) of variable $f$ & \S\ref{subsub:linearized} \\ 
 \hline\hline
 $\Omega$, $U_{K}$  &   Keplerian rotation frequency, velocity & \S\ref{sub: local approximation} \\
   \hline
      $\mu$  &    Dust-to-gas continuum density ratio $\rho_{d,0}/\rho_{0}$ & \S\ref{subsub:linearized} \\ 
 \hline
   $\rho$, $\rho_{d}$  &    Gas, dust continuum density & \S\ref{sec: disk model} \\ 
\hline
  $\bm{u}$, $\bm{v}$  &    Gas, dust flow velocity &  \S\ref{sec: disk model} \\ 
 \hline
   $t_{s}$  &   Dust stopping time (drag time) & \S\ref{sub: stopping time}  \\ 
 \hline
  $\tau_{s}$  &   $t_{s}$ in disk units (Stokes number), $\tau_{s} = \Omega\,t_{s}$ & \S\ref{sub: stopping time}  \\ 
 \hline
   $\driftvel$  & Dust-gas drift, $\bm{v}_{0}-\bm{u}_{0}=(\driftvelx,\driftvely,\driftvelz)$ &  \S\ref{subsub: streaming} \\ 
 \hline
    $\driftvelhat$, $\driftvelmag$  & Direction, magnitude of drift, $\driftvel = \driftvelmag\driftvelhat$ &  \S\ref{subsub: streaming} \\ 
 \hline
    $R_{d}$, $\bar{\rho}_{d}$  & Dust grain radius, internal density & \S\ref{sub: stopping time} \\ 
 \hline
   $c_{s}$, $\gamma_{\mathrm{gas}}$ & Gas sound speed, adiabatic exponent & \S\ref{sec: disk model} \\ 
    \hline
  $h_{g}$  &   Disk scale height (of gas) & \S\ref{sub: local approximation}  \\ 
    \hline
 $\eta$  & Gas pressure support parameter, $\eta\sim (h_{g}/r)^{2}$  & \S\ref{sub: local approximation} \\ 
    \hline
    $P$, $T$, $S$  & Gas pressure, temperature, entropy & \S\ref{sub: linear stratified} \\ 
      \hline
    $L_{0} $, $\Lambda_{S}$ & Pressure, entropy stratification parameters & \S\ref{sub: linear stratified} \\ 
  \hline
    $N_{BV}$  &  \BV\ frequency of gas & \S\ref{sub: linear stratified} \\ 
 \hline
  $\zeta_{\rho}$, $\zeta_{P}$, $\zeta_{\bm{w}}$  &  Parameterizations of drag dependence on gas   &  \S\ref{sub: stopping time}\\ 
   \hline\hline
  $\bm{k}$  & Wavenumber of  mode $\bm{k}=(k_{x},0,k_{z})$ & \S\ref{subsub:linearized} \\ 
 \hline
 $\hat{\bm{k}}$, $k$  & Wavenumber direction, magnitude, $\bm{k}=k\hat{\bm{k}}$ & \S\ref{subsub:linearized} \\ 
 \hline
   $\theta_{k} $  & Mode angle in $r$-$z$ plane, $\tan^{-1}(k_{x}/k_{z})$ &  \S\ref{subsub:linearized} \\ 
 \hline
    $\Im(\omega)$  & Growth rate of RDI mode (frequency $\Re(\omega)$) & \S\ref{subsub:linearized}  \\ 
     \hline
    $\omega_{\mathcal{F}}(\bm{k})$  & Frequency of gas wave/oscillations & \S\ref{subsub: RDIs} \\ 
 \hline
     $\bm{k}_{\mathrm{res}}$  & Resonant wavenumber of RDI, $\bm{k}_{\mathrm{res}}\cdot\driftvel = \omega_{\mathcal{F}}$ &  \S\ref{subsub: RDIs} \\ 
 \hline
 \end{tabular}
\end{center}
\caption{Important symbols used throughout this article.}
\label{tab:}
\end{table}

\subsection{Local approximation}\label{sub: local approximation}

As standard in most previous works, to keep the analysis analytically feasible, we use a local approximation.
This involves expanding about a small patch of the disk that is corotating with the background Keplerian flow velocity, ${U}_{K} = \Omega(r) r $, where $\Omega(r)\propto r^{-3/2}$ is the angular rotation frequency and $r$ is the radial coordinate. 
This transformation modifies the gas and dust momentum equations (Eqs.~\eqref{eq: gas NL u} and \eqref{eq: dust NL v}) to,
\begin{equation}
\mathcal{D}_{t}\bm{u} + \bm{u}\cdot \nabla\bm{u} +2\Omega \,\hat{\bm{z}}\times \bm{u} = \frac{3}{2}\Omega\, u_{x}\,\hat{\bm{y}} -\frac{\nabla P}{\rho}-\frac{\rho_{d}}{\rho}\frac{\bm{u}-\bm{v}}{t_{s}} + \bm{g},\label{eq: gas NL u local kep}
\end{equation}
and 
\begin{equation}
\mathcal{D}_{t}\bm{v} + \bm{v}\cdot \nabla\bm{v} +2\Omega \, \hat{\bm{z}}\times \bm{v} = \frac{3}{2}\Omega \, v_{x}\, \hat{\bm{y}} -\frac{\bm{v}-\bm{u}}{t_{s}} +\bm{F}_{d},\label{eq: dust NL v local kep}
\end{equation}
respectively. Here, $\hat{\bm{x}}$, $\hat{\bm{y}}$, and $\hat{\bm{z}}$ are the local radial ($\hat{\bm{r}}$), azimuthal ($\hat{\bm{\phi}}$), and vertical ($\hat{\bm{z}}$)
directions respectively,  $\bm{u}$ and $\bm{v}$  now denote  the deviation from the background Keplerian shear flow $\bm{U}_{K} = -(3/2)\,\Omega\, x \, \hat{\bm{y}}$, and $\mathcal{D}_{t}\equiv \partial_{t} - (3/2)\, \Omega\, \partial_{y}$. The 
density and pressure equations in the local frame are simply Eqs.~\eqref{eq: gas NL rho}, \eqref{eq: gas NL P}, and \eqref{eq: dust NL rho} with $\partial_{t}$ replaced by $\mathcal{D}_{t}$.

We consider a thin disk, with (gas)  vertical scale height $h_{g}/r \sim c_{s}/U_{K}\ll 1$, where $c_{s}^{2} = \gamma_{\mathrm{gas}}P/\rho$ is the local 
sound speed in the gas. We shall study stability 
away from the midplane of the disk by simply specifying a gas  equilibrium scale height 
$\partial_{z}\ln P_{0}\sim h_{g}^{-1}$ (where $P_{0}$ is the equilibrium gas pressure), and working in a  local frame with  background gradients treated as constant (some subtleties and uncertainties regarding this approximation are discussed in \S\ref{subsub: local linear BV caution}).
In addition to the vertical stratification, the disk is radially stratified.
The most important effect of this radial stratification is to cause the gas (in the absence of dust) to rotate slightly 
more slowly than the local Keplerian velocity, with velocity difference (in the local frame of Eqs.~\ref{eq: gas NL u local kep}),
\begin{equation}
-\eta U_{K} \equiv u_{0,y}\approx \frac{\partial P_{0}}{\partial\ln r}\frac{1}{2\,\rho\, U_{K}}.\label{eq: eta definition}\end{equation}
The support parameter $\eta \sim c_{s}^{2}/U_{K}^{2}\sim (h_{g}/r)^{2}$ is small, 
of order $\eta\sim 10^{-3}$ for the commonly used Minimum Mass Solar Nebula (MMSN) model \citep{1977Ap&SS..51..153W,Chiang:2010ew} relevant to protoplanetary disks. We see that $\partial_{r}\ln P_{0} \sim \eta^{1/2}\partial_{z} \ln P_{0}$; i.e., the stratification in the radial direction is  small compared to 
that in the vertical direction. 

Throughout this work we shall use $\eta = 0.001$ for the purposes of plotting and simple estimates. 
In  the MMSN model of \citet{Chiang:2010ew}, $\eta \approx 8\times10^{-4} (r/\mathrm{AU})^{4/7}$,  and we see that $\eta=0.001$ at $r\approx 1.5 \mathrm{AU}$; however, since most results in this work are analytic, with $\eta $ as a free parameter, they are straightforward  to extend to other regions of the disk.

\subsection{Gas-dust drag}\label{sub: stopping time} 

The interaction of a particular grain species with the gas is determined by its 
stopping time $t_{s}$,  which is the characteristic time required for a dust particle to come to rest in the frame of the gas. The dependence  of $t_{s}$ on the gas density and relative streaming velocity $|\bm{u}-\bm{v}|$ is determined by the grain size $R_{d}$ and the gas mean free path $\lambda_{\mathrm{mfp}}$. If $R_{d}\lesssim 9\lambda_{\mathrm{mfp}}/4$ the grains are in 
the Epstein regime \citep{Epstein,1965MNRAS.130...63B,1979ApJ...231...77D}, with 
\begin{equation}
t_{s}(\rho, P,|\bm{u}-\bm{v}|) = \sqrt{\frac{\pi\gamma_{\mathrm{gas}}}{8}}\,\frac{ \bar{\rho}_{d}\,R_{d}}{ \rho\, c_{s} }\left( 1+a_{\gamma}\frac{|\bm{v}-\bm{u}|^{2}}{c_{s}^{2}}\right)^{-1/2},\label{eq: epstein drag}\end{equation}
where  $a_{\gamma} \equiv{9\pi\gamma_{\mathrm{gas}}}/{128}$, $R_{d}$ is the grain radius (assuming spherical particles), and $ \bar{\rho}_{d}$ is the solid density of grain material. It is worth noting that the sound speed $c_{s}$ in Eq.~\eqref{eq: epstein drag} is for perturbations with polytropic index $\gamma_{\mathrm{gas}}$, and is related to the background temperature $T$ through $ c_{s}^{2} = \gamma_{\mathrm{gas}} k_{B}T/m_{\mathrm{eff}}$, where $m_{\mathrm{eff}}$ is the mass of gas particles (and $k_{B}$ is Boltzmann's constant). 

If  $R_{d}\gtrsim 9\lambda_{\mathrm{mfp}}/4$ but $\mathrm{Re}_{\mathrm{d}}=R_{d}|\bm{u}-\bm{v}|/(\lambda_{\mathrm{mfp}}c_{s})\lesssim 1 $ ($\mathrm{Re}_{\mathrm{d}}$ is the Reynolds number of the flow over the dust), the grains are in the Stokes regime
and $t_{s}$ in Eq.~\eqref{eq: epstein drag} should be multiplied by $4R_{d}/(9\lambda_{\mathrm{mfp}})$. This gives 
\begin{equation}
t_{s}(\rho, P) \approx \frac{\sqrt{{2 \pi\gamma_{\mathrm{gas}}}}}{9}\,\frac{ \bar{\rho}_{d}\,R_{d}^{2}\,\sigma_{\mathrm{gas}}}{  c_{s} },\label{eq: stokes drag}\end{equation}
for the subsonic flow regime in which Stokes drag is relevant (here $\sigma_{\mathrm{gas}}$ is the gas collision cross section, $\lambda_{\mathrm{mfp}} = (\rho \,\sigma_{\mathrm{gas}})^{-1}$).
Yet larger grains, with $\mathrm{Re}_{\mathrm{d}}\gtrsim 1$, will create a turbulent wake and a simple drag law is no longer applicable.

For completeness, we note that over the range of densities, temperatures, and ionization fractions of disks, other dust-gas momentum exchange terms such as Lorentz forces on grains, Coulomb drag, and photo-electric or photo-desorption processes are sub-dominant to Epstein or Stokes drag by large factors (see \S\ref{sec: other effects} and \citealt{lee:dynamics.charged.dust.gmcs}, \papertwo\ for some further discussion).

 \subsection{Units}\label{subsub: units}

In what follows, we will usually quote timescales in units of the disk dynamical time $\Omega^{-1}$ and length scales in units of $\eta r$ (meaning that the characteristic velocity unit is $ \eta r\,\Omega = \eta U_{K}$). In these units the  {gas} scale-height is $h_{g}/r\sim \eta^{1/2}$, or $h_{g} \sim \eta^{-1/2} (\eta r)$, so a wavenumber $k\eta r \sim 1$ implies there are $\sim\! \eta^{-1/2}\gg 1$ wavelengths within a scale height.

We will also define the dimensionless stopping time or rotation Stokes number $\tau_{s} \equiv \Omega\, t_{s}$, which will be used as a proxy for dust particle size.  In general, the motion of particles with $\tau_{s}\ll 1$ is dominated by gas drag, while those with  $\tau_{s}\gg1 $ are 
weakly coupled to the gas and dominated by their Keplerian orbital motion. For reference, at $r\approx1.5 \mathrm{AU}$ within the MMSN model, a grain of density $\bar{\rho}_{d}\approx 1 \,\mathrm{g}\,\mathrm{cm}^{-3}$ with $\tau_{s}\approx 1$ has a size $R_{d}\approx 40\,\mathrm{cm}$. We focus on grains with $\tau_{s}\lesssim 1$ (the fluid approximation
may be questionable for grains much larger than this). For $r\gtrsim 5 \mathrm{AU}$ such grains are always 
in the Epstein
regime (Eq.~\eqref{eq: epstein drag}), but closer to the central protostar, the higher gas density suggests some of these 
grains are in the Stokes regime. For reference, 
using the MMSN values of \citet{Chiang:2010ew},
the boundary  between the Epstein and Stokes regimes 
occurs for grains of size $R_{d,\mathrm{bound}} \approx 1.1 (r/\mathrm{AU})^{39/14}\mathrm{cm}$, or 
 $\tau_{s,\mathrm{bound}}\approx1.3 \times 10^{-3} (r/\mathrm{AU})^{30/7}$. If $\tau_{s}<\tau_{s,\mathrm{bound}}$, grains are in the Epstein drag regime;  if $\tau_{s}>\tau_{s,\mathrm{bound}}$ grains are in the Stokes drag regime.
In practice, there are only minor differences between the
Epstein and Stokes regimes for the  instabilities we study (specifically, the $\zeta$ parameters; see discussion around Eq.~\eqref{eq: eta and zeta expressions} below). Thus,  keeping in mind that a single value of $\tau_{s}$ is relevant 
to grains across a range of  physical sizes, our results can be applied to
any region of the disk with only minor changes.

\subsection{Linearized system}\label{subsub:linearized}

Throughout the majority of this work, we consider only axisymmetric linear instabilities of the coupled dust-gas systems (Eqs.~\eqref{eq: gas NL rho}--\eqref{eq: dust NL v local kep}). We shall also assume an homogenous 
background equilibrium, or equivalently, linear instabilities with wavelengths that are short compared to 
the global scales of the system (WKBJ approximation). We thus decompose each variable in the
standard way,
\begin{equation}
\delta f(\bm{x},t) =  f_{0} + \delta f e^{i\bm{k}\cdot \bm{x}-i\omega t},\label{eq: linearization}
\end{equation}
where $f=\rho_{d}, \bm{v}, \rho,$ etc.,  $f_{0}$ denotes a spatial average in the local region being considered (i.e., the homogenous part of a variable), and $\bm{k}$ is the wavenumber. (Note that we  normalize the density and pressure perturbations to their equilibrium values, $\delta \rho/\rho_{0}$, $\delta P/P_{0}$, and $\delta \rho_{d}/\rho_{d0}$,  for notational convenience.)
Inserting Eq.~\eqref{eq: linearization} into Eqs.~\eqref{eq: gas NL rho}--\eqref{eq: dust NL v local kep} leads to an
eigenvalue problem  for the mode frequency $\omega$, where  $\Im(\omega)>0$ implies linear instability.
For notational purposes, it is helpful to define $k = |\bm{k}|$, $\hat{\bm{k}}\equiv \bm{k}/k$, and the standard polar
coordinate system $\bm{k}=(k_{x},0,k_{z}) = k\, (\sin\theta_{k},0,\cos\theta_{k})$ in the local frame. We study only axisymmetric perturbations, with $k_{y}=0$, because otherwise a time-dependent, or nonmodal, treatment is necessary \citep{Goldreich:1965tg,Trefethen:1993bb,Squire:2014es}. A correct
treatment of non-axisymmetric perturbations 
would significantly complicate the analysis and require extensions to the RDI formalism.

The relative dust-gas streaming velocity (see \S\ref{subsub: streaming} below) is a 
key parameter of our stability analysis due to the importance of resonance (\S\ref{sec: resonance}).
For the sake of clarity, in our  analyses we will usually work in a frame where the background dust and gas velocities are given by
\begin{equation}
 \bm{v}_{0} = \driftvel,\quad  \bm{u}_{0}=0,\label{eq: equilibrium drift is ws}\end{equation}
where $\driftvel$ is the relative streaming velocity, with magnitude $\driftvelmag=|\driftvel|$ and direction $\driftvelhat = \driftvel/\driftvelmag$ (see \S\ref{subsub: streaming} below).
Of course, the equilibrium gas  velocity in the Keplerian frame is not identically zero (even without dust; see Eq.~\eqref{eq: eta definition}); however, it is easily verified that the shift into the frame where $ \bm{u}_{0}=0$ simply shifts  $\omega$ to $\omega - \bm{k}\cdot  \bm{u}_{0}$ and does not change 
the stability properties of the system.\footnote{The same is generally true for dust in other physical situations. For instance, when dust is radiatively accelerated through 
a gas, accelerating the gas because of the drag force, the linear stability of the accelerating quasi-equilibrium can be computed from the relative drift velocity $\driftvel$, without
considering the global acceleration of the dust and gas together (see App.~B of \papertwo). The exception, of course, is when the frame's acceleration is not constant, for instance the rotating frame described above (\S\ref{sub: local approximation}).} The choice \eqref{eq: equilibrium drift is ws} allows for simpler discussion
and isolation of the key  physics of the problem, and thus will be used throughout most of this work.

The ratio of dust to gas mass density,
\begin{equation}
\mu \equiv \frac{ \rho_{d,0}}{ \rho_{0} },\end{equation}
 is another important parameter in the problem, as is the average stopping time $ t_{s0}=t_{s}( \rho_{0}, P_{0},\driftvelmag)$.
 Where necessary, we  parameterize the linear dependence of $t_{s}$ on the perturbed density, pressure, and velocity fields through 
 \begin{equation}
\frac{\delta t_{s}}{ t_{s0}} = -\zeta_{\rho}\frac{\delta \rho}{\rho_{0}} -\zeta_{P}\frac{\delta P}{P_{0}} - \zeta_{\bm{w}} \driftvelhat\cdot \frac{\delta \bm{v}-\delta \bm{u}}{\driftvelmag},\label{eq: ts linearization}\end{equation}
where $\zeta_{\rho}= -d\ln t_{s}/d\ln\rho$, $\zeta_{P}= -d\ln t_{s}/d\ln P$ etc.,  are parameters that depend on the equilibrium, the drag law, and 
$\driftvelmag$.  
For example, from the Epstein drag expression \eqref{eq: epstein drag}, one finds
 \begin{equation}
 \zeta_{\rho} = \frac{1+2 a_{\gamma} \bar{\driftvelmag}^{2}}{2 +2 a_{\gamma} \bar{\driftvelmag}^{2}},\quad \zeta_{P} = \frac{1}{2+2 a_{\gamma} \bar{\driftvelmag}^{2}},\quad\zeta_{\bm{w}} = \frac{a_{\gamma} \bar{\driftvelmag}^{2}}{1+ a_{\gamma} \bar{\driftvelmag}^{2}},\label{eq: eta and zeta expressions}
\end{equation}
where $\bar{\driftvelmag} = \driftvelmag/c_{s}$.
Aside from $\zeta_{\rho}$, the expressions for larger particles that are in the Stokes regime are generally similar,
\begin{equation}
\zeta_{\rho}\approx -\frac{1}{2} -\frac{d\ln \sigma_{\mathrm{gas}}}{d\ln \rho}+ \frac{d\ln \sigma_{\mathrm{gas}}}{d\ln T},\quad \zeta_{P}\approx \frac{1}{2}- \frac{d\ln \sigma_{\mathrm{gas}}}{d\ln T},\quad \zeta_{\bm{w}}\approx0,\label{eq: eta and zeta stokes}\end{equation}
depending on the form of $\sigma_{\mathrm{gas}}$ (e.g., for a neutral gas, this is simply constant and $\zeta_{\rho}\approx -1/2$).
Overall, we see that in both regimes
$|\zeta_{\rho}|\approx \zeta_{P} \approx 1/2$,
while $\zeta_{\bm{w}}\sim \driftvelmag^{2}/c_{s}^{2}\ll 1$ when $\driftvelmag\ll c_{s}$. Note that a constant $t_{s}$, which does not correspond to a physical drag law but is a common approximation in the literature, corresponds to $\zeta_{\rho}=\zeta_{P}=\zeta_{\bm{w}}=0$.

\subsection{Equilibrium dust-gas streaming velocity}\label{subsub: streaming}

While the radial pressure support of the gas causes it to rotate slightly slower than the Keplerian velocity, the dust component has no equivalent pressure support. Nonetheless, 
due to its drag interaction with the gas, the equilibrium dust orbits are also modified, causing a relative streaming 
velocity ($\driftvel$) between the dust and gas. This is the origin of the YG streaming instability and other RDIs studied here. 
Inserting the gas pressure support (Eq.~\eqref{eq: eta definition}) and dust-gas coupling into the local  equations (Eqs.~\eqref{eq: gas NL u local kep} and \eqref{eq: dust NL v local kep}, with Eqs.~\eqref{eq: gas NL rho}, \eqref{eq: gas NL P}, and \eqref{eq: dust NL rho}), one
solves for the equilibrium velocities of gas  and dust, obtaining (in the Keplerian frame): 
\begin{gather}
 \bm{u}_{0}= \frac{2\,\mu \, \tau_{s}\eta\, U_{K}}{(1+\mu)^{2}+\tau_{s}^{2}}\hat{\bm{x}} -\left(1+  \frac{\mu\, \tau_{s}^{2}}{(1+\mu)^{2}+\tau_{s}^{2}}\right)\frac{\eta\, U_{K}}{1+\mu}\,\hat{\bm{y}},\label{eq: full NSH drift U}\\
 \bm{v}_{0}= -\frac{2\, \tau_{s}\eta\, U_{K}}{(1+\mu)^{2}+\tau_{s}^{2}}\hat{\bm{x}} -\left(1-  \frac{ \tau_{s}^{2}}{(1+\mu)^{2}+\tau_{s}^{2}}\right)\frac{\eta\, U_{K}}{1+\mu}\,\hat{\bm{y}}\label{eq: full NSH drift V},
\end{gather}
which is known as  the Nakagawa-Sekiya-Hayashi (NSH) drift \citep{1986Icar...67..375N,Chiang:2010ew}.
Equations \eqref{eq: full NSH drift U}--\eqref{eq: full NSH drift V} lead to the relative streaming velocity
\begin{equation}
\driftvel = \bm{v}_{0} -  \bm{u}_{0}= -2\, \frac{\eta\, U_{K}(1+\mu)\,\tau_{s}}{(1+\mu)^{2}+\tau_{s}^{2}}\,\hat{\bm{x}}  + \frac{\eta\, U_{K}\tau_{s}^{2}}{(1+\mu)^{2}+\tau_{s}^{2}}\,\hat{\bm{y}}.\label{eq: NSH drift}
\end{equation}
We see that small, strongly coupled particles, with $\tau_{s}\lesssim 1$ (i.e., when the gas drag dominates the gravitational forces) drift predominantly inwards in the radial direction, while larger, weakly 
coupled particles with $\tau_{s}\gtrsim 1$ drift predominantly in the azimuthal direction.
The drift speed  peaks at $\driftvelmag\sim \eta U_{K}\sim \eta^{1/2}c_{s}$ for $\tau_{s}\gtrsim 1$, implying that this horizontal relative drift velocity is always much less than the  sound speed.

If grains are separated from the midplane of the disk---either during the early evolution phases, or if they are transiently thrown out of the midplane by turbulence \citep{2017arXiv171006007F} or other effects---there is also a vertical dust streaming  velocity that arises from the vertical gravity force.
Modeling the motion of grains as a damped harmonic oscillator caused by gas drag and the vertical gravity force $F_{\mathrm{grav}}\sim m_{d} h\Omega^{2}$ (for particles of mass $m_{d}$ at  height $h$), and assuming 
 that large particles start at height $\sim h_{g}$, one
finds \citep{Chiang:2010ew},
\begin{equation}
(\driftvelmag)_{\mathrm{settle}} \sim  \driftvelz \approx c_{s}\,\frac{\tau_{s}}{1+\tau_{s}}  \sim c_{s}\min (\tau_{s},1),\label{eq: vertical settling drift}
\end{equation}
which is $\driftvelz/(\eta U_{K})\sim \eta^{-1/2}\min (\tau_{s},1)$ in the disk units of \S\ref{subsub: units}. 
This form arises because the motion of small particles ($\tau_{s}\lesssim 1$) is dominated by gas drag as they 
sink towards the midplane, while large particles ($\tau_{s}\gtrsim 1$) oscillate about the midplane as a 
weakly damped harmonic oscillator.
Of course, such motion is transient---it stops once the particles settle near the midplane---and, for  larger particles, the drift velocity depends on their initial height above the disk midplane. It is, however, larger than the
NSH drift (Eq.~\eqref{eq: NSH drift}) by a factor $\sim\! \eta^{-1/2}$, because the 
radial stratification length is $\sim\! \eta^{-1/2}$ times larger than the vertical stratification length. 
For small particles, the settling time is $\Omega\, t_{\mathrm{settle}}\sim \Omega\,h_{g}/\driftvelz\sim  \tau_{s}^{-1}$.

\section{Resonance instabilities}\label{sec: resonance}

In this section, we outline the resonant drag instability formalism from \paperone, which will be used to study specific RDIs in \S\S\ref{sec: resonance epicyclic}--\ref{sec: MHD and others}. The method is based on matrix perturbation theory, and enables simple, accurate identification of instabilities and computation of their maximum growth rates, subject to certain assumptions (e.g.\ $\mu \ll 1$). Here we give a general overview and the relevant formulae, referring the reader to \paperone\ for more  discussion. 

We emphasize that all numerical results plotted in this paper are exact solutions to the {\em full} dispersion relation of the coupled gas-dust system---e.g.,\ the ninth-order coupled dust-gas equations for $\delta\rho,\delta P,\,\delta{\bf u},\,\delta\rho_{d},\,\delta{\bf v}$---without any assumption about small values of $\mu$ (although there are, of course,  approximations involved in writing down a local dispersion relation; see \S\ref{subsub: local linear BV caution}). However, these full dispersion relations are very complex and  uninformative to write down explicitly,  requiring numerical solutions
that do not yield any obvious criteria for the maximum growth rates as a function of wavenumber. In most figures, numerical results are plotted using 
thick, colored lines, while  analytic approximations, derived using the methods outlined in this section, are shown 
with black or gray crosses and/or dashed lines.  We see that our
simple analytic expressions provide excellent approximations to these exact results, even 
for values of $\mu$  approaching unity (i.e., the theory is generally accurate for $\mu\lesssim 1$).
Moreover they give us considerable additional intuition about the nature of the instabilities (see \S\ref{sub: toy model}).

 \subsection{Resonant drag instabilities}\label{subsub: RDIs}
 
 In \paperone\ we presented a simple algorithm for computing the fastest-growing instabilities 
 of coupled dust-gas fluid systems, such as Eqs.~\eqref{eq: gas NL rho}--\eqref{eq: dust NL v}, when $\mu\ll1$. The core  concept is that of a \emph{resonance} between 
 the dust and gas systems. We termed the resulting class of instabilities the ``Resonant Drag Instability,'' or RDI. 
The general idea---that resonances lead to instabilities---is related to a wide variety of well-known systems, for instance, shear-flow instabilities (e.g., \citealp{1994JFM...276..327B,2016ApJ...830...95U}), kinetic plasma instabilities and Landau damping (e.g., \citealp{1965pfig.book.....S,1967JPlPh...1...75K,2016PhPl...23g2111Z,Hopkins:2018rdi}), and a diverse 
array of industrial and engineering applications (e.g., \citealp{dobson2001strong,doi:10.1146/annurev.fluid.35.101101.161151}). 
The connection between  these more general applications and the formalism introduced in \paperone\  will be explored in detail in future work.

A linearized set of equations for a single Fourier mode can always be written as a linear eigenvalue equation with some eigenvalue $\omega$ and linear matrix operator $\mathbb{T}$. Specifically, the linearized version of Eqs.~\eqref{eq: gas NL rho}--\eqref{eq: dust NL v}
can be written in the form,
 \begin{equation}
\omega \left(\begin{array}{c}
\bm{a}\\ \bm{f}  
\end{array}\right)=\mathbb{T}_{0}\left(\begin{array}{c}
\bm{a}\\ \bm{f} 
\end{array}\right) + \mu\, \mathbb{T}^{(1)} \left(\begin{array}{c}
\bm{a}\\ \bm{f} 
\end{array}\right),\quad \mathbb{T}_{0}=\left(\begin{array}{cc}
\mathcal{A} & \mathcal{C} \\ {0} & \mathcal{F}
\end{array}\right),\label{eq: general linear blocks}
\end{equation}
where $\bm{a}$ and $\bm{f}$ denote the dust and fluid variables respectively; e.g.\ $\bm{a} = (\delta\rho_{d}/\rho_{d,0},\,\delta{\bf v})$, $\bm{f} = (\delta\rho/\rho_{0},\,\delta{\bf u},\,\delta P/P_{0},\,\delta {\bf B}/|{\bf B}_{0}|,\,...)$. 
Here $\mathbb{T} = \mathbb{T}_{0} + \mu\,\mathbb{T}^{(1)}$ is the full linearized system of equations, which can be decomposed into the block form of Eq.~\eqref{eq: general linear blocks}, in terms of $\mathcal{F}$, $\mathcal{A}$, $\mathcal{C}$, and $\mu\,\mathbb{T}^{(1)}$. The $\mathcal{F}$ operator contains the fluid (gas) equations of motion, in the absence of dust (i.e., Eqs.~\eqref{eq: gas NL rho}--\eqref{eq: gas NL P} with $\mu = 0$). Likewise,   $\mathcal{A}$ represents  the direct effect of a dust perturbation on the dust (Eqs.~\eqref{eq: dust NL rho}--\eqref{eq: dust NL v} including the equilibrium drift \eqref{eq: equilibrium drift is ws}). 
The  $\mathcal{C}$ matrix represents the coupling {from the gas onto the dust}; i.e.,  the dependence of  dust motion on the gas variables, encapsulated in the drag term, $\bm{u}/t_{s}$. The ``back-reaction'' {from the dust onto the gas}, $-(\rho_{d}/\rho)(\bm{u}-\bm{v})/t_{s}$, is separated here in the $\mu\,\mathbb{T}^{(1)}$ term. This separation is completely general: we decompose in this manner because, at small $\mu \ll 1$, the $\mu\,\mathbb{T}^{(1)}$ term can be treated using perturbation theory.

Because the dust is pressure free (its bulk velocity perturbation $\delta{\bf v}$ does not depend on density perturbations $\delta \rho_{d}$), and $t_{s}$ is independent of $\rho_{d}$, the terms $\mathcal{A}$ and ${C}$ must have the form,
\begin{equation}\mathcal{A} = 
\left(\begin{array}{cc}
\bm{k}\cdot \driftvel & \bm{k}^{T} \\ \bm{0} &(\bm{k}\cdot \driftvel)\,  {\mathbb{I}} + \mathcal{D}_{\bm{v}},\end{array}\right), \quad C = \left(\begin{array}{c} \bm{0} \\ \mathcal{C}_{\bm{v}},
\end{array}\right), \label{eq: dust matrix form}
\end{equation}
where $\mathbb{I}$ is the identity matrix.
The top row of $\mathcal{A}$ is simply the continuity equation, $\omega\,\delta\rho_{d} = \bm{k}\cdot\driftvel\,\delta\rho_{d} + \rho_{d,0}\,\bm{k}\cdot\delta\bm{v}$.
The operators $\mathcal{D}_{\bm{v}}$ and $\mathcal{C}_{\bm{v}}$ are determined by  $\bm{F}_{d}$ (assumed to depend only on $\delta \bm{v}$, and not $\rho_{d}$) and $t_{s}$ in Eq.~\eqref{eq: dust NL v}.  We will calculate their actual form, which depends on the specific problem, below (see Eqs.~\eqref{eq: D matrix epi} and \eqref{eq: T and C matrix epicyclic}).
Importantly, this  form of $\mathcal{A}$ {always}  has the eigenvalue $\omega_{0}=\bm{k}\cdot \driftvel$ (regardless of $\mathcal{D}_{\bm{v}}$). Physically, this represents a  density perturbation being advected by the background dust flow $\driftvel$.

\paperone\ showed that when $\mathcal{A}$ (the dust operator) and $\mathcal{F}$ (the gas operator, absent dust) both share an eigenvalue $\omega_{0}$---i.e.,\ when there is a \emph{resonance} between the dust and gas systems---the linear system (Eq.~\eqref{eq: general linear blocks})  is generically unstable to an RDI, at any finite $\mu\ll1$. Noting that $\bm{k}\cdot \driftvel$ is always an eigenvalue of $\mathcal{A}$, we see that this resonance occurs when $\bm{k}\cdot \driftvel = \omega_{0} = \omega_{\mathcal{F}}(\bm{k})$, where $\omega_{\mathcal{F}}(\bm{k})$ is any eigenvalue of $\mathcal{F}$; i.e.,\ any linear oscillation frequency, or normal mode, of the gas {\em without} dust. 

More specifically, this result comes from applying perturbation theory in  $\mu \ll 1$ to Eq.~\eqref{eq: general linear blocks}. One finds that the perturbation $\mu\, \mathbb{T}^{(1)}$ splits $\omega_{0}$, which is a degenerate\footnote{As shown in \paperone, $\omega_{0}$ is not just degenerate, but also \emph{defective}, meaning there is  only one associated eigenvalue of $\mathbb{T}_{0}$. This is 
the cause of the $\omega^{(1)}\sim \mu^{1/2}$ (rather than  $\omega^{(1)}\sim \mu$) scaling in Eq.~\eqref{eq: dust eval pert}.} eigenvalue of $\mathbb{T}_{0}$, into two eigenvalues, with the   
 lowest-order (in $\mu$) correction,
\begin{equation}
\omega = \omega_{0}+\omega^{(1)} = \omega_{0}\pm i\, \mu^{1/2 }\left[ (\bm{\xi}_{\mathcal{F}}^{L} \mathcal{T}^{(1)}_{\rho_{d}})\,(\bm{k}^{T}\mathcal{D}_{\bm{v}}^{-1}\mathcal{C}_{\bm{v}}\,\bm{\xi}_{\mathcal{F}}^{R})\right]^{1/2}.\label{eq: dust eval pert}
\end{equation}
Here $\mathcal{T}^{(1)}_{\rho_{d}}$ is the left-most column vector of the bottom-left block of $\mathbb{T}^{(1)}$, which physically  represents how the perturbed gas variables $\bm{f}=(\delta\rho/\rho_{0},\,\delta{\bf u},\,...)$ depend on dust density perturbations $\delta\rho_{d}$.\footnote{For example, if the coupling of dust onto gas, $\mu\,\mathbb{T}^{(1)}$ takes the form of drag back-reaction, $\partial {\bm u}/\partial t = (\rho_{d}/\rho)\,({\bm v}-{\bm u})/t_{s} + ...$, then the linear perturbation of the gas from  $\delta \rho_{d}$ (i.e., the $\mathcal{T}^{(1)}_{\rho_{d}}$ part of $\mathbb{T}^{(1)}$) is $-i\,\omega\, \delta{\bm u} = (\delta\rho_{d}/\rho_{0})\,(\bm{v}_{0}-\bm{u}_{0})/t_{s,0} +\cdots= \mu\,(\delta\rho_{d}/\rho_{d0})\,\driftvel/t_{s,0}+\cdots$. Thus, if we consider, for example,  the gas variables $\bm{f}=(\delta\rho/\rho_{0},\,\delta{\bm u}_{\bot},\,\delta{\bm u}_{\|}$)---where $\delta{\bm u}_{\|}$ is the component of $\delta\bm{u}$ parallel to $\driftvel$ and $\delta{\bm u}_{\bot}$ is perpendicular---then we obtain $\mathcal{T}^{(1)}_{\rho_{d}} = (0,\,0,\,i\,\driftvelmag / t_{s,0}$).}
The  symbols $\bm{\xi}^{R}_{\mathcal{F}}$ and $\bm{\xi}^{L}_{\mathcal{F}}$ denote the right and left eigenvectors of $\mathcal{F}$, which are defined by  $(\mathcal{F}-\omega_{0}\,{\mathbb{I}})\,\bm{\xi}^{R}_{\mathcal{F}} = \bm{0}$ and $\bm{\xi}^{L}_{\mathcal{F}}\,(\mathcal{F}-\omega_{0}\,\mathbb{I}) = \bm{0}$, with the normalization constraint $\bm{\xi}^{L}_{\mathcal{F}}\,\bm{\xi}^{R}_{\mathcal{F}}=1$. Physically, these determine  the structure of the fluid modes   that resonate with the dust motion.

Equation \eqref{eq: dust eval pert} has several important consequences. First, we see that the only way to \emph{not} get an instability is if the term in square brackets in Eq.~\eqref{eq: dust eval pert} is purely real and negative. Because the individual matrices and vectors, $\bm{\xi}_{\mathcal{F}}$, $\mathcal{T}^{(1)}_{\rho_{d}}$, $\mathcal{D}_{\bm{v}}$ etc., are generically complex valued (see, for example, Eqs.~\eqref{eq: T and C matrix epicyclic}--\eqref{eq: epicyclic evalues and modes} below),   this implies that resonances generically cause instabilities. Second, we see that $\omega^{(1)}$ scales as $\mathcal{O}(\mu^{1/2})$, rather than the usual perturbation theory expectation $\mathcal{O}(\mu)$. This implies that when $\mu\ll1$, resonant instabilities
will grow faster than instabilities at other $\bm{k}$, $\driftvel$, etc., and will thus  (presumably) be the most dynamically important. Third, Eq.~\eqref{eq: dust eval pert} is often much simpler to evaluate 
than an expansion of the dispersion relation, and can thus significantly decrease the algebraic complexity of the analysis  for the relevant ($\mu\ll1$) regime. Note that in practice (see, e.g., Fig.~\ref{fig: 2D epicyclic from YG}) we find that the dominance of 
the resonant wavenumber, and the results of Eq.~\eqref{eq: dust eval pert}, are generally valid for even  relatively large $\mu\lesssim1$, as often occurs in perturbation theories.

\subsection{How to find an instability}\label{sub:How to find an instability}

Practically speaking, Eq.~\eqref{eq: dust eval pert} gives us a  simple algorithm for finding the most-unstable wavenumbers of dust-gas streaming instabilities (RDIs) and calculating their growth rates.   The steps are: 
\begin{enumerate}
\item Choose a wave in the fluid system of interest and calculate its frequency $\omega_{0} = \omega_{\mathcal{F}}(\bm{k})$, as 
well as the corresponding left and right eigenvectors, $\bm{\xi}^{L}_{\mathcal{F}}$ and $\bm{\xi}^{R}_{\mathcal{F}}$.
\item A resonance occurs when the dust streaming frequency matches $\omega_{\mathcal{F}}$, \emph{viz.}, when \begin{equation}
\bm{k}\cdot \driftvel = \omega_{\mathcal{F}}(\bm{k}).\label{eq: resonant condition}\end{equation}
Because $\omega^{(1)}\sim \mathcal{O}(\mu^{1/2})$ at resonant wavenumbers, whereas $\omega^{(1)}\sim \mathcal{O}(\mu)$ at all other wavenumbers,  the solution of Eq.~\eqref{eq: resonant condition} automatically 
tells us what wavenumbers $\bm{k}$  have the fastest growth rates at $\mu\ll 1$, unless  $\omega^{(1)}$ is real or zero. 
We denote this resonant wavenumber $\bm{k}_{\mathrm{res}}$
\item Insert $\bm{k}_{\mathrm{res}}$ and the coupling terms $\mathcal{C}_{\bm{v}}$ and $\mathcal{T}^{(1)}_{\rho_{d}}$ into Eq.~\eqref{eq: dust eval pert}, to confirm that the system is unstable at the resonant wavenumber \eqref{eq: resonant condition}, and obtain the growth rates of the RDI. 
\end{enumerate}
\vspace{-1.5ex}

\begin{figure}
\begin{center}
\vspace{-0.15cm}
\hspace{-0.2cm}\includegraphics[width=0.99\columnwidth]{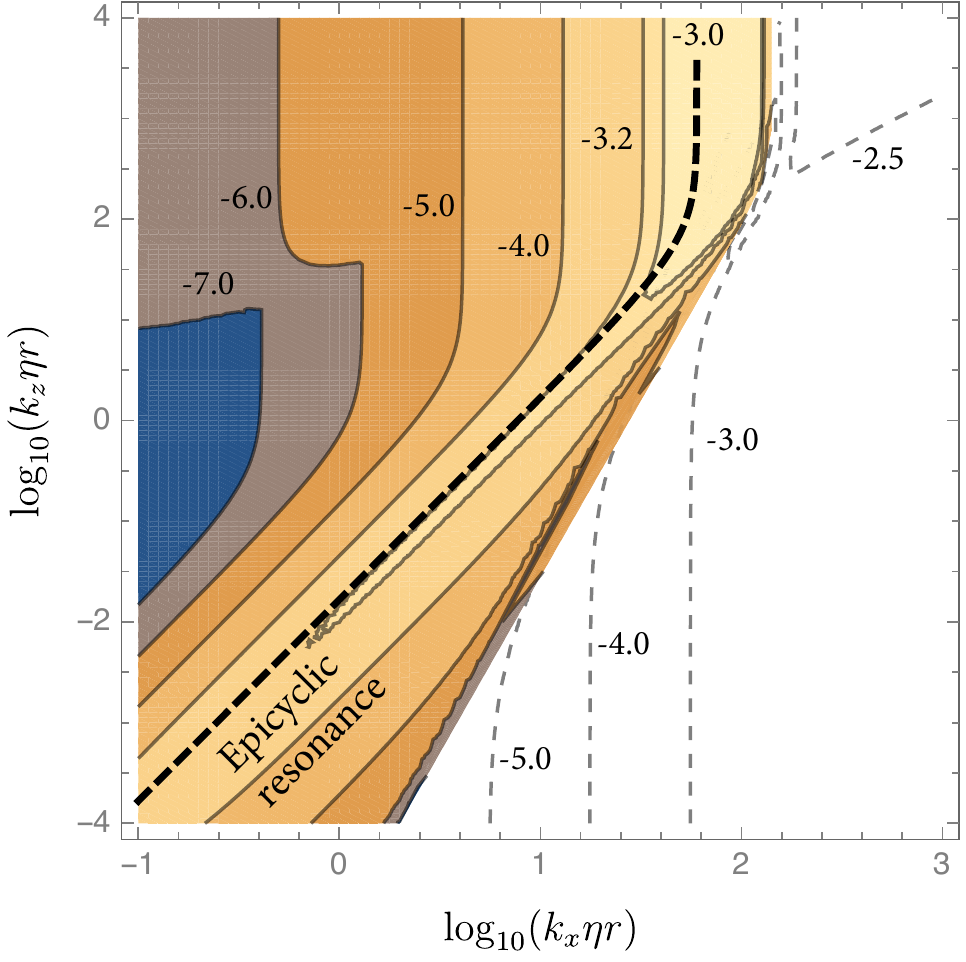}
\vspace{-0.25cm}
\caption{Contours of the growth rate of the YG streaming instability (epicyclic RDI with NSH drift velocities) as a function 
of  the radial ($k_{x}$) and vertical ($k_{z}$) wavenumbers (in units of  $\eta r$, where $\eta\sim(h_{g}/r)^{2}$ at radius $r$), calculated from  numerical solutions of the full coupled dust-gas dispersion relation. The parameters and range of this plot are identical to Fig.~2 of \citet{2005ApJ...620..459Y}, with stopping time/Stokes number $\tau_{s}= t_{s}\,\Omega =0.01$, dust-to-gas ratio $\mu = \rho_{d}/\rho_{\rm gas} =0.2$, and the NSH expressions (Eq.~\eqref{eq: NSH drift}) for 
dust drift velocities. Colored regions and solid contours indicate regions of instability ($\Im(\omega)>0$), dashed contours show stable regions ($\Im(\omega)<0$), and the contour labels indicate $\log_{10}[|\Im(\omega)|/\Omega]$. The thick dashed 
line is the epicyclic resonance line, Eq.~\eqref{eq: NSH epi resonant condition}; i.e.,  those wavenumbers $\bm{k}$ that  satisfy $\bm{k}\cdot\driftvel = \hat{k}_{z}\Omega$, where the drift velocity $\driftvel$ (projected along $\hat{\bm{k}}$) is resonant with the phase velocity of epicyclic oscillations in the gas. This predicts  the fastest-growing modes nearly perfectly, even at this relatively high $\mu$.}
\label{fig: 2D epicyclic from YG}
\end{center}
\vspace{-0.25cm}
\end{figure}

This paper is simply an application of this algorithm to waves and dust streaming motions of interest in protoplanetary disks. 
Before getting lost in the complexity of a full analysis, let us walk through  a couple of examples:
\vspace{-1.5ex}
\begin{description}
\item[\textbf{Sound waves:}] As studied in detail in \papertwo, one of the simplest choices is 
to take the fluid wave as a sound wave in a neutral fluid. Sound waves   satisfy $\omega_{\mathcal{F}}(\bm{k}) = \pm c_{s}\,k$, so the resonance condition is simply $\hat{\bm{k}}\cdot \driftvel = \pm c_{s}$. Taking, for simplicity, $\driftvelhat=\hat{\bm{z}}$, this becomes  $\hat{\bm{k}}\cdot \driftvelhat =\cos\theta_{k}= \pm c_{s}/\driftvelmag$. It is thus possible to find a resonant mode for any $\driftvelmag> c_{s}$, and 
the particular mode angle is resonant for all $k$. Application of Eq.~\eqref{eq: dust eval pert} shows that $\Im(\omega)$ continues to grow without bound as $k\rightarrow \infty$, and analysis of the full dispersion relation 
shows that, while a wide variety of modes are unstable, those at the resonant angle are the fastest growing (by a large margin).

\item[\textbf{Epicyclic oscillations:}] Axisymmetric epicyclic oscillations, which will be treated in detail in \S\ref{sec: resonance epicyclic}, satisfy $\omega_{\mathcal{F}}(\bm{k})=\pm \hat{k}_{z}\, \Omega=\pm \cos\theta_{k} \,\Omega$. For some chosen mode 
angle, the resonant wavenumber is $k_{\mathrm{res}} = \Omega \cos\theta_{k}/(\hat{\bm{k}}\cdot \driftvel)$. Thus, 
we expect that $\Im(\omega)$ will peak at some particular $k=k_{\mathrm{res}}$, which depends on $\driftvel$ and the 
chosen mode angle $\theta_{k}$.  The fastest growing wavenumbers will thus trace the contour $k = \Omega \cos\theta_{k}/(\hat{\bm{k}}\cdot \driftvel)$ in $(k_{x},k_{z})$ space, which indeed occurs (see  Fig.~\ref{fig: 2D epicyclic from YG}). With little algebraic effort, Eq.~\eqref{eq: dust eval pert} yields the growth rate of the instability at these particular (fastest-growing) wavenumbers. Note that the RDI analysis, as formulated, can only apply to axisymmetric modes because of the background shear (see \S\ref{subsub:linearized}).
\end{description}
\vspace{-1.5ex}

Because all RDIs arise from the resonance with the dust density perturbation, we know that such instabilities act to clump grains, and thus may be generically of interest to the planetesimal formation process.In this work, we focus on the epicyclic RDI (streaming instability; \S\ref{sec: resonance epicyclic}) and the effects
of gas stratification (\S\ref{sec: adding stratification}), which can also cause a \BV\ RDI. 
We shall also briefly discuss MHD-related RDIs, including the resonance with slow/fast waves and the
Whistler/Alfv\'en RDI in Hall MHD, in \S\ref{sec: MHD and others}.

 Finally, we note that the formula \eqref{eq: dust eval pert} is only valid in the regime when $\mathcal{D}_{\bm{v}}$ is not dominated by $\bm{k}^{T}$ in Eq.~\eqref{eq: dust matrix form}; otherwise the RDI is still present (with the same resonance condition and wavenumbers) but the expression for the growth rate is slightly different (see \paperone\ and \papertwo). Because this condition is always satisfied for the Epicyclic RDI and \BV\ RDIs in the regimes of interest in this work, we will not derive these alternative expressions here.

\section{Epicyclic RDI (Streaming Instability)}\label{sec: resonance epicyclic}

Our first application of the RDI theory from \S\ref{sec: resonance} is to the streaming instability \citep{2005ApJ...620..459Y}. This results from the resonance between streaming dust and  epicyclic oscillations of the gas and could thus be termed the ``epicyclic RDI'' within our nomenclature. The streaming instability has been studied extensively in recent years, both in the linear \citep{2005ApJ...620..459Y,2007ApJ...662..613Y,Jacquet:2011cy,2013MNRAS.434.1460K,0004-637X-817-2-140} and nonlinear regimes (e.g., \citealp{Johansen:2009ih,Bai:2010gm,Johansen:2015he,2016ApJ...822...55S,Schafer:2017hu}). 
However, there are several features of our analysis that are (so far as we are aware)  novel.
Firstly,  the 
 origin of the standard YG streaming instability as a resonance between dust streaming and gas epicycles has not been recognized previously, although other interesting aspects of its physical mechanism have been discussed in various of works (see, e.g., \citealt{Chiang:2010ew,Jacquet:2011cy} as well as  \citealt{2000Icar..148..537G} for more general discussion 
 of secular dust-gas instability).
 Secondly,  we know of no previous works that give simple closed-form expressions for its growth rate with a clear range of validity, 
which may be important for constructing simplified models and general understanding of the instability.
Thirdly, and most importantly, we include in our analysis the  vertical streaming motion, or settling, of dust grains.
We find that this  increases the growth rate of the instability dramatically for small grains, 
and, given it differs in character from the YG streaming instability, we  term this  the disk  ``\si.''

In this section, we treat the low-metallicity $\mu<1$ limit, when Eq.~\eqref{eq: dust eval pert} is applicable. In App.~\ref{app: high mu streaming}, we derive analytic expressions for growth rates at $\mu> 1$, when there is no longer a clear concept of resonance and the instability changes character. We also give a brief discussion of the mechanism for this instability and its necessary ingredients in App.~\ref{appsub: high mu mechanism}; however, 
given our focus on RDIs in this work, our analysis is somewhat less detailed than that given here for the $\mu<1$  instability.

\subsection{General derivation}\label{subsub: general epi derivation}

As in \citet{2005ApJ...620..459Y}, we take the gas to be incompressible at this stage; the compressible (and stratified) case will be treated below (\S\ref{sub: BV epi RDI}). Noting that the streaming velocities 
of interest (Eq.~\eqref{eq: NSH drift}) are highly subsonic, we also neglect the velocity dependence of $t_{s}$ in the dust and gas drag,\footnote{The velocity dependence of $t_{s}$ can easily be accounted for if so desired, but the effect on growth rates is very minor and not worth the added complexity.} which amounts to setting $\zeta_{\bm{w}}=0$. Further, because $\delta \rho =0$
(the gas is incompressible), the dependence of $t_{s}$ on $\rho$, which was parameterized through $\zeta_{\rho}$ in Eq.~\eqref{eq: ts linearization}, has no effect. 
The linearized dust equations are then given by Eq.~\eqref{eq: dust matrix form} with 
\begin{equation}
\mathcal{D}_{\bm{v}} = -i\,\Omega\left(\begin{array}{ccc}
\tau_{s}^{-1} & -2 & 0 \\ 
1/2 & \tau_{s}^{-1}  & 0 \\ 
0 & 0 & \tau_{s}^{-1} 
\end{array}\right).\label{eq: D matrix epi}
\end{equation}
We use the vorticity variables,
\begin{equation}
\varpi_{x}\equiv ik_{y}u_{z}- i k_{z}u_{y}=- i k_{z}u_{y}, \quad \varpi_{y} \equiv i k_{z}u_{x}-i k_{x}u_{z},\end{equation}
to enforce incompressibility, which impies $\mathcal{F}$ operates on $\bm{f}=(\varpi_{x},\,\varpi_{y})$ instead of $(\delta\rho/\rho_{0},\,\delta{\bf u},\,...)$.
The linearized gas equations are then
\begin{equation}
\mathcal{F}=i\,\Omega\left(\begin{array}{cc}
0 & {\hat{k}_{z}^{2}}/2\\ -2 & 0
\end{array}\right),\label{eq: gas F epicyclic}
\end{equation}
while the coupling terms are,
\begin{equation}
\mathcal{T}^{(1)}_{\rho_{d}} = \frac{\Omega}{\tau_{s}}\left(\begin{array}{c}
k_{z} \driftvely\\ k_{x}\driftvelz-k_{z}\driftvelx
\end{array}\right),\quad\mathcal{C}_{\bm{v}} = -\frac{\Omega}{k^{2}\tau_{s}} \left(\begin{array}{cc}
0 & -k_{z} \\  {k^{2}}/{k_{z}} & 0 \\ 0 & k_{x} 
\end{array}\right).\label{eq: T and C matrix epicyclic}
\end{equation}
The gas eigenmodes are epicyclic oscillations 
with
\begin{equation}
\omega_{\mathcal{F}} =\pm\omega_{\mathrm{epi}}= \pm \hat{k}_{z}\Omega,\quad \bm{\xi}^{R}_{\mathcal{F}} = 
\left(\begin{array}{c}
\pm i\,\hat{k}_{z}/2\\ 1
\end{array}\right),\quad
\bm{\xi}^{L}_{\mathcal{F}} = 
\left(
\mp i\, \frac{1}{\hat{k}_{z}}\,\, \quad \frac{1}{2}
\right).\label{eq: epicyclic evalues and modes}
\end{equation}
From Eq.~\eqref{eq: epicyclic evalues and modes}, we see that the condition 
for resonance is
\begin{equation}
\bm{k}\cdot \driftvel=\pm\hat{k}_{z}\Omega\quad \mathrm{or}\quad k_{\mathrm{res}} = \pm\frac{\hat{k}_{z}}{\hat{\bm{k}}\cdot \driftvel}\Omega,\label{eq: epicyclic resonance condition}\end{equation}
which sets the magnitude of the resonant wavenumber $k_{\mathrm{res}}$ for a chosen $\hat{k}_{x}$ and $\hat{k}_{z}$ (or equivalently,  mode angle $\theta_{k}$).

\begin{figure}
\vspace{-0.15cm}
\hspace{-0.25cm}\plotonesize{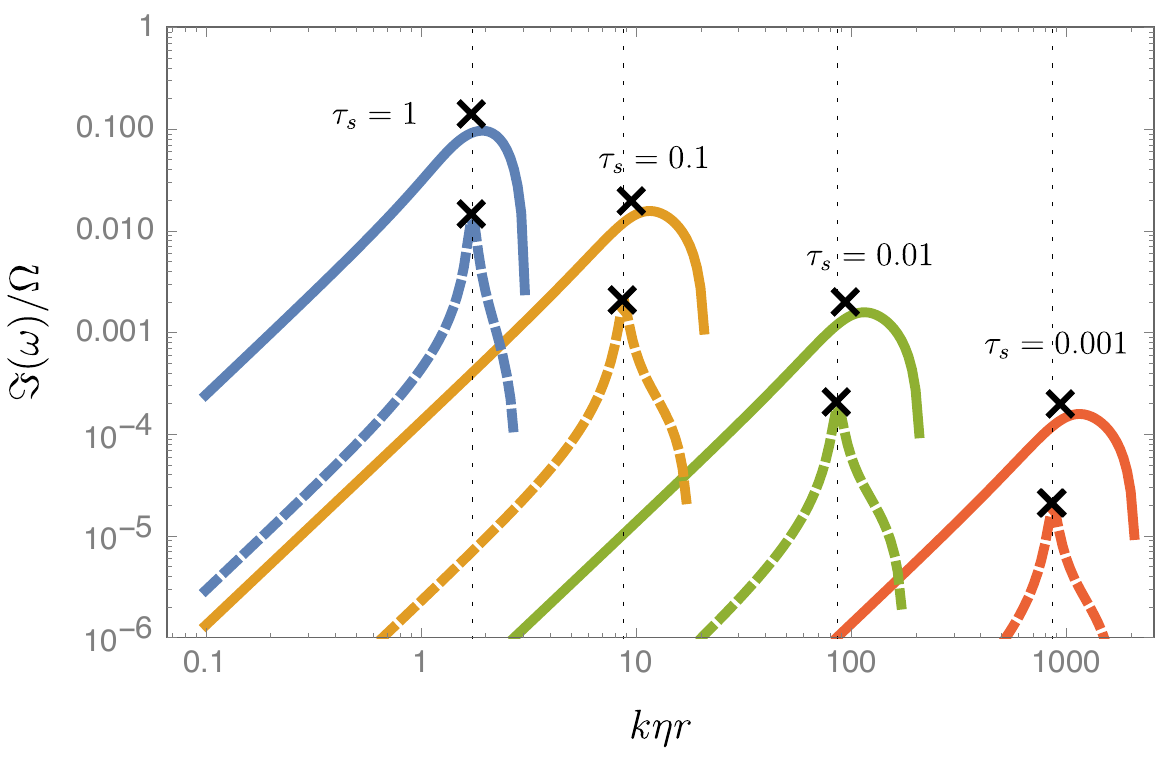}{1.0}
\vspace{-0.25cm}
\caption{Normalized growth rate $\Im(\omega)/\Omega$  of the YG streaming instability (epicyclic RDI) as a function 
of $k\eta r$, for a variety of physically interesting $\tau_{s}$  as
labeled on the figure (blue, $\tau_{s}=1$; yellow, $\tau_{s}=0.1$; green, $\tau_{s}=0.01$; red, $\tau_{s}=0.001$). 
In each case, solid curves show the numerically calculated growth rate at $\mu=0.1$ and dashed curves 
show the growth rate at $\mu=0.001$. We set the mode angle at $\theta_{k}=30^{\circ}$. The black crosses show our simple analytic expression for the maximum growth rates (Eq.~\eqref{eq: epicyclic RDI growth neg}) corresponding to each curve, while the vertical 
dotted lines show the resonant $k$ from Eq.~\eqref{eq: epicyclic resonance condition} in the $\mu\rightarrow0$ limit (there is a minor $\mu$ dependence in $\driftvel$; see Eq.~\eqref{eq: NSH drift}). Growth rates ($\Im(\omega)$) and characteristic wavelengths ($\sim k^{-1}$) of the instability scale proportionally to $ \tau_{s}$. 
We see very good agreement between the analytic predictions and the full numerical result, although 
there are minor discrepancies for larger $\mu$, as might be expected. 
}
\label{fig: epicyclic 1D}
\vspace{-0.25cm}
\end{figure}

We can then use Eq.~\eqref{eq: dust eval pert} to calculate the growth 
rate of resonant modes. For resonance with the positive frequency mode ($\omega_{\mathcal{F}}=\hat{k}_{z}\,\Omega$), a straightforward calculation gives $\omega = \hat{k}_{z}\,\Omega + \omega^{(1)}$ with 
\begin{equation}
\omega^{(1)} \approx \pm\left(\frac{\mu\, \Omega\, k_{\mathrm{res}}}{1+\tau_{s}^{2}}\right)^{1/2}\left[\frac{\hat{k}_{x}}{2}\left(2i \driftvely + \hat{k}_{z}\driftvelx - \hat{k}_{x}\driftvelz\right)\left(1- i\hat{k}_{z}\tau_{s}\right)\right]^{1/2}.\label{eq: epicyclic RDI growth pos}
\end{equation}
With the negative frequency mode ($\omega = -\hat{k}_{z}\,\Omega + \omega^{(1)}$), the  frequency perturbation is
\begin{equation}
\omega^{(1)} \approx \pm\left(\frac{\mu\, \Omega\, k_{\mathrm{res}}}{1+\tau_{s}^{2}}\right)^{1/2}\left[\frac{\hat{k}_{x}}{2}\left(2i \driftvely - \hat{k}_{z}\driftvelx + \hat{k}_{x}\driftvelz\right)\left(1+i\hat{k}_{x}\tau_{s}\right)\right]^{1/2}.\label{eq: epicyclic RDI growth neg}
\end{equation}
In Eqs.~\eqref{eq: epicyclic RDI growth pos} and \eqref{eq: epicyclic RDI growth neg}, $k_{\mathrm{res}}$ should be inserted from the resonant condition \eqref{eq: epicyclic resonance condition}, which  varies with the chosen $\hat{k}_{x}$ and $\hat{k}_{z}$.

\subsection{NSH drift velocities: the YG streaming instability}\label{subsub: epicyclic horizontal streaming}

Here, we derive growth rates and properties of the standard YG streaming instability (at $\mu\lesssim1$), which 
results from inserting the NSH drift velocities \eqref{eq: NSH drift}  into equations~\eqref{eq: epicyclic RDI growth pos} or \eqref{eq: epicyclic RDI growth neg}.
The resonance condition \eqref{eq: epicyclic resonance condition} depends only on $\driftvelx$ because $k_{y}=0$, and is 
\begin{equation}
k_{\mathrm{res}} = \left| \frac{\hat{k}_{z}}{\hat{k}_{x} \driftvelx}\right| \Omega= \left| \frac{\hat{k}_{z}}{\hat{k}_{x}}\right|  \frac{(1+\mu)^{2}+\tau_{s}^{2}}{2(1+\mu)\tau_{s}}(\eta r)^{-1}\approx \left| \frac{\hat{k}_{z}}{\hat{k}_{x}}\right|  \frac{1}{2\tau_{s}}(\eta r)^{-1}, \label{eq: NSH epi resonant condition}
\end{equation}
where the latter approximate equality assumes $\mu \ll 1$, $\tau_{s} \ll 1$. 
In Fig.~\ref{fig: 2D epicyclic from YG}, which is a reproduction of Fig.~2a from \citet{2005ApJ...620..459Y}, we overlay this resonance condition on a contour plot of exact numerical solutions of the full $6^{\mathrm{th}}$-order coupled dust-gas dispersion relation. As expected from the general arguments put forth in \S\ref{subsub: RDIs}, the resonance condition, Eq.~\eqref{eq: NSH epi resonant condition}, nicely predicts the wavenumbers of the fastest growing modes.

In Fig.~\ref{fig: epicyclic 1D}, we compare the analytic prediction, Eq.~\eqref{eq: epicyclic RDI growth neg}  
to numerical  solutions of
the full dispersion relation, for a variety of $\tau_{s}$ and $\mu$ (we take $k_{x}/k_{z}>0$, meaning the resonance is with the negative frequency epicycle). The analytic 
result, shown with black crosses, predicts the maximum growth rate very accurately at $\mu=0.001$, although there are some minor discrepancies at $\mu=0.1$ (since Eq.~\eqref{eq: epicyclic RDI growth neg} is a leading-order expression for low-$\mu$). Growth rates at larger values of $\tau_{s}$ are also well captured by Eqs.~\eqref{eq: epicyclic RDI growth pos}--\eqref{eq: epicyclic RDI growth neg}, although the relative errors  increase somewhat (for the same $\mu$) because various terms in the matrices (Eqs.~\eqref{eq: D matrix epi}--\eqref{eq: T and C matrix epicyclic}) become small compared to $k$.

A simple expression for the growth rate when $\tau_{s}\lesssim1$ is obtained by
inserting $\driftvel$ (from Eq.~\eqref{eq: NSH drift}) and  $k$ (from Eq.~\eqref{eq: NSH epi resonant condition}) into Eq.~\eqref{eq: epicyclic RDI growth pos} or Eq.~\eqref{eq: epicyclic RDI growth neg},  and expanding   in  $\tau_{s}\ll 1$. This yields,
\begin{equation}
\frac{\omega}{\Omega}\approx \pm\hat{k}_{z} \left[1\mp \left(\frac{\mu }{2}\right)^{1/2}\right] \pm i \sqrt{\frac{\mu}{8}}\,( 2\,\hat{k}_{z}^{2}+\hat{k}_{x}^{2})\,\tau_{s} + \mathcal{O}(\tau_{s}^{2},\mu),\label{eq: epicyclic RDI growth small tau}\end{equation}
which shows the linear scaling of the maximum growth rate with $\tau_{s}$ \citep{2005ApJ...620..459Y}.
We also see that the growth rate is largest for modes with $k_{z}\gg k_{x}$.

\subsection{Including the vertical settling drift: the disk \SI}\label{subsub: epicyclic vertical streaming}

In this section, we also include the vertical settling drift of dust grains (Eq.~\eqref{eq: vertical settling drift}) in our calculation of the
epicyclic RDI, yielding the disk ``\si'' (or more formally, the vertical-epicyclic RDI). Although
this drift is necessarily transient---it halts once the particles reach the midplane---we see 
that it causes very significant changes to the dispersion relation, increasing the growth rate for small dust particles by orders of magnitude. 
Further,  for modes at a particular ``double-resonant'' angle $\theta_{k}=\theta_{\mathrm{res}}$ where $\bm{k}\cdot \driftvel=0$, the growth rate of the instability increases \emph{without bound} with $k$,  surpassing $\Im(\omega)\sim \Omega$ even when $\tau_{s}\ll 1$ and $\mu\ll 1$. In addition, 
across a broad range of $\theta_{k}$, $\Im(\omega)$  no longer scales proportionally to $\tau_{s}$ in the $\tau_{s}\ll 1$ limit, and grows
much faster than the settling time $t_{\mathrm{settle}}\sim (\Omega \tau_{s})^{-1}$ for small particles. 
This suggests that
significant clumping of smaller grains could occur as they settle towards the midplane, with potentially 
important consequences for planetesimal formation (see \S\ref{sub: discussion epi rdi}).
For simplicity, in this section we introduce the \si\ without considering the dynamical effect of 
the stratification that induces the drift in the first place (which allows \BV\ oscillations in the gas). This omission is rectified in \S\ref{sub: BV epi RDI}, where 
we treat the joint epicyclic-\BV\ RDI, finding very  similar properties to the simpler case treated here.

\begin{figure*}
    \vspace{-0.15cm}
\includegraphics[width=1.05\columnwidth]{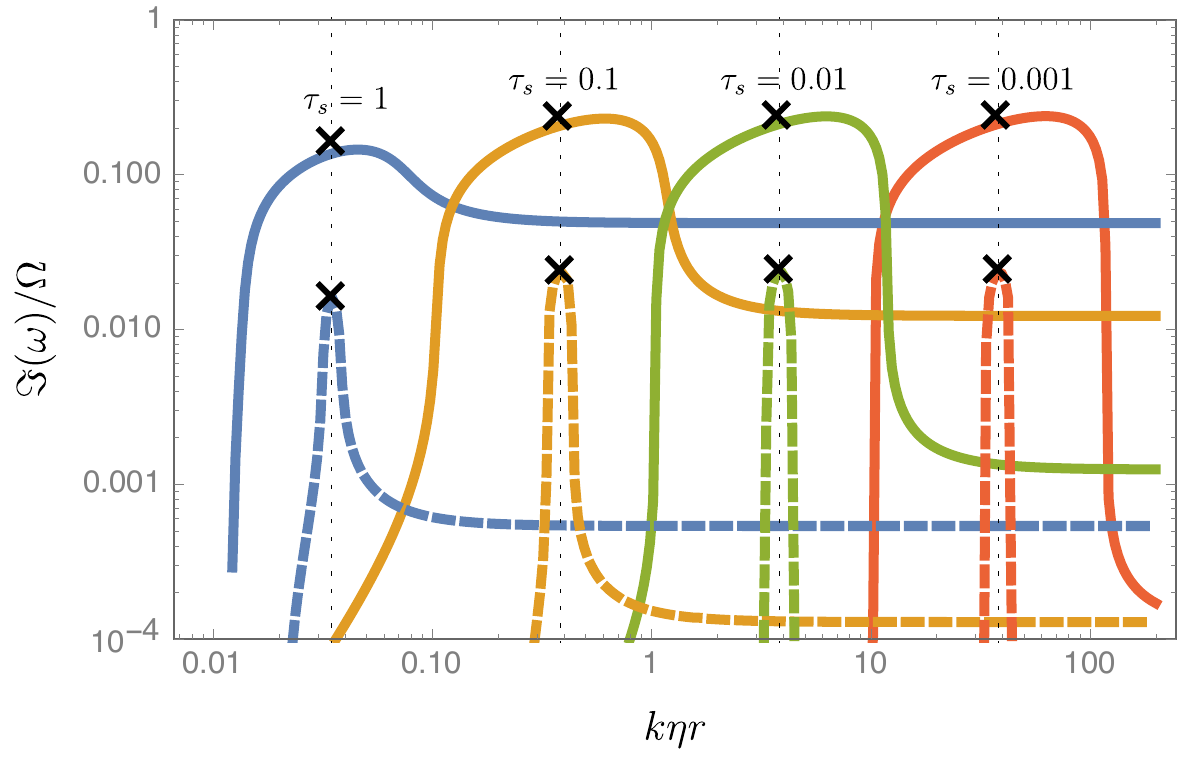}~~
\includegraphics[width=1.03\columnwidth]{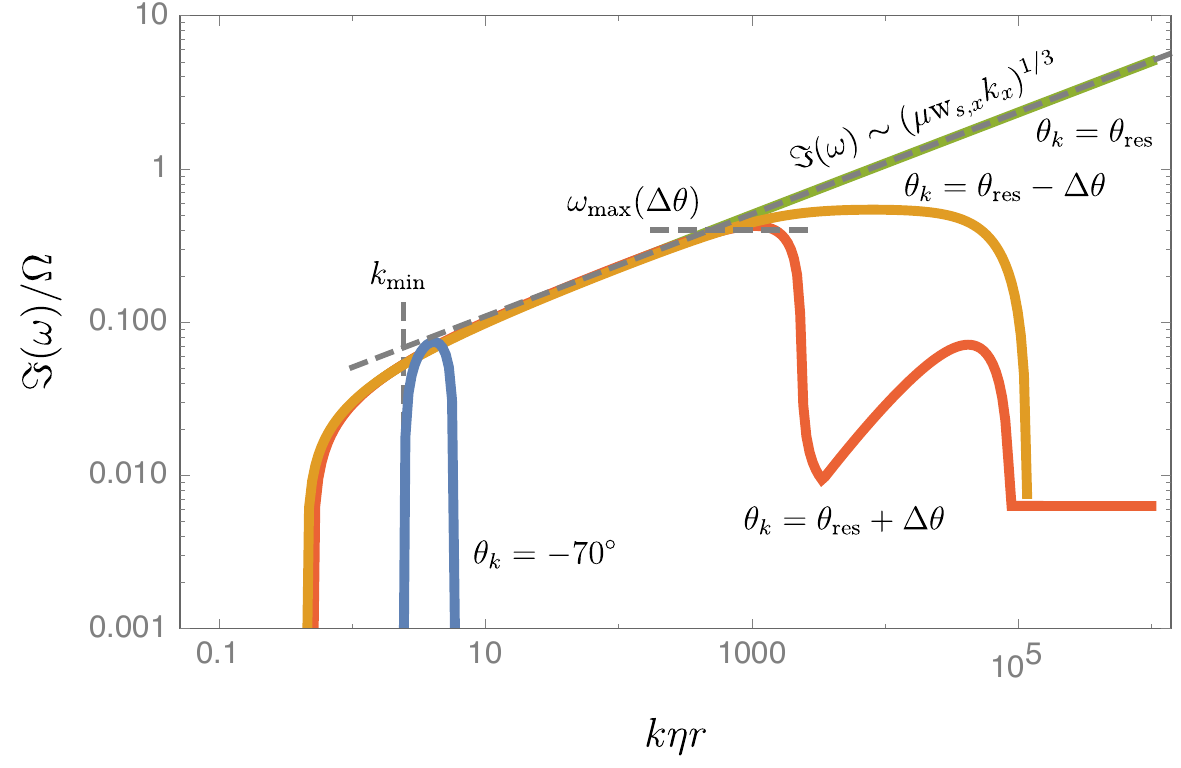}
    \vspace{-0.35cm}
\caption{The \si\ (vertical-epicylic RDI)---i.e., the streaming instability  including a vertical settling velocity of the dust towards the midplane. The left-hand panel is of the same form 
as Fig.~\ref{fig: epicyclic 1D}, but includes the settling drift $\driftvelz$ from Eq.~\eqref{eq: vertical settling drift} with $\eta =10^{-3}$. The solid curves show $\mu=0.1$, dashed curves show $\mu=0.001$, 
  black crosses show the analytic estimate of the maximum growth rates (Eq.~\eqref{eq: epicyclic RDI growth neg}), and we take
 $\theta_{k}=70^{\circ}$ for the mode angle (this angle  is chosen so as to show the basic RDI behavior, away from the double-resonant mode). Remarkably, maximum growth rates are large ($\Im(\omega)/\Omega\approx\!\! \sqrt{\mu/2}$, reaching an appreciable fraction of $\Omega$ for larger $\mu$) and {\em independent} of the stopping time  $\tau_{s}$ (i.e., they are independent of grain size). The vertical settling time is $t_{\mathrm{settle}}\sim 1/(\Omega\,\tau_{s})$, which is much longer than the growth timescale for $\tau_{s} \ll 1$ grains (radial drift timescales are longer still). Also note that the characteristic, maximally unstable wavelengths of the \si\ are larger than the \YG\ streaming instability by a factor $\eta^{-1/2} \sim 30$  (c.f., Fig.~\ref{fig: epicyclic 1D}).
 The right-hand panel shows the behavior of the mode near
 the ``double-resonant'' solution at $\theta_{k}=\theta_{\mathrm{res}}$ when $\bm{k}\cdot \driftvel =0$.  The solid  lines each show numerically 
 calculated dispersion relations with $\mu=0.01$ and $\tau_{s}=0.01$, evaluated at the labeled angles $\theta_{k}=70^{\circ}$ (blue curve), $\theta_{k}=-\theta_{\mathrm{res}}\approx 86.4^{\circ}$ (green curve), $\theta_{k}=\theta_{\mathrm{res}} - 0.002 \approx \theta_{\mathrm{res}} - 0.1^{\circ}$ (red curve), and $\theta_{k}=\theta_{\mathrm{res}} + 0.002$ (orange curve). 
 The dashed gray curves illustrate the various analytic results from the text: the growth rate  of the double-resonant mode (Eq.~\eqref{eq: epi double res init solution}), the low-$k$ cutoff wavenumber (Eq.~\eqref{eq: epi double res low k cutoff }), and the maximum growth rate (Eq.~\eqref{eq: epi double res max omega}) (i.e., the high-$k$ cut off) for the chosen $\Delta \theta = 0.002$. The growth rate of double-resonant modes, at mode angle  $\theta_{k}=\theta_{\mathrm{res}}$, increases without bound with $k$ as $\Im(\omega)\propto k^{1/3}$. 
     \vspace{-0.25cm}
\label{fig: vertical streaming epicyclic}}
\end{figure*}


It is necessary to account for two changes in our results from \S\ref{subsub: epicyclic horizontal streaming}: first, we now have $\driftvelz \ne 0$ in the growth rate, Eq.~\eqref{eq: epicyclic RDI growth pos} or Eq.~\eqref{eq: epicyclic RDI growth neg}; second, $\driftvelz \ne 0$ in the resonant condition, $\bm{k}\cdot\driftvel = \pm \hat{k}_{z}\,\Omega$, so that Eq.~\eqref{eq: NSH epi resonant condition} is modified to 
\begin{equation}
k = \left| \frac{\Omega}{\tan \theta_{k}\driftvelx + \driftvelz}\right| .
\label{eq: epicyclic res condition with vertical}
\end{equation}
For concreteness, we shall consider the $z<0$ region of the disk, where $\driftvelz>0$, and set $k_{z}>0$ (i.e., $-\pi/2<\theta_{k}<\pi/2$; results with $\driftvelz<0$ are effectively identical).  
Noting that  $|\driftvelx|\ll \driftvelz>0$, we see that the resonance condition is satisfied for positive-frequency epicycles ($\omega_{\mathcal{F}}=\hat{k}_{z}\Omega$) with $\hat{k}_{x}>0$. We  then simply insert the resonant $k$ (Eq.~\eqref{eq: epicyclic res condition with vertical}) and the drift velocities (Eqs.~\eqref{eq: NSH drift}--\eqref{eq: vertical settling drift}) into 
the growth rate expression (Eq.~\eqref{eq: epicyclic RDI growth pos}), and expand in  $\tau_{s}\ll 1$ to obtain,
\begin{equation}
\frac{\omega}{\Omega} = \hat{k}_{z} \pm i\left(\frac{\mu}{2}\right)^{1/2}\left(\frac{1+2\eta^{1/2}\cot \theta_{k}}{1-2\eta^{1/2}\tan \theta_{k}}\right)^{1/2}\sin\theta_{k} + \mathcal{O}(\tau_{s},\mu).\label{eq: vertical epicyclic resonant omega}
\end{equation}
If we simplify, for the moment, to mode angles where $|k_{x}|\sim k_{z}$, specifically $\eta^{1/2}\ll |\theta_{k}| \ll \pi/2-\eta^{1/2}$, the 
growth rate of the RDI mode is simply 
\begin{equation}
\Im(\omega) \approx \Omega\left(\frac{\mu}{2}\right)^{1/2}\left| \hat{k}_{x}\right| + \mathcal{O}(\tau_{s}, \eta^{1/2},\mu)\quad \text{at}\quad   k\eta r\approx \frac{\eta^{1/2}}{\tau_{s}},\label{eq: vertical epicyclic resonant omega mid angles}
\end{equation}
which can also be obtained by setting $\driftvelx=\driftvely=0$ in both the resonant condition (Eq.~\eqref{eq: epicyclic res condition with vertical}) and growth rate (Eq.~\eqref{eq: epicyclic RDI growth neg}; this should be expected, since $|\driftvelx| \approx 2\tau_{s}\,\eta U_{K}\ll |\driftvelz|\approx \eta^{-1/2}\tau_{s}\, \eta U_{K}$).

While the result \eqref{eq: vertical epicyclic resonant omega mid angles} appears very different to the YG streaming instability,
examination of Eq.~\eqref{eq: epicyclic RDI growth small tau} shows that the standard streaming instability 
does in fact  have an $\mathcal{O}(\tau_{s}^{0})$ term in the $\mathcal{O}(\mu^{1/2})$ perturbation to $\omega$: the $(\mu/2)^{1/2}$ term in square brackets in Eq.~\eqref{eq: epicyclic RDI growth small tau}. However, this term is purely real when $\driftvelz=0$.  In contrast, when a vertical streaming  dominates the drift velocity, the symmetry that caused this term to be real is broken, and the instability has a $\tau_{s}$-independent part.
In the left panel of Fig.~\ref{fig: vertical streaming epicyclic}, which is of the same form as Fig.~\ref{fig: epicyclic 1D}, we compare the numerically 
calculated growth rates to the analytic expression (Eq.~\eqref{eq: epicyclic RDI growth pos}) for a variety of $\tau_{s}$ and $\mu$.

Examining the eigenmodes of the \si, we see that the linear mode contains a substantial dust density perturbation. This
is also the case for the YG streaming instability, and, in fact, must be true for any instability in the RDI family, because  RDIs
arise due to the gas wave resonance with  the density perturbation  of the dust (see \S\ref{sec: resonance}). The size 
of the dust density perturbation (compared to other components of the eigenmode) scales as $\mu^{-1/2}$: thus, at
decreasing grain concentrations, the relative perturbation of the dust density increases, more directly seeding large dust-to-gas ratio
fluctuations without stirring up the gas. This behavior is expected and very similar to that seen in other RDIs (see, e.g.,  \S3.9 of \papertwo\ for further discussion).

\subsubsection{The double-resonant $\theta_{k}$}\label{subsub:double resonant theta vertical epi}

A careful examination of Eqs.~\eqref{eq: epicyclic res condition with vertical} and \eqref{eq: vertical epicyclic resonant omega} uncovers  an interesting effect that is not captured by Eq.~\eqref{eq: vertical epicyclic resonant omega mid angles}: the resonant 
wavenumber and growth rate {approach infinity} as $\bm{k}\cdot \driftvel$ approaches zero (or equivalently $2\eta^{1/2}\tan\theta_{k}$ approaches $1$). This can also be seen in the full RDI expression, Eq.~\eqref{eq: epicyclic RDI growth neg}, which increases  as $k_{\mathrm{res}}$ increases, but does not contain $\bm{k}\cdot \driftvel$ in the numerator. As we now show,  at this ``double-resonant'' angle,
\begin{equation}
\theta_{\mathrm{res}} = \arctan\left( -\frac{\driftvelz}{\driftvelx}\right) \approx \arctan\left( \frac{1}{2\eta^{1/2}}\right)\approx 86^{\circ}\text{ (at }\eta=10^{-3}),\end{equation}
the growth rate increases \emph{without bound} with $k$, scaling as $\Im(\omega)/\Omega\! \sim\! \mu^{1/3}  \tau_{s}^{1/3}k^{1/3}$ (of course, we are neglecting viscosity, turbulence, and other dissipative effects; see \S\ref{sec: other effects}). Although we derive its 
properties here in an unstratified incompressible gas, this mode survives  the addition of dust and gas stratification  and a  compressible treatment (see \S\ref{sub: BV epi RDI}).
The numerically calculated dispersion relation is shown in the right-hand panel of Fig.~\ref{fig: vertical streaming epicyclic} for a variety of angles near $\theta_{k}=\theta_{\mathrm{res}}$.
 
Properties of the double-resonant mode are simplest to derive from the dispersion relation for the
full coupled dust-gas system.  This is found from the matrix operators, Eqs.~\eqref{eq: D matrix epi}--\eqref{eq: T and C matrix epicyclic}, as  the characteristic polynomial of $\mathbb{T}_{0}+\mu \mathbb{T}^{(1)}$ (Eqs.~\eqref{eq: general linear blocks}--\eqref{eq: dust matrix form}) after inserting $\bm{k}\cdot \driftvel =0$. We then insert the ansatz $\omega/\Omega = \varpi\,\mu^{1/3}\tau_{s}^{1/3}(k\eta r)^{1/3}$, insert equation~\eqref{eq: NSH drift} for $\driftvel$, and expand in high $k$ ($k\eta r\sim \epsilon^{-1}$) and small $\mu$ and $\tau_{s}$ ($\mu\sim \epsilon^{\nu}$, $\tau_{s}\sim \epsilon^{1-\nu}$, with $0 < \nu<1$), yielding the polynomial,
\begin{equation}
\varpi^{3}- \varpi \frac{\cos^{2}\!\theta_{k}}{(\mu\,\tau_{s}\,k\eta r)^{2/3}}+ 2 \sin\theta_{k}=0.\label{eq: epi double res init polynomial}\end{equation}
When $\omega/\Omega = \varpi\mu^{1/3}\tau_{s}^{1/3}(k\eta r)^{1/3}\gg \cos\theta_{k}$, the middle term in Eq.~\eqref{eq: epi double res init polynomial} is negligible, giving the unstable  root,
\begin{equation}
\frac{\omega}{\Omega} \approx \frac{1}{2}(-1+i\sqrt{3})(2 \tau_{s}\,\mu\, k_{x}\eta r)^{1/3},\label{eq: epi double res init solution}\end{equation}
which justifies our original ansatz for $\omega$ and shows that $\Im(\omega)\rightarrow \infty $ as $k\rightarrow \infty$. The middle term in equation~\eqref{eq: epi double res init polynomial} is important when 
$\omega/\Omega \lesssim  \cos\theta_{k}\approx \pi/2-|\theta_{k}|$. This gives the minimum
 wavenumber for which the solution  \eqref{eq: epi double res init solution} is valid,
 \begin{equation}
k_{\mathrm{min}}\eta r\sim \frac{(\pi/2-|\theta_{\mathrm{res}}|)^{3}}{\mu\,\tau_{s}}\label{eq: epi double res low k cutoff },\end{equation}
which is shown in the lower panel of Fig.~\ref{fig: vertical streaming epicyclic}. We note that this estimate for $k_{\mathrm{min}}$ is modified in the presence of gas stratification, because the gas oscillations are modified; see \S\S\ref{subsub: double resonant mode with strat}--\ref{subsub: epi BV compared to epi} and Eq.~\eqref{eq: epi double res init polynomial comp}.

In addition to dissipative effects not included here (see \S\ref{sec: astrophysics}), the instability is cut off at high wavenumbers due to  misalignment of $\theta_{k}$ from $\theta_{\mathrm{res}}$. Because in reality (or numerical simulations) not \emph{all} mode angles are
necessarily possible, it is helpful to understand how this cutoff scales with the misalignment $\Delta \theta = \theta_{k}-\theta_{\mathrm{res}}$. To do this, we recalculate the dispersion relation from  $\mathbb{T}_{0}+\mu \mathbb{T}^{(1)}$, but now with $\bm{k}\cdot \driftvel = \kappa\, k\, \tau_{s}$, where $\kappa\sim -\eta^{-1/2}\Delta \theta$ is a small parameter. Repeating the expansion described above using the same ordering  and $\kappa\, \tau_{s}\sim 1$ yields the additional terms $(\kappa/\mu)\cos^{2}\theta_{k}+(\kappa/\mu)(\mu\,\tau_{s} k\eta r)^{2/3}\varpi^{2}$ in equation~\eqref{eq: epi double res init polynomial}. The 
former term has no effect, but the latter term is important to the solutions for $\varpi$ unless $(\kappa/\mu)(\mu\,\tau_{s} k\eta r)^{2/3}\ll 1$. Asserting that this term be negligible, we obtain the cutoff growth rate,
\begin{equation}
\frac{\omega_{\mathrm{max}}}{\Omega}\sim \left( \frac{\mu\,\eta^{1/2}}{|\Delta \theta|}\right)^{1/2}.\label{eq: epi double res max omega}\end{equation}
In the lower panel of Fig.~\ref{fig: vertical streaming epicyclic}, we also show modes with $\theta_{k}=\theta_{\mathrm{res}}\pm 0.002$, 
illustrating  nice agreement with equation~\eqref{eq: epi double res max omega}. The cutoff in Eq.~\eqref{eq: epi double res max omega} also 
helps to clarify the connection between the double-resonant solution (Eq.~\eqref{eq: epi double res init solution}) and the RDI solution (Eq.~\eqref{eq: vertical epicyclic resonant omega}): as $\theta_{k}$ approaches $\theta_{\mathrm{res}}$ (from below $|\theta_{k}|<|\theta_{\mathrm{res}}|$), the predicted RDI growth rate, obtained by expanding  Eq.~\eqref{eq: vertical epicyclic resonant omega} in $\theta_{k}$ about $\theta_{\mathrm{res}}$, is $\Im(\omega) =(\mu\,\eta^{1/2}/|\Delta \theta|)^{1/2}\sin\theta_{k}$. This matches the cutoff growth rate  of the double-resonant mode (Eq.~\eqref{eq: epi double res max omega}). Put differently, the RDI solution in Eq.~\eqref{eq: vertical epicyclic resonant omega} correctly predicts the maximum of $\Im(\omega)$, although it cannot predict the $\Im(\omega)\sim k^{1/3}$ scaling of the double-resonant mode.

\section{Gas stratification}\label{sec: adding stratification}

In the previous section, we studied the YG streaming instability and epicyclic RDI more generally. The most interesting
result of this section was that
the instability becomes significantly faster-growing when the dust also undergoes vertical streaming motion
(i.e., settling towards the midplane of the disk)---a new instability that we termed the ``\si.'' However, those regions of the disk away from the midplane 
are also stratified, which (if stable) allows for buoyancy oscillations that can  cause  another RDI 
(the {\BV\ RDI}). With this in mind, the purpose of this section is twofold: first, we examine the resonance with \BV\ (BV) oscillations and the resulting instability; second, we verify that the behavior of the disk \si\ described in \S\ref{subsub: epicyclic vertical streaming} is
robust to the addition of gas/dust  stratification and compressibility. To do this, we derive the ``vertical-epicyclic-\BV\ RDI,'' which results from the resonance with joint epicyclic-BV oscillations in the gas. The properties 
of this RDI are very similar to the pure epicyclic RDI, so, in the astrophysical discussion of \S\ref{sec: astrophysics}, we will simply term this RDI the disk \si\ also.

After introducing useful variables and the local formulation in \S\ref{sub: linear stratified}, we shall examine a simple stratified fluid (i.e., without rotation) in \S\ref{sub: BV RDI}. This allows us  to better 
understand the properties of the resulting \BV\ (BV) RDI,
which was briefly introduced in \paperone,
without undue complications. This instability may be interesting in its own right for other (non-disk) applications and has likely been observed in previous numerical simulations (\citealp{Lambrechts:2016fg}; see \S\ref{sub: discussion BV rdi}). We then treat the full, stratified, rotating, compressible 
problem in \S\ref{sub: BV epi RDI}, deriving the epicyclic-BV RDI,  illustrating how 
this reduces to  the epicyclic and BV RDIs separately in the relevant limits (i.e., both the pure epicyclic and 
BV RDIs are special cases of the epicyclic-BV RDIs), and discussing the influence 
of stratification on the disk \si\ (\S\ref{subsub: epi BV compared to epi}).

Finally, we note that, formally, the local treatment of background gradients used throughout this section may  not be
 appropriate. In principle, for this problem, it may be necessary to embark on a fully global treatment or a spatially dependent WKBJ expansion \citep{bender1978advanced,white2010asymptotic}, which is quite complicated and beyond the scope of this work. While this could potentially yield corrections to the growth rates  presented here (e.g., from second-order gradients of background quantities), it is likely that the local treatment  correctly  captures most aspects of the \BV\ and vertical-epicyclic-BV RDIs. In any case,  the main purpose of this section is to show that the stratification has only a relatively weak
effect on the  \si\ in disks, and, given the uncertainty that surrounds the actual stratification profile in disks,   nonlinear simulations are obviously required to study the instability in significantly more detail. Brief discussion of these mathematical issues is given below in \S\ref{subsub: local linear BV caution}.

\subsection{The linear system to be solved}\label{sub: linear stratified}

We shall examine a local patch of disk with  
 an arbitrary  background pressure and temperature gradient 
in the $\hat{\bm{x}}$ (radial) and $\hat{\bm{z}}$ (vertical) directions.  
The fluid equations are Eqs.~\eqref{eq: gas NL rho}--\eqref{eq: gas NL u} with 
the pressure gradient  balancing the combination of gravity ($\bm{g}=g\hat{\bm{g}}$) and the  drag force parallel to $\bm{g}$; i.e., 
$\rho_{0}^{-1} \nabla P_{0}= - \mu  ({\driftvel\cdot \hat{\bm{g}}})/{t_{s0}}\,\hat{\bm{g}} + \bm{g}$.
As described above (see \S\ref{subsub:linearized} and \papertwo), an additional force perpendicular to $\hat{\bm{g}}$  (e.g,. from 
radiation pressure on the grains) could in principle cause a perpendicular $\driftvel$ also, accelerating
the dust and gas together once the dust reaches its terminal velocity (the analysis is then carried out in the free-falling frame). Thus, in 
our derivation of the  \BV\ (BV) RDI  in \S\ref{sub: BV RDI}, we allow for a nonzero perpendicular drift  for completeness.

Rather than the pressure $P$ (Eq.~\eqref{eq: gas NL P}), it is easier to work with the entropy, $S \equiv \gamma_{\mathrm{gas}}^{-1} \ln (P/\rho^{\gamma_{\mathrm{gas}}})$, which evolves according to
${\partial_{t} S}+ \bm{u}\cdot \nabla S = 0$.
The gas equilibrium is then determined by $\bm{g}$, $P_{0}$ and $\rho_{0}$ (through $c_{s0}^{2}=\gamma_{\mathrm{gas}}P_{0}/\rho_{0}$), and $\nabla S_{0}$, and it is helpful to define the following variables to describe this:
\begin{gather}
L_{0}^{-1}\equiv \gamma_{\mathrm{gas}}^{-1}\,\frac{1}{P_{0}}\,\frac{\partial P_{0}}{\partial z},\quad L_{0R}^{-1}\equiv \gamma_{\mathrm{gas}}^{-1}\,\frac{1}{P_{0}}\,\frac{\partial P_{0}}{\partial r}\sim \eta^{1/2}L_{0}^{-1},\nonumber \\
c_{s0}=\gamma_{\mathrm{gas}}\frac{P_{0}}{\rho_{0}},\quad\tilde{\bm{g}} \equiv \frac{1}{\rho_{0}} \nabla P_{0}= c_{s0}^{2}(L_{0R}^{-1}\,\hat{\bm{x}}+L_{0}^{-1}\,\hat{\bm{z}}),\nonumber \\
- \Lambda_{S}\equiv {L_{0}} \frac{\partial S_{0}}{\partial z}\approx {L_{0R}} \frac{\partial S_{0}}{\partial r}.\label{eq: stratified equilibrium defs}
\end{gather}
In these definitions, we have neglected  a background dust density or $\driftvel$ stratification, which 
is  treated in App.~\ref{app: dust stratification} and results in minor modifications to the RDI 
growth rates.\footnote{While the addition of dust stratification does not add significant complexity 
to the analysis, when $\nabla\cdot \driftvel\neq 0$, we can no longer formally apply the block-matrix RDI analysis method without modification. For this reason,
 we relegate its explanation to App.~\ref{app: dust stratification}.} We also assume for simplicity
that the stratification direction  of $S_{0}$  is the same as that of $P_{0}$ (i.e., we need only the parameter $\Lambda_{S}$, rather than a separate parameter for the vertical and radial directions separately).
Relaxing this assumption does not fundamentally modify the RDIs studied here, but can also lead to baroclinic instabilities, which we do not wish to study  (see, e.g., \citealt{0004-637X-788-1-21,2016MNRAS.457L..54L,2017arXiv170802945L}).
The definitions in equation~\eqref{eq: stratified equilibrium defs} give $\nabla\ln \rho_{0} = (L_{0R}^{-1}\,\hat{\bm{x}}+L_{0}^{-1}\,\hat{\bm{z}})\,(1+ \Lambda_{S})$ and yield the vertical \BV\ frequency 
$N_{BV}^{2}=c_{s0}^{2}(L_{0}^{-2}+L_{0R}^{-2})\, \Lambda_{S}$ (see below). 
Note that, because we expand in $\mu$ to $\mathcal{O}(\mu^{1/2})$, there is no need  to distinguish 
between $\bm{g}$ and
$\tilde{\bm{g}}= - \mu  ({\driftvel\cdot \hat{\bm{g}}})/{t_{s0}}\,\hat{\bm{g}} + \bm{g}$  in our analytic analysis below (the full terms are retained in our numerical solutions).
For concreteness, we shall set $L_{0}>0$, as appropriate for  regions
below the midplane. The natural direction for the settling velocity---i.e., dust streaming towards the midplane---is thus
$\driftvelz>0$, as used in \S\ref{subsub: epicyclic vertical streaming} (regions above and below the midplane behave  identically, we specify the direction only for notational clarity).

We construct the local equations by taking  $k_{z}L_{0} \gg 1$ and $k_{x}L_{0R} \gg 1$, and  assuming the 
background gradients of $P_{0}$, $S_{0}$, and $\rho_{0}$ are constant so as to Fourier 
analyze the equations in the $x$ and $z$ directions. This is nearly equivalent to a formal WKBJ expansion 
to lowest order in $(kL_{0})^{-1}$, and is discussed in more detail below (\S\ref{subsub: local linear BV caution}). Also assuming axisymmetric perturbations ($k_{y}=0$),
we obtain the linearized gas and dust  equations,
\begin{align}
-i\omega  \frac{{\delta\rho}}{\rho_{0}} &+ i \bm{k}\cdot \delta \bm{u} + \delta \bm{u}\cdot  (L_{0R}^{-1}\hat{\bm{x}}+L_{0}^{-1}\hat{\bm{z}})\,(1+\Lambda_{S}) =0,\label{eq: BV rho}\\[1ex]
-i \omega \delta \bm{u} &=-i  c_{s0}^{2} \bm{k}\left(\delta S +\frac{\delta \rho}{\rho_{0}}\right)+ \frac{\delta \rho}{\rho_{0}}\tilde{\bm{g}}+ \frac{3}{2}\Omega\delta u_{x}\hat{\bm{y}} \nonumber \\
& -2\Omega \hat{\bm{z}}\times \delta \bm{u}- \mu \frac{\delta\bm{u}-\delta \bm{v}}{t_{s0}}+\mu \frac{\driftvel}{t_{s0}} (\delta t_{s} +\frac{\delta \rho}{\rho_{0}}  -\frac{\delta \rho_{d}}{\rho_{d0}} ),\label{eq: BV u}\\[1ex]
-i\omega \delta S& - \delta \bm{u}\cdot (L_{0R}^{-1}\hat{\bm{x}}+L_{0}^{-1}\hat{\bm{z}})\,\Lambda_{S} =0,\label{eq: BV p}\\[1ex]
(-i\omega+&i\bm{k}\cdot \driftvel)  \frac{{\delta\rho_{d}}}{\rho_{d0}} + i \bm{k}\cdot \delta \bm{v} =0,\label{eq: BV rho d}\\[1ex]
(-i\omega+&i\bm{k}\cdot \driftvel )\,  \delta \bm{v} = -2\Omega \hat{\bm{z}}\times \delta \bm{v}+ \frac{3}{2}\Omega\delta v_{x}\hat{\bm{y}}- \frac{\delta\bm{v}-\delta \bm{u}}{t_{s0}}- \driftvel\frac{\delta t_{s}}{t_{s0}}.\label{eq: BV v d}
\end{align}
As appropriate for subsonic streaming velocities $\driftvelmag \ll c_{s0}$, we neglect the velocity dependence of $t_{s}$, taking $\delta t_{s}/t_{s0} = -\zeta_{\rho}\, \delta {\rho}/\rho_{0} -\zeta_{P}\,\delta {P}/P_{0} = 
 -(\zeta_{\rho}+\gamma_{\mathrm{gas}}\zeta_{P})\, \delta {\rho}/\rho_{0} -\gamma_{\mathrm{gas}}\zeta_{P}\,\delta {S} $. 
  
 \subsubsection{A cautionary note about the local approximation}\label{subsub: local linear BV caution}
 As mentioned  in the introduction above, caution should be used in interpreting the solutions 
 to Eqs.~\eqref{eq: BV rho}--\eqref{eq: BV v d}, because it is possible that there are neglected terms that could modify 
 the growth rate. In this section we briefly discuss this subtlety, and how it can be remedied in future work. Those readers uninterested in these somewhat esoteric mathematical details should feel free to skip to \S\ref{sub: BV RDI}. 
  
Formally, a local ``dispersion relation'' should be derived from the linearized fluid 
 equations through a WKBJ expansion, without assuming anything about the background $\rho_{0}(\bm{x})$, $P_{0}(\bm{x})$. This 
 involves expanding in  $\epsilon \sim (kL_{0})^{-1}$, assuming that the linear fields  ($\delta \rho$, $\delta \bm{u}$, etc.) have the form $\exp[i\, \epsilon^{-1} \sum_{i=0}^{\infty}\epsilon^{j}Q_{j}(\bm{x})]$. The lowest-order expression in $\epsilon$ yields a ``dispersion relation'': more formally, a local 
 relationship between $\nabla Q_{0}$ and $\omega$, which (for a given $\omega$) specifies how the wavelength varies with background quantities. For example, applying such a procedure to a pure stratified gas (i.e., Eqs.~\eqref{eq: gas NL rho}--\eqref{eq: gas NL P} with $\mu=0$), one finds either the \BV\ dispersion relation or the sound-wave dispersion 
 relation, depending on the choice for the asymptotic scaling of $\omega$ (i.e., the choice of dominant balance). The $\omega$ one obtains, and the eigenmodes---i.e., the local relationship between $\delta \rho(\bm{x})$, $\delta \bm{u}(\bm{x})$, and $\delta S(\bm{x})$---are  identical,  at lowest order in $(kL_{0})^{-1}$,
 to those obtained through an expansion  of the local equations  (Eqs.~\eqref{eq: BV rho}--Eqs~\eqref{eq: BV p} with $\mu=0$). These are also identical to a standard (unstratified) sound wave, and an analysis using the Boussinesq approximation (indeed, this amounts to a formal derivation of the linear Boussinesq approximation). However, this exact agreement between the local and formal WKBJ result is only valid at lowest order in $(kL_{0})^{-1}$, and it is not logically consistent to expand the local solutions (Eqs.~\eqref{eq: BV rho}--\eqref{eq: BV u}) to  higher order.  More precisely, while the the WKBJ dispersion relation is  unmodified at the next order $(kL_{0})^{-1}$ in the WKBJ expansion  (more generally, the dispersion relation is only modified at every second order in $\epsilon$; see \citealt{bender1978advanced}), the WKBJ eigenmodes have corrections that appear at order $(kL_{0})^{-1}$.
 
 In the coupled dust-gas system, the potential for a problem arises because $\driftvelmag/c_{s}$ is also a small parameter, which
 is of the same order as $(k L_{0})^{-1}$ (this must be the case due to the resonance condition; see Eq.~\eqref{eq: BV resonant wavenumber} below). As will become clear below, this mixes  the $\mathcal{O}[(kL_{0})^{-1}]$ correction
 to the eigenmodes---which is not captured correctly by Eqs.~\eqref{eq: BV rho}--\eqref{eq: BV p}---into the lowest-order result for the RDI, and may cause the RDI growth rate to depend on, for instance, second derivatives of  $P_{0}$ and $\rho_{0}$. However, it is also possible that the resonant-mode growth rates derived from Eqs.~\eqref{eq: BV rho}--\eqref{eq: BV v d} are correct, if the intuition above---that the  lowest-order WKBJ dispersion relation is captured correctly by the local equations---also holds for this much more complicated coupled dust-gas system.\footnote{In fact, it transpires RDI growth 
 rates derived from Eqs.~\eqref{eq: BV rho}--\eqref{eq: BV v d} do not depend on the exact form of the local equations, even though the eigenmodes do, which suggests the dispersion relation may be relatively robust.} Unfortunately, checking this explicitly is not a trivial task, 
 and is beyond the scope of this work. We  will address this issue in future work with a fully global analysis.
 
One clear regime of validity for our results, and for Eqs.~\eqref{eq: BV rho}--\eqref{eq: BV v d} in general, is that $\mu$ must be sufficiently large such that the perturbation on the gas modes
from the dust is larger than the higher-order WKBJ corrections. Equivalently, noting that the correction to the \BV\ (or epicyclic-BV) mode arises at $\mathcal{O}[(kL_{0})^{-2}]$, the perturbed eigenmode ($\omega^{(1)}$) should satisfy  
$\omega^{(1)}\gtrsim (kL_{0})^{-2} N_{BV}$,  which becomes $\mu^{1/2}\gtrsim (kL_{0})^{-1}$ using Eq.~\eqref{eq: final compressible BV mode} below (this can also 
be seen directly from the gas-dust equations, noting that the RDI theory of \S\ref{sec: resonance} shows that $\omega^{(1)}$, the $\mathcal{O}(\mu^{1/2})$ correction to $\omega$, depends only on the coupling of dust to gas). 
This condition is not particularly stringent, and well  satisfied for smaller grains in disks (see \S\ref{sub: discussion BV rdi} for further discussion).

 Finally, it is worth noting that the \BV\ RDI has likely already been seen in simulations in \citet{Lambrechts:2016fg}.  As discussed in \S\ref{subsub: BV RDI in simulations}, the 
 observed growth rates are comparable to our predictions, although a detailed comparison is not possible.

\subsection{\BV\ RDI}\label{sub: BV RDI}

In this section, we treat a stratified fluid in the absence of rotation.
As discussed above, while this situation
is not directly applicable to thin disks (the rotation is always dynamically important), the 
treatment is helpful to isolate the 
different character of the RDI that arises due to \BV\ (BV) oscillations.
As we show below (\S\ref{sub: BV epi RDI}), 
the instability is effectively a special case of the joint epicyclic-BV RDI for $\Omega=0$.
We consider Eqs.~\eqref{eq: BV rho}--\eqref{eq: BV v d} without the influence of rotation ($\Omega=0$), 
also setting $L_{0R}=0$ because the stratification direction is arbitrary  if $\Omega=0$.
However, even though we set $\Omega =0$ in the dynamical equations, for clarity and consistency with the rest of the
paper, we  quote results in terms of $\tau_{s} = t_{s}\Omega$ (i.e., $\Omega$ is effectively an arbitrary frequency scale) and the pressure support parameter $\eta$. Thus we consider 
a vertical stratification profile appropriate for a disk, with  $L_{0}\sim h_{g} \sim \eta^{1/2} r$ and $c_{s0}\sim h_{g} \Omega$, leading to the natural scaling for the 
 settling drift velocity (in the absence of an external acceleration on the dust), $\driftvel  \approx \tau_{s}c_{s0}\hat{\bm{z}}$ (see \S\ref{subsub: streaming}).
Preemptively noting that the resonance condition gives $\driftvelmag/c_{s}\sim (k_{\mathrm{res}}L_{0})^{-1} $, as well as the fact that $(kL_{0})^{-1} \ll 1$ is required for any sort of local treatment, our analysis shall proceed by expanding all expressions in $\epsilon\ll1$, with $\epsilon\sim \tau_{s}\sim (kL_{0})^{-1}\sim \driftvelmag/c_{s0}$. As discussed above, we require $\mu^{1/2}\gg \epsilon$ for the consistency of the expansion.

\begin{figure}
\begin{center}
\includegraphics[width=1.0\columnwidth]{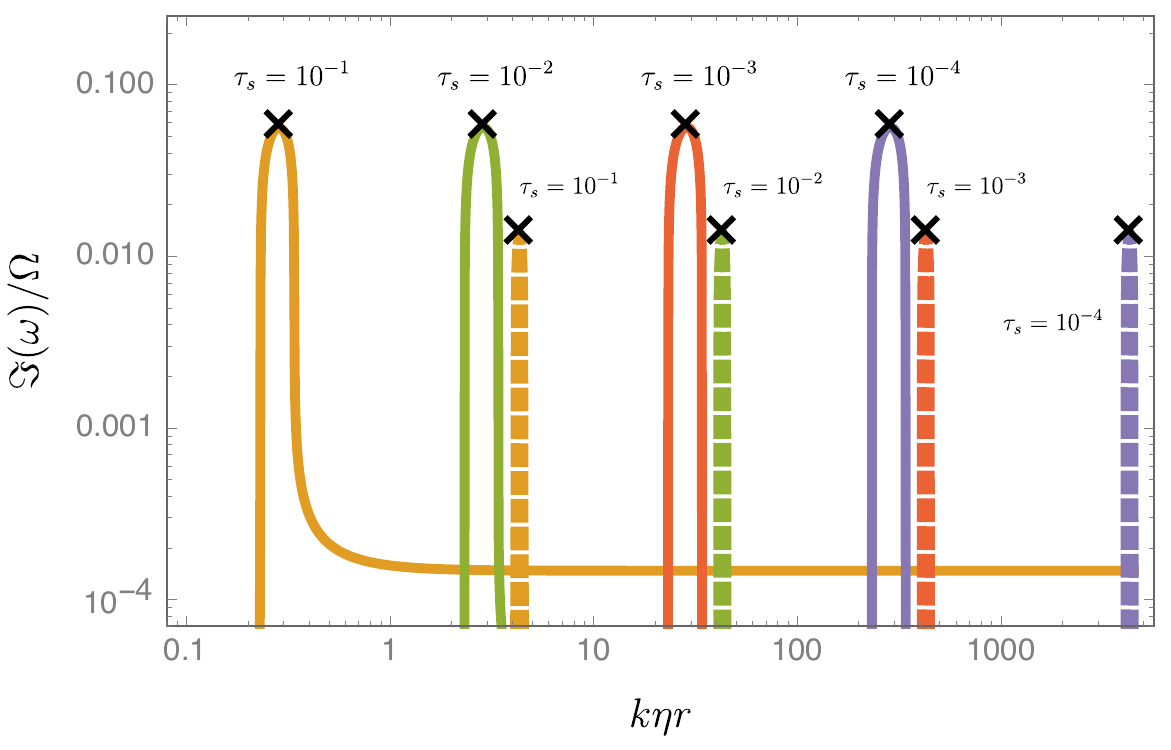}
\end{center}
\caption{
Illustration of the growth rate $\Im(\omega)$ of the pure \BV\ RDI including compressibility but
neglecting disk rotation. This is a distinct dust-gas instability which arises from resonance with BV oscillations. Each  curve is calculated through numerical solution of the local 
dispersion relation from Eqs.~\eqref{eq: BV rho}--\eqref{eq: BV v d} with $\Omega=0$, and uses  $\mu=0.01$, Epstein drag ($\zeta_{\rho}=\zeta_{P}=1/2$), stratification parameter $\Lambda_{S}=2$, equation-of-state $\gamma_{\mathrm{gas}}=5/3$, and 
$\tau_{s}$ as labeled (orange curves, $\tau_{s}=0.1$; green curves, $\tau_{s}=0.01$; red curves, $\tau_{s}=10^{-3}$; purple curves, $\tau_{s}=10^{-4}$). Even though rotation is not included in the calculation, we use disk units (\S\ref{subsub: units}) to allow direct comparison to other figures, taking the vertical pressure scale-length $L_{0}=h_{g}\approx\eta^{-1/2}(\eta r)$ and sound speed $c_{s0}=\eta^{-1/2} (\eta U_{K})$ with $\eta=0.001$ (ignoring the radial pressure-scale length, i.e.\ $L_{0R}^{-1}=0$, because the geometry here is arbitrary).  The solid curves 
show $\Im(\omega)$ including the vertical settling drift of the dust (Eq.~\eqref{eq: vertical settling drift}), while dashed  
curves show $\Im(\omega)$ with $\driftvelz=0$ (only radial/azimuthal drift) for comparison. For both cases, we take the mode angle to be $\hat{k}_{x}=\hat{k}_{z}$ ($\theta_{k}=\pi/4$), which maximizes the growth rate for the  $\driftvelz=0$ RDI (with the settling drift included, the growth 
rate is maximized near $\theta_{k}=\pi/2$). The maximum growth rates reach an appreciable fraction of $\Omega$ (even at low dust-to-gas ratios) and are independent of grain size ($\tau_{s}$), although the resonance is narrower in wavenumber compared to the (disk) \si\ shown in Fig.~\ref{fig: vertical streaming epicyclic}.
\label{fig: BV RDI}}
\end{figure}

\subsubsection{Gas oscillations}

With $\mu =0$ (i.e., no dust), Eqs.~\eqref{eq: BV rho}--\eqref{eq: BV p} have five eigenmodes:
$\omega=0$ (this represents an undamped zonal flow in $u_{y}$),  two sound-wave eigenmodes, with
\begin{equation}
\omega_{\mathcal{F}} = \pm  \epsilon^{-1}c_{s0}\, k + \dots \end{equation}
and  two BV eigenmodes,
\begin{equation}
\omega_{\mathcal{F}} = \pm \hat{k}_{x}N_{BV} +\dots,\label{eq: BV eigenmodes} \end{equation}
where $N_{BV}^{2} = \Lambda_{S}c_{s0}^{2}L_{0}^{-2}$ is the BV frequency. We focus 
on resonance with the BV modes (Eq.~\eqref{eq: BV eigenmodes}) and refer to \papertwo\ App.~C for an analysis of stratified sound waves, which introduce their own, additional RDIs (the acoustic RDI, discussed separately below). The BV modes  are stable so long as $\Lambda_{S}>0$ and represent slow (compared to sound waves)  oscillations, with the restoring force provided by buoyancy. As expected, the BV eigenmode is approximately incompressible, $\bm{k}\cdot \delta \bm{u}\approx 0$ to lowest order in $\epsilon$, and the density fluctuations
dominate over pressure fluctuations (i.e., $\delta \rho/\rho_{0} + \delta S \approx 0$). This is the linear manifestation of the  the Boussinesq 
approximation, which stipulates $\nabla\cdot \delta \bm{u}\approx 0$ and $\delta \rho/\rho_{0}\approx -\delta T/T_{0}$ (where $\delta T/T_{0}$ is the relative temperature 
fluctuation). At next order in $\epsilon \sim (kL_{0})^{-1}$, the BV eigenmode has a small compressive component (the form of this can be found, for example, from a WKBJ analysis; see \S\ref{subsub: local linear BV caution})

\subsubsection{Resonant drag instability}

A BV RDI can occur when the drift frequency $\bm{k}\cdot \driftvel$ matches the BV oscillation frequency, \emph{viz.,}
 at the resonant wavenumber,
\begin{equation}
k_{\mathrm{res}} \approx\pm \frac{\sin\theta_{k} N_{BV}}{\hat{\bm{k}}\cdot \driftvel}=  \frac{c_{s0}}{\driftvelmag}\Lambda_{S}^{1/2}L_{0}^{-1}\frac{\sin\theta_{k} }{\hat{\bm{k}}\cdot \driftvelhat}.\label{eq: BV resonant wavenumber}\end{equation}
We see,  as mentioned earlier, that $(k_{\mathrm{res}}L_{0})^{-1}\sim \driftvelmag/c_{s}$, justifying the $\epsilon$ ordering used for the expansion.
In the same way as for the derivation of the streaming instability in \S\ref{subsub: general epi derivation}, we then insert the  eigenmodes corresponding to BV oscillations (Eq.~\eqref{eq: BV eigenmodes}) into the RDI growth rate formula (Eq.~\eqref{eq: dust eval pert}), to 
obtain an expression for the growth rate of the BV RDI at $k=k_{\mathrm{res}}$.
For the
positive frequency BV mode this yields (to lowest order in $\epsilon$), $\omega = \hat{k}_{x} N_{BV}+\omega^{(1)}$ with 
\begin{align}
&(\omega^{(1)})^{2} =   \mu\frac{\zeta_{\rho}N_{BV}\,(\hat{\bm{k}}\cdot \driftvel) \,(\hat{\bm{k}}\times\driftvel)_{y}}{2 c_{s0}^{2} t_{s0}}(k_{\mathrm{res}}L_{0}) +\mu\frac{\hat{k}_{x} N_{BV} (\hat{\bm{k}}\times \driftvel)_{y}}{2c_{s0} t_{s 0}\sqrt{\Lambda_{S}}},
\label{eq: compressible BV mode full}\end{align}
or, inserting $k_{\mathrm{res}}$ from Eq.~\eqref{eq: BV resonant wavenumber},
\begin{equation}
\omega \approx \hat{k}_{x}N_{BV} + \mu^{1/2}\left[\frac{\hat{k}_{x}(\hat{\bm{k}}\times\driftvel)_{y}}{2t_{s0}L_{0}}\Theta_{S}\right]^{1/2},\label{eq: final compressible BV mode}\end{equation}
where $\Theta_{S} = 1+\zeta_{\rho}\Lambda_{S}$.
The $\zeta_{\rho}\Lambda_{S}$ term in  $\Theta_{S}$ arises from the lowest-order (Boussinesq) contribution 
to the gas BV eigenmode. Because this part of the oscillation is incompressible, the contribution to the RDI depends directly on
the dependence of $t_{s}$ on $\delta \rho $ (through $\zeta_{\rho}$). The  $1$ term in $\Theta_{S}$ arises from the
next-order (in $\epsilon\sim (kL_{0})^{-1}$) contribution to the BV eigenmode, and has entered at the same order in
the RDI growth rate because this compressible part of the BV oscillation interacts more strongly with the dust. As discussed
above in \S\ref{subsub: local linear BV caution}, the exact form of this contribution could be modified (e.g., 
by second derivatives of the background) if a true WKBJ treatment is carried out, and this  
value should be treated with some skepticism. However, the general physical picture---that the instability is enhanced
due to the interaction with the compressive part of the BV mode---is likely quite general, because the 
first-order compressive part of the BV eigenmode (the correction to $\delta \rho$) is correctly captured by Eqs.~\eqref{eq: BV rho}--\eqref{eq: BV v d}.  This picture also fits well into the toy model outlined in \S\ref{sub: toy model} and Fig.~\ref{fig: toy model},
in which gas pressure perturbations play a particularly important role in the RDI's mechanism.
The global analysis necessary to treat this compressive contribution more formally will be considered in future work.

As shown in App.~\ref{app: dust stratification}, the addition of dust and/or $\driftvel$ stratification causes the $\Theta_{S}=1+\zeta_{\rho}\Lambda_{S}$ factor to become $1+\zeta_{\rho}\Lambda_{S}-\Lambda_{\rho_{d}}$, where $\Lambda_{\rho_{d}}= L_{0}^{-1}d\ln\rho_{d0}/dz$ (this result is again subject to  the caveats of the local model outlined in \S\ref{subsub: local linear BV caution}; see also App.~C of \papertwo).   

\subsubsection{Properties of the \BV\ RDI}\label{sub:sub: BV RDI properties}

It is worth briefly commenting on some properties on the BV RDI, Eq.~\eqref{eq: final compressible BV mode}, and how 
this depends on the sign of the $\Theta_{S} = 1+\zeta_{\rho}\Lambda_{S}$ factor (or, more precisely, whatever modified version of $\Theta_{S}$ appears due to dust stratification or a  more formal WKBJ treatment). Noting that 
$(\hat{\bm{k}}\times\driftvel)_{y}=\hat{k}_{z}\driftvelx - \hat{k}_{x}\driftvelz$, and that the term in square 
brackets in Eq.~\eqref{eq: final compressible BV mode} must be negative to cause an RDI, we see that when $\Theta_{S}>0$ and $\driftvelx\ll\driftvelz$,
an RDI occurs if $\driftvelz$ and $\nabla\ln P_{0}$ (or $\nabla\ln \rho_{0}$) have the same sign.
This is the ``natural'' direction for particles to drift when the gas is pressure supported and dust is not, i.e., in the  direction of gravity, towards the midplane of the disk.  In contrast, if $\Theta_{S}<0$, the RDI is most unstable when $\nabla\ln P_{0}$ and $\driftvelz$
have opposite signs, \emph{viz.}, when the dust in streaming in the direction opposite to gravity (this case is of course less physical but could occur, e.g., due to radiation pressure or another external force). If the dust has a substantial drift   perpendicular to the stratification direction ($\driftvelx \sim \driftvelz$), the BV RDI growth rate is comparable for either $\Theta_{S}>0$ or $\Theta_{S}<0$ (with different signs of $\hat{k}_{x}$).

Assuming, for the sake of discussion, that we have little dust stratification ($\Lambda_{\rho_{d}}\lesssim 1$) and that the
possible corrections to $\Theta_{S}$ in a more formal WKBJ treatment are minor, we see that the sign of $\Theta_{S}$ depends primarily on the drag regime  (Epstein or Stokes). Because $\Lambda_{S}>0$ for the system to be hydrodynamically stable,  grains in the Epstein regime ($\zeta_{\rho}\approx 1/2$) always satisfy $\Theta_{S}>0$ and so are unstable when $\driftvel$ and $\nabla P_{0}$ have the same orientation. 
As discussed further below in \S\ref{sub: discussion BV rdi}, this makes the BV RDI rather generic: it will occur whenever grains settle through a stratified atmosphere (see also \citealp{Lambrechts:2016fg}).
Grains in the Stokes regime, with $\zeta\approx -1/2$ can cause $\Theta_{S}$ to have either sign, depending on the strength of the entropy stratification $\Lambda_{S}$, so the instability will be slower growing and less generic for these larger grains. We illustrate the behavior 
of the dispersion relation as $\Theta_{S}$ flips sign in Fig.~\ref{fig:app: dust stratification}.

Finally, it is worth reiterating that our treatment here has suggested that the BV RDI is somewhat 
more robust, and faster growing, than predicted using the Boussinesq approximation (albeit with the caveats that come with assuming linear background gradients; see \S\ref{subsub: local linear BV caution}). This occurs because gas pressure perturbations are particularly 
important to the mechanism of the RDI (see \S\ref{sub: toy model}), but these are neglected in the Boussinesq 
treatment of BV oscillations. The two results agree for a gas that is very stably stratified, with $\Lambda_{S}\gg 1$ (e.g., strong temperature stratification in the direction opposite to the 
density stratification).

\subsection{Stratified epicyclic instability (the \SI\ including stratification)}\label{sub: BV epi RDI}

In this section, we calculate the RDI for the full stratified, rotating system. As discussed above, this
procedure yields an instability that is, in most regimes, very similar to the pure vertical-epicyclic RDI (\S\ref{subsub: epicyclic vertical streaming}), and we shall also term this instability the disk ``\si.'' Despite 
the complexity of the equations, we derive a relatively compact 
expression for the vertial-epicyclic-BV RDI to lowest order in $\tau_{s}$. The primary
purpose of this derivation is to highlight the relevance of the  results derived in \S\ref{sec: resonance epicyclic} and \S\ref{sub: BV RDI}. In particular, 
we find that the RDI of the full system---including joint epicyclic-BV gas oscillations, gas compressibility, a general dust drag 
law, and dust and gas stratification---behaves very similarly to  the unstratified epicyclic RDI (\si; \S\ref{sec: resonance epicyclic}),  with a slightly larger growth rate. 
We shall also see  that the double-resonant behavior studied in \S\ref{subsub:double resonant theta vertical epi}, which caused the streaming instability  growth rate to approach $\infty$ at $k\rightarrow \infty$,
is not pathological; i.e., the fast growth rates of the disk \si\ still exist in stratified regions of disks where the  vertical streaming velocity has a
clear physical origin. 

The same caveats regarding the local approximation apply here, specifically to those terms in the
epicyclic-BV RDI that arise from directly from the gas stratification. In particular, as outlined in \S\S\ref{subsub: local linear BV caution} and \ref{sub:sub: BV RDI properties}, the $\Theta_{S}$ term may be modified in a more formal WKBJ treatment. However, since this causes only minor modifications to the settling instability growth rate,  any minor modifications to $\Theta_{S}$ would have little effect on our general conclusions.

\begin{figure*}
\begin{center}
\includegraphics[width=1.05\columnwidth]{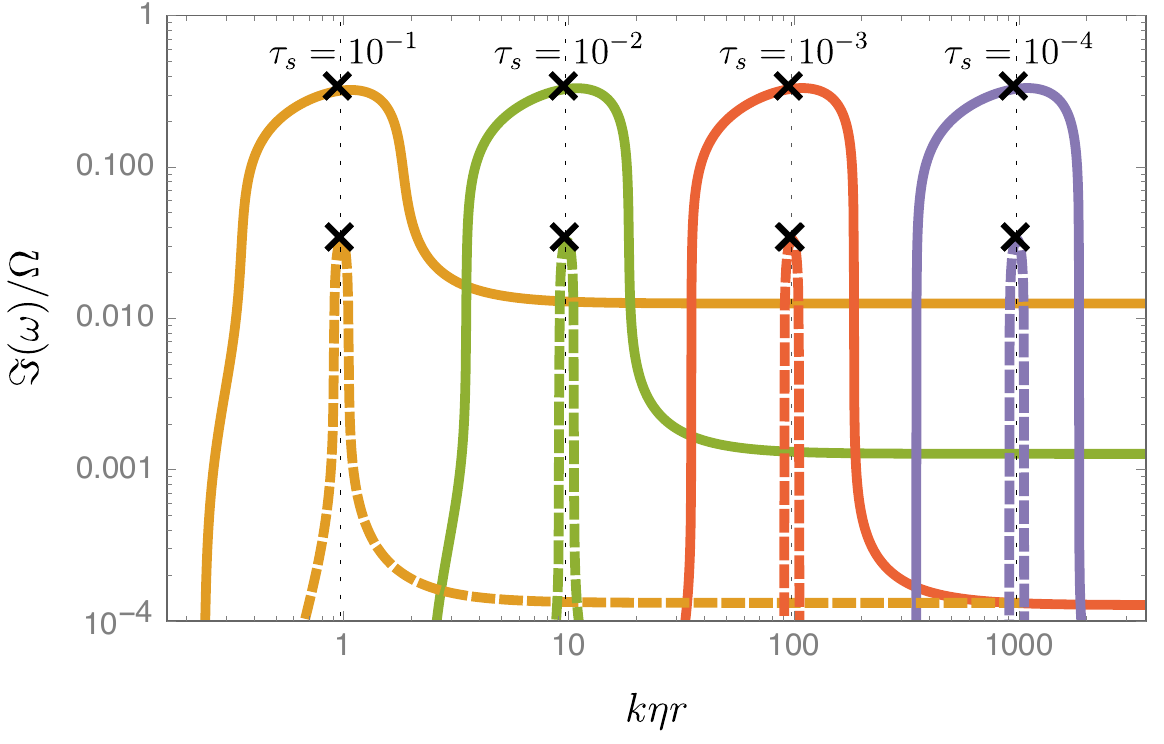}~~~
\includegraphics[width=1.03\columnwidth]{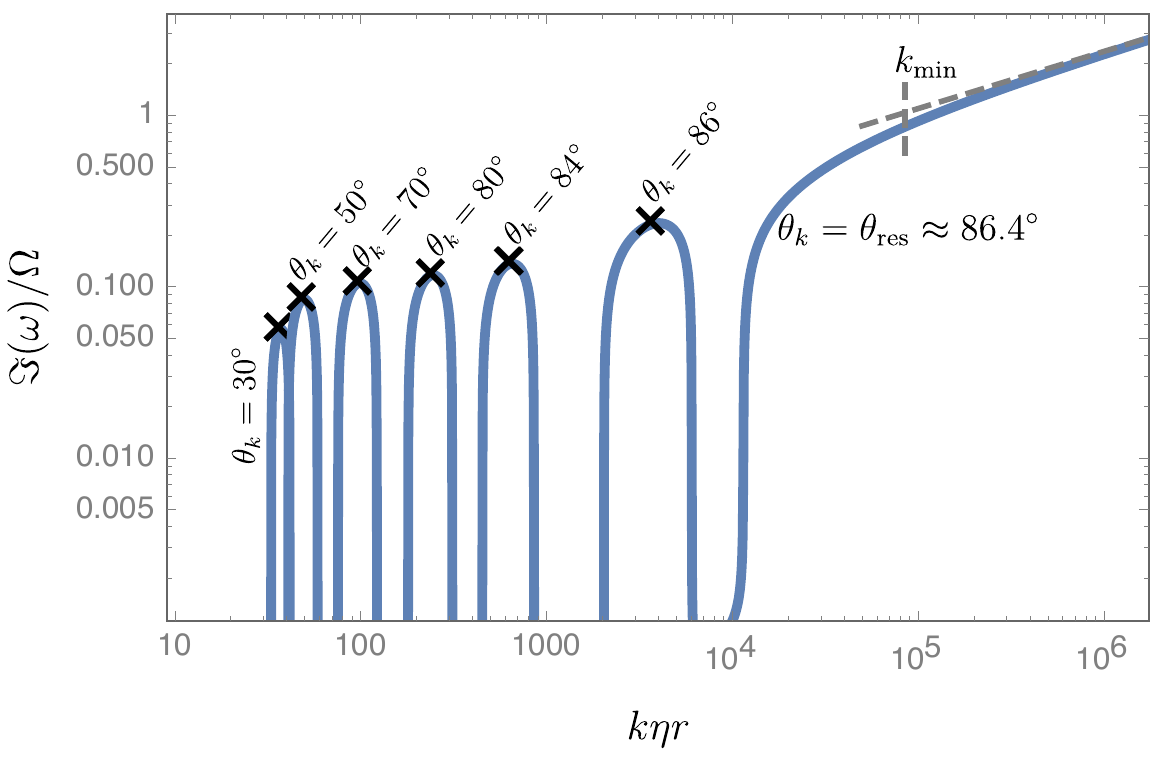}
\caption{
Illustration of the growth rate $\Im(\omega)$ of the vertical-epicyclic-\BV\ RDI (\si\ including gas stratification), which arises from the resonance of the dust with epicyclic-Brunt-V\"ais\"al\"a oscillations of the gas. The calculation includes gas compressibility and stratification in the vertical and
radial directions.  All curves are calculated through numerical solution of the local
dispersion relation from Eqs.~\eqref{eq: BV rho}--\eqref{eq: BV v d}, and use $\eta=0.001$ and $c_{s0}=\eta^{-1/2} (\eta U_{K})$, $L_{0}=h_{g}=\eta^{-1/2}(\eta r)$, $L_{0R}=\eta^{-1/2}L_{0}$, $\Lambda_{S}=2$, $\zeta_{\rho}=\zeta_{P}=1/2$ (Epstein drag),  and $\gamma_{\mathrm{gas}}=5/3$. 
The left-hand panel is of the same form as Fig.~\ref{fig: epicyclic 1D} and the left panel of Fig.~\ref{fig: vertical streaming epicyclic}, and shows the numerically calculated growth rate (curves) and analytic result (Eq.~\eqref{eq: epi BV RDI}; crosses) for a range of dust sizes as labeled: $\tau_{s}=10^{-1}$ (orange curves), $\tau_{s}=10^{-2}$ (green curves), $\tau_{s}=10^{-3}$ (red curves), and $\tau_{s}=10^{-4}$ (purple curves).   The solid curves show $\mu = 0.1$, the dashed curves show $\mu=0.001$, and we keep the mode angle fixed at $\theta_{k} = 70^{\circ}$ (this angle is chosen so as to show the characteristic RDI behavior, away from the double-resonant mode). The dispersion relations are very similar to those in Fig.~\ref{fig: vertical streaming epicyclic}, with the biggest differences being the slightly higher growth rates and resonant wavenumbers (see \S\ref{subsub: epi BV compared to epi}), and the slightly narrower resonances for otherwise identical parameters (this may change in a true global treatment, however).
The right-hand panel shows the dispersion relation at $\tau_{s}=10^{-3}$ and $\mu=0.01$ for a variety of angles  as labeled, from $\theta_{k}=30^{\circ}$ up to the double-resonant angle  where $\bm{k}\cdot \driftvel =0$ ($\theta_{k}\approx86.4^{\circ}$). In this panel, the curve formed by joining the maximum growth rate at each angle (black crosses)  effectively  illustrates the maximum disk \si\  growth rate as a function of wavenumber $k\eta r$. In addition, we show the double-resonant mode growth rate ($\Im(\omega)\propto k^{1/3}$) and minimum wavenumber $k_{\mathrm{min}}$ with the dashed gray lines (see Eq.~\eqref{eq: epi double res init polynomial comp}). It is worth noting that value   of the resonant $k$ as a function of $\theta_{k}$ is determined by $\omega_{\mathrm{EBV}}$, which implies that the resonant $k$ values decrease for a system that is closer to being neutrally stratified ($\Lambda_{S}=0$, such that $\omega_{\mathrm{EBV}}=\omega_{\mathrm{epi}}$); see \S\ref{subsub: epi BV compared to epi} for discussion. 
}
\label{fig: Epi BV}
\end{center}
\end{figure*}

For simplicity, we shall neglect radial stratification in our analytic derivations below; i.e., we set $L_{0R}^{-1}=0$ in Eqs.~\eqref{eq: BV rho}--\eqref{eq: BV v d}. In 
numerical results (i.e., Fig.~\ref{fig: Epi BV}), we include a radial stratification $\partial_{r}\ln P_{0}=\eta^{1/2}\partial_{z}\ln P_{0}$
and note that it makes very little difference to the results,  because  the BV RDI depends only weakly on slight differences between the streaming direction and stratification direction so long as $\Theta_{S}=1+\zeta_{\rho}\Lambda_{S}>0$ (see \S\ref{sub:sub: BV RDI properties}).

\subsubsection{Expansion in $\tau_{s}$}\label{subsub: expand in tau}

As in \S\ref{sub: BV RDI}, we carry out the expansion in $\epsilon\sim \tau_{s} $, which incorporates the smallness 
of $(kL_{0})^{-1}$ and $\driftvelmag/c_{s}$ (specifically $\driftvelmag/c_{s}\sim (kL_{0})^{-1}\sim \tau_{s}\sim \epsilon\ll1$) and  leads to relatively simple and physically intuitive expressions that are easily analyzed.
We do not feel that this restriction to $\tau_{s}\ll1$ is a severe limitation on the applicability of our results: grains with $\tau_{s}\sim 1$  settle out of stratified regions  quickly
with velocities approaching the sound speed (see \S\ref{subsub: streaming}), so are more naturally treated in 
the midplane region anyway (i.e., the YG streaming instability; see e.g., Fig.~\ref{fig: epicyclic 1D}).

The expected drift velocity from Eqs.~\ref{eq: NSH drift}--\eqref{eq: vertical settling drift},
is \begin{equation}
\frac{\driftvel}{c_{s0} }\approx (-2\eta^{-1/2}\tau_{s},\,\eta^{-1/2}\tau_{s}^{2},\,\tau_{s})+ \mathcal{O}(\tau_{s}^{3}).\label{eq: drift velocity at small tau}\end{equation}
To lowest order in $\tau_{s}$, this motion is simply due to the gas stratification, \emph{viz.,} it is the grain settling drift that would arise in a {stationary} gas with the pressure stratification that we have assumed for the disk ($\partial_{r}\ln P_{0} \sim \eta^{1/2}\partial_{z} \ln P_{0}$).

\subsubsection{Gas oscillations}

As in the non-rotating  case, there are five gas eigenmodes (Eqs.~\eqref{eq: BV rho}--\eqref{eq: BV p} with $\mu=0$), 
which are $\omega=0$ (a zonal mode in $u_{y}$),  two sound-wave eigenmodes,
\begin{equation}
\omega_{\mathcal{F}} = \pm \epsilon^{-1}c_{s0}k  + \dots,
\end{equation}
and two eigenmodes for  epicyclic-BV, or inertia gravity, oscillations,
\begin{equation}
\omega_{\mathcal{F}} = \pm \omega_{\mathrm{EBV}} = \pm \left(\Omega^{2}\hat{k}_{z}^{2} + N_{BV}^{2}\hat{k}_{x}^{2}\right)^{1/2} + \dots,\label{eq: epi-BV gas frequency}\end{equation}
where  $N_{BV}^{2} = \Lambda_{S}c_{s0}^{2}L_{0}^{-2}$. As 
was the case in a stratified gas without rotation (\S\ref{sub: BV RDI}), the 
epicyclic-BV oscillations are Boussinesq in character, and  incompressible  to lowest order. We note that when $N_{BV}\sim \Omega$, as occurs in a disk, the buoyancy force (stratification) most strongly modifies the epicyclic oscillations for radially directed modes ($\hat{k}_{x}>\hat{k}_{z}$).

\subsubsection{Resonant drag instability}\label{subsub: RDI for BV epi}

In the now familiar procedure, our next step is to evaluate the RDI growth 
rate (Eq.~\eqref{eq: dust eval pert}) using Eq.~\eqref{eq: drift velocity at small tau} for the drift velocity, and insert the resonant wavenumber,
\begin{equation}
k_{\mathrm{res}}\approx \pm \frac{\omega_{\mathrm{EBV}}}{\hat{\bm{k}}\cdot \driftvel} .\label{eq: kres for epi BV}\end{equation}
Taking the positive root and expanding the resulting expression in $\epsilon$ (i.e., in $\tau_{s}$), this yields, to lowest order, 
\begin{align}
\frac{\omega^{(1)}}{\Omega} &\approx \pm\mu^{1/2}\left[  \hat{k}_{x} (\hat{\bm{k}}\times\driftvel)_{y}\frac{\tau_{s} \hat{k}_{z} + \Omega^{-1}L_{0}^{-1}\Theta_{S}\driftvel\cdot \hat{\bm{k}}}{2\tau_{s }\driftvel\cdot \hat{\bm{k}}}\right]^{1/2} \nonumber \\
&\approx \pm i\left(\frac{\mu}{2}\right)^{1/2}\sin\theta_{k} \left( \frac{1+2\eta^{1/2}\cot\theta_{k}}{1-2\eta^{1/2}\tan\theta_{k}} \right)^{1/2}\nonumber \\ &\qquad \times \left[1+\frac{c_{s}}{\Omega L_{0}}\Theta_{S}\left(1-2\eta^{1/2}\tan\theta_{k}\right)\right]^{1/2},\label{eq: epi BV RDI}
\end{align}
where $\Theta_{S} = 1+\zeta_{\rho}\Lambda_{S}$ is the coefficient from the BV RDI (see Eq.~\eqref{eq: final compressible BV mode}), which becomes $\Theta_{S} = 1+\zeta_{\rho}\Lambda_{S}-\Lambda_{\rho_{d}}$ in the 
presence of dust stratification $d\ln\rho_{d}/dz=\Lambda_{\rho_{d}} L_{0}^{-1}$ (see App.~\ref{app: dust stratification}). The second line of Eq.~\eqref{eq: epi BV RDI} arises from inserting $\driftvel$ from Eq.~\eqref{eq: drift velocity at small tau}; note also that $c_{s}/(\Omega L_{0})\sim 1$ when $L_{0}\sim h_{g}$.
This expression contains aspects of both the incompressible epicyclic RDI, as discussed in \S\ref{subsub: epicyclic vertical streaming}, and the BV RDI discussed in \S\ref{sub: BV RDI}.
In particular, noting that $\driftvelz \gg \driftvelx$, setting $(\hat{\bm{k}}\times\driftvel)_{y}\approx -k_{x}\driftvelz$ and 
neglecting  stratification ($L_{0}\rightarrow\infty$), one obtains the epicyclic RDI (\si), Eq.~\eqref{eq: vertical epicyclic resonant omega mid angles}. Similarly, when the stratification
part, $\Omega^{-1}L_{0}^{-1}\Theta_{S}\driftvel\cdot \hat{\bm{k}}$, dominates, which occurs when  $\theta_{k}$ is sufficiently 
close to $\pi/2$ and $\hat{\bm{k}}\cdot \driftvel\sim \driftvelmag$ (i.e., we are not close to the double-resonant angle where $\hat{\bm{k}}\cdot \driftvel=0$),
the expression becomes identical to the BV RDI, Eq.~\eqref{eq: final compressible BV mode}.
However,  for most mode angles ($\eta^{1/2}\ll \theta_{k}\ll \pi/2-\eta^{1/2}$), the addition of the 
stratification has little qualitative effect, simply increasing the growth rate  compared to the pure epicyclic case by a factor $\sim (1+L_{0}^{-1} c_{s}/\Omega\,\Theta_{S})^{1/2}$, as well as changing the resonant wavenumber.
We illustrate this  in Fig.~\ref{fig: Epi BV} (note that these numerical solutions also include the radial stratification) and discuss the overall importance of stratification in more detail below (\S\ref{subsub: epi BV compared to epi}).

\subsubsection{Double resonant $\theta_{k}$}\label{subsub: double resonant mode with strat}

As was the case for the 
incompressible epicyclic RDI, we see that the RDI growth rate approaches infinity at the ``double-resonant'' angle, $\theta_{k}=\theta_{\mathrm{res}}$, where $\hat{\bm{k}}\cdot \driftvel=0$.
As in  \S\ref{subsub:double resonant theta vertical epi}, we study the properties of this by 
 inserting the ansatz
$\omega/\Omega = \varpi\,\mu^{1/3}\tau_{s}^{1/3}(k\eta r)^{1/3}$ into the characteristic polynomial 
of $\mathbb{T} = \mathbb{T}_{0}+\mu \mathbb{T}^{(1)}$ (Eqs.~\eqref{eq: BV rho}--\eqref{eq: BV v d}). An expansion in 
 in high $k$ ($k\eta r\sim \epsilon^{-1}$) and small $\mu$ and $\tau_{s}$ ($\mu\sim \epsilon^{\nu}$, $\tau_{s}\sim \epsilon^{1-\nu}$, with $0 < \nu<1$) yields (to lowest order in $\epsilon$) the polynomial 
\begin{equation}
\varpi^{3}- \varpi \frac{\omega_{\mathrm{EBV}}^{2}}{\Omega^{2}(\mu\,\tau_{s}k\eta r)^{2/3}}+ 2 \sin\theta_{k}=0.\label{eq: epi double res init polynomial comp}\end{equation}
For $\omega/\Omega = \varpi\mu^{1/3}\tau_{s}^{1/3}(k\eta r)^{1/3}\gg \omega_{\mathrm{EBV}}/\Omega$, Eq.~\eqref{eq: epi double res init polynomial comp} has the unstable root $\omega/\Omega\approx (-1/2+i\sqrt{3}/2)(2 \tau_{s}\mu\, k_{x}\eta r)^{1/3}$. This is identical to the incompressible epicyclic double-resonant solution (Eq.~\eqref{eq: epi double res init solution}),
aside from a modified low-$k$ cutoff. In particular, the solution \eqref{eq: epi double res init solution} is now 
valid for $\omega\gg \omega_{\mathrm{EBV}}$, or $k\eta r \gg k_{\mathrm{min}}\eta r\sim (\omega_{EBV}/\Omega)^{3}/(\mu\tau_{s})$, rather than for $\omega\gg \omega_{\mathrm{epi}}=\Omega \cos\theta_{k}$ (see Eq.~\eqref{eq: epi double res low k cutoff }). Because $\driftvelz\gg\driftvelx$ and $\theta_{\mathrm{res}}\approx \pi/2$, $\omega_{\mathrm{EBV}}>\omega_{\mathrm{epi}}$ and the double-resonant solution is cut off at higher a higher value of $k_{\mathrm{min}}$ than in the case without stratification. 
This behavior can be straightforwardly understood: the frequency of the double-resonant mode 
must be larger than that of the background gas oscillations to follow the simple $\Im(\omega)\sim k^{1/3}$ scaling, and pure epicycles have a lower frequency than epicyclic-BV oscillations when $\hat{k}_{z}\gg \hat{k}_{x}$ (compare Eqs.~\eqref{eq: epicyclic evalues and modes} and \eqref{eq: epi-BV gas frequency}). 
We illustrate the double-resonant mode in the right-hand panel of
Fig.~\ref{fig: Epi BV}.
These results show that very large growth rates---such that the instability grows very fast compared to the time required for particles to settle ($\Omega\, t_{\mathrm{settle}}\sim  \tau_{s}^{-1}$)---are possible when small grains settle through rotating stratified regions towards the midplane of the disk. Astrophysical implications and the effects of gas viscosity are discussed further in \S\ref{sub: discussion epi rdi}.

\subsubsection{The dependence on stratification}\label{subsub: epi BV compared to epi}

The vertical temperature stratification profile, which determines $\Lambda_{S}$, is uncertain  in disks, depending on details of the environment and central object (e.g., heating from radiation).  Further, as outlined in \S\ref{subsub: local linear BV caution}, there are uncertainties related to the details of the theoretical treatment of stratification, which may change $\Theta_{S}$. For this reason, in this section we summarize how the properties of the \si\ (vertical-epicyclic-BV RDI) depend on gas stratification (in particular $\Lambda_{S}$, which changes the BV frequency at constant pressure gradient), the regime of 
dust drag (see also \S\ref{sub:sub: BV RDI properties}), and possible modifications to $\Theta_{S}$. 
For the purposes of planetesimal formation, 
we are most interested in the maximum growth rate of the RDI  over mode angles, as a function of
$k \eta r$, since long-wavelength instabilities are likely more dynamically important than short-wavelength 
instabilities when each have similar growth rates. 
There are then two separate modifications due to the stratification: the first, the modification of the resonant wavenumber due to the different gas oscillations; the second, the change 
in the growth rate itself through $\Theta_{S}$; i.e.,  Eq.~\eqref{eq: epi BV RDI}. 

Let us first consider the modification of the resonant wavenumber. As discussed in \S\ref{subsub: epicyclic vertical streaming}, with pure epicyclic modes, the resonant wavenumber is always $k_{\mathrm{res}} \approx \Omega/\driftvelz$, with no dependence on angle (when $\eta^{1/2}\ll |\theta_{k}| \ll \pi/2-\eta^{1/2}$).
This is because $\bm{k}\cdot \driftvel \approx k_{z}\driftvelz$ has the same angular dependence as
the gas epicyclic frequency $\omega_{\mathrm{epi}}=\hat{k}_{z}\Omega$. 
This is no longer the case when there also 
exists buoyancy force, because $k_{\mathrm{res}} \approx \omega_{\mathrm{EBV}}/(\hat{k}_{z}\driftvelz)$
and $ \omega_{\mathrm{EBV}} \approx N_{BV} \hat{k}_{x}$ when  $N_{BV} \hat{k}_{x} \gg \Omega \hat{k}_{z}$ (see Eq.~\eqref{eq: epi-BV gas frequency}). The scale of the faster-growing modes, which have a primarily radial orientation ($\hat{k}_{x}>\hat{k}_{z}$), 
thus decreases somewhat as $\Lambda_{S}$ increases. This 
can be  seen in the right-hand panel of Fig.~\ref{fig: Epi BV}, which shows how the resonant mode increases in 
$k$ as the mode angle is increased. If we choose a lower value for $\Lambda_{S}$, corresponding to a stratification profile that is closer to being buoyantly unstable, there is less dependence on the mode angle, and the double-resonant solution (which is valid once $\omega\gg \omega_{\mathrm{EBV}}$; see Eq.~\eqref{eq: epi double res init polynomial comp}) applies for a smaller value of $k$.

The growth rate itself also depends on $\Lambda_{S}$ through $\Theta_{S} = 1+\zeta_{\rho}\Lambda_{S}-\Lambda_{\rho_{d}}$. As clear from the second equality in Eq.~\eqref{eq: epi BV RDI},  the
growth rate increases (decreases)  for $\Theta_{S}>0$ ($\Theta_{S}<0$). Thus, in the Epstein drag regime, with 
$\zeta_{\rho}\approx 1/2$, the growth rate increases with increasing temperature stratification (increasing $\Lambda_{S}$), while opposite occurs for grains in the Stokes drag regime ($\zeta_{\rho}\approx -1/2$). If  $\Theta_{S}\lesssim -1$ ---as could occur in the Stokes drag regime with large $\Lambda_{S}$, or with very strong dust stratification (assuming our expressions for $\Theta_{S}$ are approximately correct) ---the epicyclic-BV RDI can be stabilized at this order in $\tau_{s}$. However, note that more extreme conditions (i.e., larger $\Lambda_{S}$ or $\Lambda_{\rho_{d}}$) are required to achieve such stabilization than for the BV RDI because of the destabilizing contribution from the epicyclic part (c.f. Eqs.~\eqref{eq: final compressible BV mode} and \eqref{eq: epi BV RDI}).

Overall, we see that in the most relevant Epstein drag regime, an increasingly stable temperature stratification (larger $\Lambda_{S}$)  causes moderate increases in the growth rates for most modes, 
while causing the scale of the faster-growing modes (with $\hat{k}_{x}>\hat{k}_{z}$) to decrease, and disrupting the double-resonant mode at higher $k$. In practice, the effect of these changes on the maximum  growth rate ($\Im(\omega)(k)$ marginalized over $\theta_{k}$) is relatively minor, because the growth rates of the RDI modes and double-resonant modes at the same wavelength differ only by order-unity factors.

\section{Other physics \&\ resonant drag instabilities of interest}\label{sec: MHD and others}

In this section, we outline various other RDIs and their relevance to protoplanetary disks and planetesimal formation. 
These include instabilities arising from resonance with sound waves, magnetosonic waves, 
and nonideal MHD waves. Our general conclusion is that due to the relatively low streaming velocity of grains and 
low ionization fraction, such instabilities are unlikely to be important for dust clumping near the disk midplane in standard  models (e.g., the MMSN model); however, far from the midplane and in winds, such instabilities could play an important role. 
Here we briefly outline the mathematical aspects of each, and astrophysical considerations are discussed further in \S\ref{sub: discussion magnetic rdi}.

\subsection{Acoustic instability}\label{sub: resonance hydro}
The acoustic RDI, explored in detail in \papertwo, involves the resonance 
between streaming dust and gas sound waves. Because sound waves satisfy $\omega_{\mathcal{F}}=\pm k c_{s}$,
and thus always have a phase velocity of $c_{s}$, the resonance condition $\bm{\hat{k}}\cdot \driftvelhat = c_{s}/\driftvelmag$ can be satisfied only if $\driftvelmag > c_{s}$.  Thus, in the bulk disk, where grains generally stream with $\driftvelmag \ll c_{s}$, we do not expect the acoustic RDI 
to be important.\footnote{One possible exception may be large grains with $\tau_{s}\gtrsim 1$ displaced from the midplane, which would oscillate about the midplane with speeds approaching $c_{s}$ (see \S\ref{subsub: streaming}). However, it is not clear what physics might cause 
large grains to reach a significant distance above the midplane in sufficient numbers such that our continuum approach is valid, so we do not consider this further.} As shown in \papertwo, there also exists  a non-resonant 
acoustic instability, which is unstable for $\driftvelmag < c_{s}$. However, the fastest growth rate of this instability 
is $t_{s}\,\Im(\omega) \sim \mu (\driftvelmag/c_{s})^{2}$, which (at $\tau_{s}\ll 1$) is $\Im(\omega)/\Omega \sim \mu\,\tau_{s}^{-1} (\driftvelmag/c_{s})^{2}\sim \mu\, \eta\, \tau_{s}$ for NSH streaming (Eq.~\eqref{eq: NSH drift}) or $\Im(\omega)/\Omega \sim \mu\, \tau_{s}$ for vertical settling (Eq.~\eqref{eq: vertical settling drift}), suggesting its growth rates are likely too small to play any  significant role in dust dynamics.

\subsection{Magnetosonic instability}\label{sub: resonance MHD}

Another, more complicated RDI, is that due to the resonance with MHD waves. 
In \paperone and \citet{Hopkins:2018rdi}, we study the ``magnetosonic RDI,'' arising from the resonance
with fast or slow magnetosonic waves.\footnote{The importance of Lorentz and electrostatic forces (e.g.\ Coulomb drag), and other forces related to grain charge (e.g.\ photo-electric or photo-desorption processes) is briefly discussed in \S\ref{sec: other effects} and extensively discussed in  \citet{Hopkins:2018rdi}. At the densities and temperatures of protoplanetary disks, these terms are completely negligible compared to Epstein/Stokes drag, although they might be important in the diffuse gas in disk winds.} Compared to the sonic instability discussed above (\S\ref{sub: resonance hydro}), the magnetosonic RDI has the potential to be more interesting for protoplanetary disk dynamics: it is possible for grains to be in resonance with the 
slow wave for any $\driftvelmag \lesssim c_{s0}$, because the phase velocity of the slow wave approaches zero perpendicular
to the magnetic field. Further, in the absence of dissipation,  the instability's growth rate formally approaches infinity at small scales \emph{for any} $\driftvelmag$ ($\Im(\omega)\sim k^{1/3}$ at very large $k$; see \paperone\ and \papertwo).  Similar instabilities also occur when the  grains are
 charged, and thus directly affected by magnetic fields as well as gas drag; these are studied in \citet{Hopkins:2018rdi}. 
 
Here, we simplify the (rather complicated) expressions from \paperone\ for uncharged grains, in the limit where the
streaming is much less than the sound speed, as relevant to protoplanetary disks. 
However, we shall see that nonideal effects, which are very strong in the bulk regions of protoplanetary disks 
due to the low ionization fraction \citep{2001ApJ...552..235B,2007Ap&SS.311...35W}, limit the growth rate of
the magnetosonic RDI to very low values. For this reason, this section is kept quite brief, and we provide only
 simplistic estimates.

\subsubsection{The  system to be solved}

To isolate the relevant physics, and because we will find that the magnetosonic RDI is unlikely to be important anyway, 
we shall neglect rotation (epicyclic oscillations and the background shear) and stratification throughout this section. We thus consider the fiducial gas system, Eqs.~\eqref{eq: gas NL rho}--\eqref{eq: dust NL v}, but with $\bm{g}=0$, an additional magnetic stress  $(4\pi)^{-1} (\nabla\times \bm{B})\times \bm{B}$ in the gas momentum equation 
\eqref{eq: gas NL u}, and the magnetic field evolution equation,
\begin{align}
\partial_{t}\bm{B} &= \nabla\times (\bm{u}\times \bm{B}) - \nabla\times( \eta_{\mathrm{Ohm}}\bm{J}) - \nabla\times (\eta_{\mathrm{Hall}} \bm{J}\times \hat{\bm{b}}) \nonumber \\& \qquad- \nabla \times \left[\eta_{\mathrm{Ambi}} \hat{\bm{b}} \times (\bm{J}\times \hat{\bm{b}})\right].\label{eq: induction with nonideal}\end{align}
Here, $\bm{J}=\nabla\times \bm{B}$ is the current density,  $\hat{\bm{b}}=\bm{B}/B$ (where $B\equiv|\bm{B}|$), and
 $\eta_{\mathrm{Ohm}}$, $\eta_{\mathrm{Hall}}$, and $\eta_{\mathrm{Ambi}}$ are the effective diffusivities for Ohmic diffusion, the Hall effect, and Ambipolar diffusion respectively  \citep{1994ApJ...421..163B,2007Ap&SS.311...35W,Lesur:2014cc}, which are the most important nonideal MHD effects that arise from the low ionization fraction in protoplanetary disks (see \S\ref{subsub: resonance nonideal MHD}).
 
  As in earlier sections, we linearize Eqs.~\eqref{eq: gas NL rho}--\eqref{eq: dust NL v} and Eq.~\eqref{eq: induction with nonideal} to apply the RDI algorithm described in \S\ref{sec: resonance}.
We assume a homogenous background and neglect stratification and rotation, making the geometry arbitrary. We specify the background field strength through the ratio of thermal to magnetic pressure $\beta= 8\pi P_{0}/B_{0}^{2}$, with $\beta>1$ expected in disks, and also define the Alfv\'en speed $v_{A0}\equiv B_{0}/\!\sqrt{4\pi \rho_{0}}$.
Neglecting nonideal corrections ($\eta_{\mathrm{Ohm}}=\eta_{\mathrm{Hall}}=\eta_{\mathrm{Ambi}}=0$) the gas supports three sets of forwards and backwards propagating waves \citep{Alfven:1942hl}: the shear-Alfv\'en wave, the slow wave, and the fast wave. The shear-Alfv\'en and slow wave each approach zero phase speed at angles perpendicular to the magnetic field ($\omega_{\mathcal{F}}\rightarrow 0$ as $\hat{\bm{k}}\cdot \bm{B}_{0}\rightarrow0$), while the fast wave behaves like a sound wave modified by the magnetic pressure.
We specify the dust streaming velocity to be at angle $\theta_{w}$ to the magnetic field (i.e., $\hat{\bm{b}}_{0}\cdot \driftvelhat = \cos\theta_{w}$).

\subsubsection{Magnetosonic RDI}

When Lorentz forces on grains are negligible compared to drag forces, the shear-Alfv\'en wave does not cause an RDI at moderate wavelengths because $\omega^{(1)}$ in Eq.~\eqref{eq: dust eval pert} evaluates to zero. This
is expected based on our toy model in \S\ref{sub: toy model} because linear shear-Alfv\'en waves do not contain a gas pressure perturbation. The fast and slow waves each cause  RDIs for waves propagating at angles $\hat{\bm{k}}\cdot \driftvel = v_{F}$ and $\hat{\bm{k}}\cdot \driftvel = v_{S}$ respectively, where
$v_{F}$  and $v_{S}$ are the fast and slow wave phase velocities  (these depend on $\hat{\bm{k}}\cdot \hat{\bm{b}}_{0}$). 
The growth rates of these magnetosonic   RDIs
can then be calculated directly from  Eq.~\eqref{eq: dust eval pert}; however, the resulting expression is complicated and unintuitive, so we do not reproduce it here (see Eq.~(15) of \paperone).

For the conditions relevant in a protoplanetary disk, we are most interested in the limit $\driftvelmag \ll c_{s0}$ and usually $\beta\gg 1$ (or equivalently $c_{s0}\gg v_{A0}$). Thus, only the slow wave resonance
is of interest, because it  can occur when $\driftvelmag <c_{s0}$. 
As discussed in detail in \citet{Hopkins:2018rdi}, this resonance is possible---i.e., it is possible to satisfy the resonance 
condition \eqref{eq: resonant condition} with $\omega_{\mathcal{F}}=kv_{S}$---for 
modes that are nearly perpendicular to the magnetic field; specifically, modes that satisfy
\begin{equation}
\arctan \left( \frac{|v_{A0}/\driftvelmag-\cos\theta_{w}|}{\sin\theta_{w}} \right)< \theta_{k}^{B}<\frac{\pi}{2},\label{eq: magnetosonic RDI theta range}\end{equation}
where $\theta_{k}^{B}$ parameterizes the angle between $\hat{\bm{b}}_{0}$ and $\hat{\bm{k}}$ ($\cos\theta_{k}^{B} = \hat{\bm{b}}_{0}\cdot\hat{\bm{k}}$).
 If $v_{A0}\ll \driftvelmag$, an RDI will usually be possible for a relatively wide range of angles (unless $\theta_{w}$ is close to zero), while if $v_{A0}\gg \driftvelmag$ there will be only  a narrow range around $\theta_{k}^{B}=\pi/2$ where the resonance is can occur. 
We note that  
 the magnetic field is likely primarily  directed in the toroidal direction in a disk \citep{2017ApJ...845...75B}, while the streaming velocity of small grains is dominated by radial or 
vertical motions for small grains (see \S\ref{subsub: streaming}). Thus $\theta_{w}\approx \pi/2$ and $\theta_{k}^{B}\approx\pi/2$ are reasonable values to use for  simplistic estimates . 
Expanding in $v_{A0}/c_{s0}\sim\driftvelmag/c_{s0}\ll 1$, and
noting  that $\zeta_{\bm{w}}\approx 0$ and 
$\zeta_{\rho}+\gamma_{\mathrm{gas}}\zeta_{P}\approx (1+\gamma_{\mathrm{gas}})/2$ (see  Eq.~\eqref{eq: eta and zeta expressions}),\footnote{Of course,
there is no requirement that $v_{A0}/c_{s0}$ and $\driftvelmag/c_{s0}$ be of similar order. A  more complete analysis is carried out in section 5.3 of \citet{Hopkins:2018rdi}, yielding similar results.}, one can simplify the slow-mode RDI growth rate from Eq.~\eqref{eq: dust eval pert} to 
\begin{align}
\omega^{(1)}_{S} \approx &\frac{1+i}{\sqrt{2}}\left(\frac{\mu  c_{s0} k}{2 t_{s0}}\right)^{1/2}\left(\frac{v_{A0}}{c_{s0}}\right)^{3/2}\nonumber\\&\times\left[\frac{1-\gamma_{\mathrm{gas}}}{2} \cos\theta_{k}^{B} \left(\cos^{2}\theta_{k}^{B}-\frac{\driftvelmag}{v_{A0}}\hat{\bm{b}}_{0}\cdot \driftvelhat \right)\right]^{1/2}.\label{eq: simp magnetosonic RDI}
\end{align}
The  noteworthy feature of Eq.~\eqref{eq: simp magnetosonic RDI} is its scaling with 
$(v_{A0}/c_{s0})^{3/2}$, which shows that the growth rate of the RDI is rather low 
in the high-$\beta$, subsonic regime, as might be expected.

\begin{figure}
\begin{center}
\includegraphics[width=1.0\columnwidth]{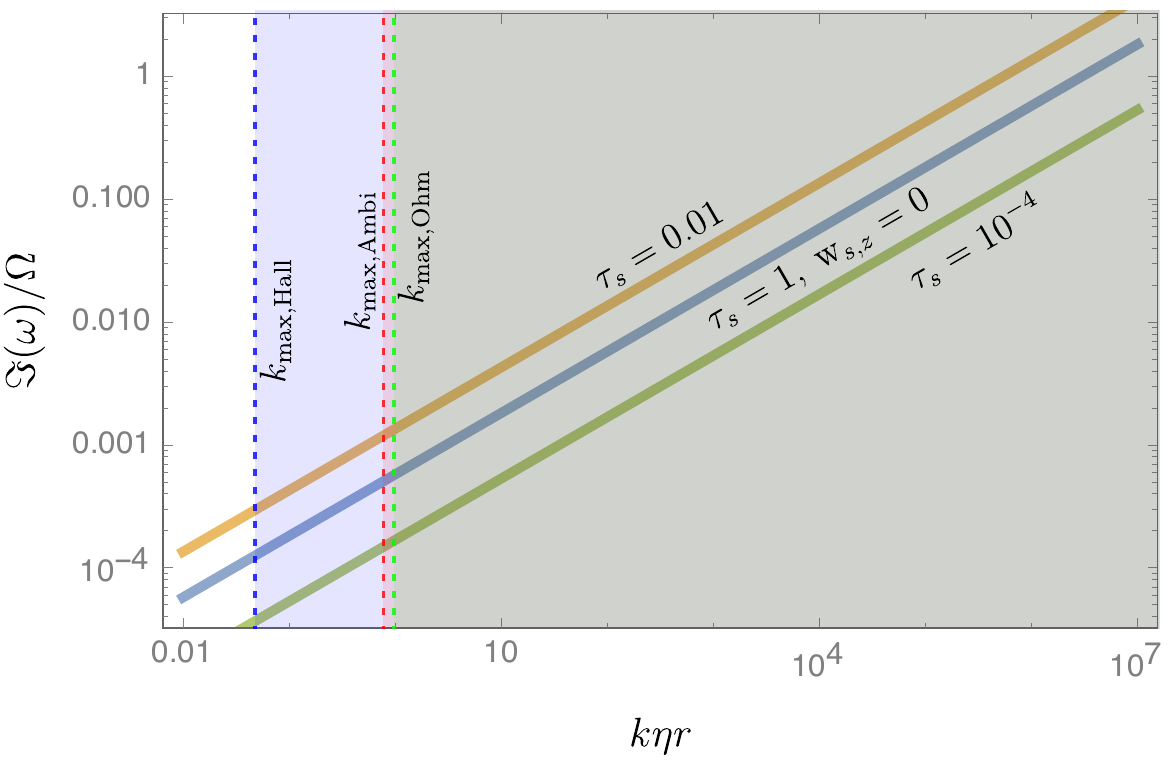}
\end{center}
\caption{The ideal magnetosonic RDI (dust-gas instabilities resonant with slow-magnetosonic waves; here neglecting rotation or stratification effects), for $\beta=100$ and grain sizes as labeled (blue solid line, $\tau_{s}=1$; orange solid line, $\tau_{s}=0.01$; green solid line, $\tau_{s}=10^{-4}$. 
We use  parameters appropriate to the MMSN model at $r\approx 1.5 \mathrm{AU}$, with  $\eta=0.001$ (as above), $\mu = 0.01$, and $\gamma_{\mathrm{gas}}=5/3$. Noting that the magnetic field might be primarily in the toroidal direction, we specify the angle between field and drift $\theta_{w}=80^{\circ}$, then chose the mode angles $\theta_{k}^{B}$, $\varphi_{k}^{B}$ such that the resonant condition is satisfied ($\hat{\bm{k}}\cdot\driftvel = v_{S}$) and the growth rate is approximately maximal for this RDI (see Eq.~\eqref{eq: magnetosonic RDI theta range}; this is $\theta_{k}^{B}\approx 69^{\circ}$, $\theta_{k}^{B}\approx 85^{\circ}$, and $\theta_{k}^{B}\approx 89.95^{\circ}$ for $\tau_{s}=1$, $\tau_{s}=0.01$, and  $\tau_{s}=10^{-4}$, respectively). 
For the $\tau_{S}=0.01$ and $\tau_{s}=10^{-4}$ curves, we evaluate $w_{s}$ as that arising from both the NSH and vertical settling drift, while for $\tau_{s}=1$ we use only the NSH drift (i.e., $\driftvelz=0$; which makes $\driftvelmag$ a factor $\sim \eta^{1/2}$
smaller). The vertical dashed lines and shaded regions show where we expect nonideal effects to become 
important (see \S\ref{subsub: resonance nonideal MHD}) due to Ohmic diffusion (green dashed line), ambipolar diffusion (red dashed line), or the Hall effect (blue dashed line). We use an  ionization fraction $x_{e}\sim 10^{-10}$ and the MMSN model  (the gray shaded region shows where all three effects are important). 
For typical expected parameters of protoplanetary disks (at least near the midplane in the regions of the disk not too close to the star), the magnetosonic RDI is suppressed by nonideal effects before reaching large growth rates.
\label{fig: standard MHD}}
\end{figure}

In Fig.~\ref{fig: standard MHD}, we plot the magnetosonic RDI growth rate for reasonable disk parameters ($\beta=100$), in  disk units, so as to allow direct comparison to previous figures. We see that rather large growth rates, approaching $\Im(\omega)\sim \Omega$,
are predicted at very small scales, for a variety of different grain sizes.  However, as discussed in the next section, 
nonideal effects are likely very important under the cold conditions expected near the disk midplane, and these 
will suppress magnetosonic waves (hence the magnetosonic RDI as well) even on relatively large spatial scales, limiting $\Im(\omega)$ to correspondingly small values. 
Note that the high-$k$ scaling, $\Im(\omega)\sim k^{1/3}$ (see \paperone\ and \papertwo), occurs at yet higher-$k$ than shown in Fig.~\ref{fig: standard MHD}, and so we have refrained from discussing this for simplicity.

\subsubsection{Nonideal MHD effects}\label{subsub: resonance nonideal MHD}

Due to the very low ionization fraction, nonideal MHD effects play a key role in protoplanetary disks
 (see, e.g., \citealp{1994ApJ...421..163B,1999MNRAS.307..849W,2001ApJ...552..235B,Kunz:2004ib,2007Ap&SS.311...35W,Bai:2013jg}). In particular, $\eta_{\mathrm{Ohm}}$, $\eta_{\mathrm{Hall}}$, and $\eta_{\mathrm{Ambi}}$ in Eq.~\eqref{eq: induction with nonideal} may {not} be assumed small throughout the
disk and can strongly influence the dynamics \citep{Bai:2013jg,Lesur:2014cc,2017ApJ...845...75B}.
For the slow-magnetosonic RDI, these terms become important when $\eta_{\mathrm{Diss}} k^{2}\bm{B} \sim \partial_{t}\bm{B}\sim k v_{S} \bm{B}$ (since $\omega\sim kv_{S}$),
suggesting that the maximum $k$ at which we can expect the standard magnetosonic RDI to operate is 
$k_{\mathrm{max,Diss}} \sim v_{S}/\eta_{\mathrm{Diss}}$, where $\eta_{\mathrm{Diss}}$ can be one of 
$\eta_{\mathrm{Ohm}}$, $\eta_{\mathrm{Hall}}$, or $\eta_{\mathrm{Ambi}}$.
Using the notation of \citet{Lesur:2014cc} (see also \citealt{2001ApJ...552..235B,2007Ap&SS.311...35W,Bai:2013jg}), 
the diffusivities are,
\begin{equation}
\eta_{\mathrm{Ohm}}\approx \frac{c^{2}m_{e}n\langle \sigma v\rangle_{e}}{4\pi e^{2}n_{e}},\:\eta_{\mathrm{Hall}}\approx\frac{B c}{\sqrt{4\pi}en_{e}}, \: \eta_{\mathrm{Ambi}}\approx \frac{B^{2}(m_{n}+m_{i})}{\langle \sigma v\rangle_{i}m_{i}m_{n}n_{i} n},\label{eq: diffusivities for MHD}\end{equation}
where $m_{e}$, $m_{i}$, and $m_{n}$ are the electron, ion, and neutral effective masses, $n_{e}$, $n_{i}$, and $n_{n}$ are the electron, ion, and neutral number densities, and  $\langle \sigma v\rangle_{e}$ and $\langle \sigma v\rangle_{i}$ are the electron-neutral and ion-neutral collision rates.
In Fig.~\ref{fig: standard MHD}, we overplot  $k_{\mathrm{max,Diss}}$ obtained from Eq.~\eqref{eq: diffusivities for MHD} from Ohmic and ambipolar diffusion and the Hall effect at an ionization fraction $x_{e}\sim10^{-10}$, showing that the ideal magnetosonic RDI  is affected by all three nonideal effects well before reaching interesting growth rates.

\begin{figure}
\begin{center}
\includegraphics[width=1.0\columnwidth]{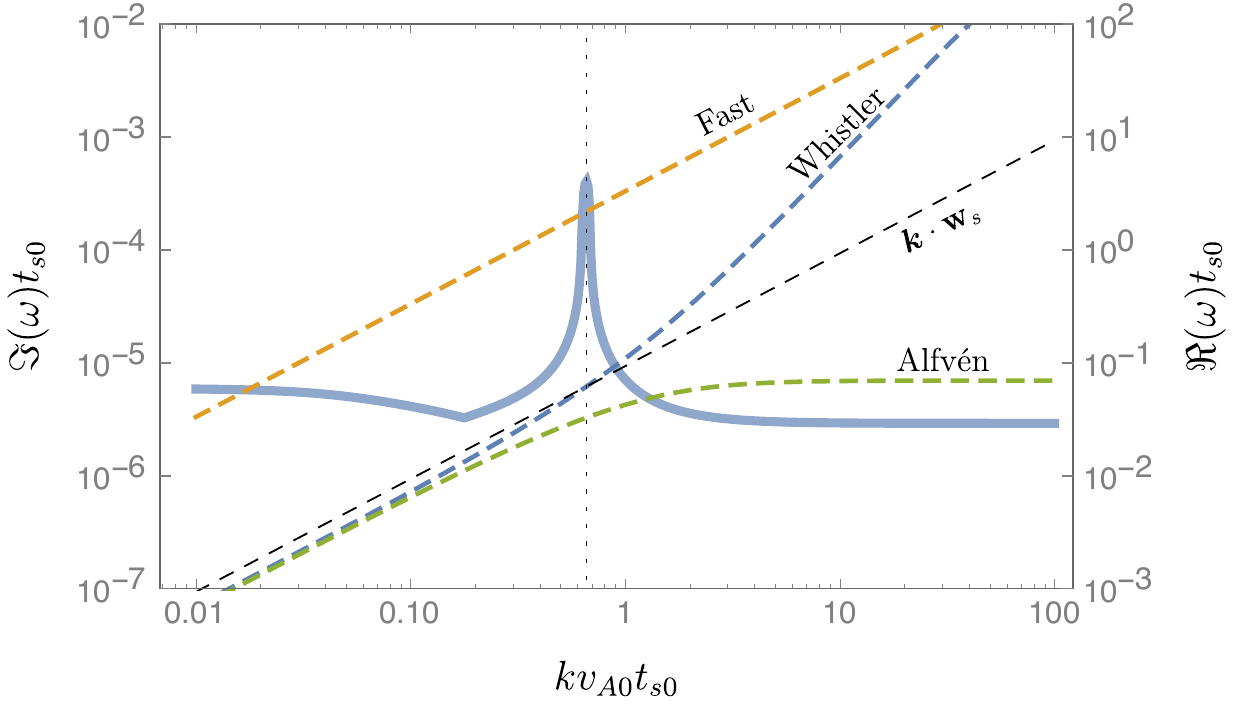}
\caption{The Hall-MHD RDI (RDI resonant with Hall-whistler waves). The solid line and left-hand axis shows $\Im(\omega)$ of the fastest-growing mode, numerically 
calculated  from Eqs.~\eqref{eq: gas NL rho}--\eqref{eq: dust NL v} and \eqref{eq: induction with nonideal} including
the Hall effect, with $k_{\mathrm{max,Hall}}\sim v_{A0}/\eta_{\mathrm{Hall}}= (v_{A0}t_{s0})^{-1}$. We use $\beta=100$, $\mu=0.01$, $\driftvelmag = 0.1 v_{A0}$, $\theta_{w}=80^{\circ}$, $\theta_{k}^{B}=86^{\circ}$, $\varphi_{k}^{B}=20^{\circ}$, and $\gamma_{\mathrm{gas}}=5/3$. Note that we use  units normalized to the dust stopping time (rather than disk units), because this 
figure is intended as a general example and we have not embarked on a full analysis of the Hall-MHD RDIs.  The 
right-hand axis and dashed lines show the dispersion relation $\Re(\omega)$ for Hall-MHD waves (the Alfv\'en, whistler, and fast branches, as labeled), while the black thin dashed line shows $\bm{k}\cdot \driftvel$. At the  
 resonance condition, $\bm{k}\cdot \driftvel=\omega_{\mathrm{whist}}$ (vertical dotted line), $\Im(\omega)$ peaks strongly---the Hall-whistler RDI. Similarly, the Hall-Alfv\'en RDI occurs in resonance with the Alfv\'en branch, although it has somewhat lower growth rates. }
\label{fig: Hall MHD}
\end{center}
\end{figure}

A secondary question then becomes: are there  different, nonideal MHD RDIs, which can operate for $k>k_{\mathrm{max,Diss}}$? For an RDI to be generically unstable when $\driftvelmag\ll c_{s}$, the fluid must support some undamped wave with similar
characteristics to the slow mode (in particular a phase velocity that approaches zero at some angle). 
Unfortunately, this immediately rules out an RDI modified by Ohmic diffusion, since for constant $\eta_{\mathrm{Ohm}}$, the relevant term in Eq.~\eqref{eq: induction with nonideal}
is simply $\eta_{\mathrm{Ohm}}\nabla^{2}\bm{B}$, which damps all $\bm{B}$ perturbations.
Similarly, ambipolar diffusion, although more complicated than standard Ohmic diffusion, does 
not allow for any undamped or weakly damped waves  other then the fast wave,  in the regime of interest. However, the Hall term, $\nabla\times (\eta_{\mathrm{Hall}}\bm{J}\times \hat{\bm{b}})$, does not dissipate waves, but 
simply acts to modify the slow and shear-Alfv\'en waves into whistler and Alfv\'en branches. These  no longer 
have a constant phase velocity for $k>k_{\mathrm{max,Hall}}$ ($\omega_{\mathrm{whist}} \sim k^{2}$ for the whistler branch; $\omega_{\mathrm{Alfv}}\sim \mathrm{const.}$ for the Alfv\'en branch; see Fig.~\ref{fig: Hall MHD}), but can still cause a ``Hall-whistler RDI'' or  ``Hall-Alfv\'en RDI'' at    the wavenumbers for which $\bm{k}\cdot \driftvel = \omega_{\mathrm{whist}} $ or $\bm{k}\cdot \driftvel = \omega_{\mathrm{Alfv}} $ respectively.
This provides an entertaining example of the RDI in a system
with a more complex wave structure, and so in Fig.~\ref{fig: Hall MHD} we show a numerical calculation of the  Hall-whistler RDI growth rate, along with the
dispersion relation of Hall-MHD waves.
We see that, exactly as predicted in \S\ref{sec: resonance}, $\Im(\omega)$ peaks strongly when $\bm{k}\cdot \driftvel = \omega_{\mathrm{whist}} $. However, even 
when the Hall terms dominate, ambipolar and Ohmic diffusion terms are generally still important
at relatively low $k$ (see Fig.~\ref{fig: standard MHD}), and will damp the Hall-MHD RDIs. For this reason we do not expect the Hall-MHD RDIs to be of particular importance in protoplanetary disks, at least under typical disk midplane conditions  at radii $r\gtrsim 0.1\rightarrow1\,\mathrm{AU}$ (where a MMSN-type model applies). 
However, it is worth noting that ideal MHD can be a reasonable approximation 
 in both the upper regions (well above the midplane; see, e.g., \citealp{2017ApJ...845...75B}) and inner regions (close to the central protostar; see, e.g., \citealp{2017ApJ...835..230F}) of the disk, which are thought to be well ionized. While 
the physics is such regions can also become more complex due to the expected higher  levels of turbulence (the
magnetorotational instability is likely to be unstable), and dust sublimation, it is worth noting that the 
magnetosonic RDI may become much more relevant under these more extreme conditions. This is discussed further below (\S\ref{sub: discussion magnetic rdi}).

\section{Physical effects not included in our analysis} \label{sec: other effects}

There are a variety of physical effects not included in the derivations and discussions above. While
the complexity of disk and dust models can be increased, nearly without bound, by including 
grain and fluid chemistry (e.g., \citealp{2009ApJ...701..737B}), nonideal magnetic effects (see \S\ref{subsub: resonance nonideal MHD}), grain charging, and radiation effects, it is beyond the scope of this work 
to consider these in any serious detail. Here 
we simply examine some simple effects that we have neglected in our model, Eqs.~\eqref{eq: gas NL rho}--\eqref{eq: dust NL v}, 
and offer some commentary on how these might affect the various RDIs studied in \S\S\ref{sec: resonance epicyclic}--\ref{sec: MHD and others}.
Throughout this section, to obtain simple order-of-magnitude estimates, we shall use the standard MMSN values for disk parameters from  \citet{Chiang:2010ew}, with a dust grain of density $\bar{\rho}_{d}\approx 1 \,\mathrm{g}\,\mathrm{cm}^{-3}$.

\begin{description}
\item[\textbf{Viscosity}]  The viscosity of the gas damps small-scale motions and becomes 
important (i.e., damps the RDI) when $\omega\, \delta u \sim \nu_{\mathrm{vis}}k^{2}\delta u$, where $\delta u$ is the velocity perturbation and
$\nu_{\mathrm{vis}}$ is the kinematic viscosity. Using $\nu_{\mathrm{vis}}\sim c_{s}\lambda_{\mathrm{mfp}}$, 
we find the maximum RDI  wavenumber (where viscosity is not important), $k_{\mathrm{max}} \sim \omega^{1/2}(c_{s}\lambda_{\mathrm{mfp}})^{-1/2}$, or using the MMSN model,
\begin{equation}
k_{\mathrm{max}}\eta r \sim 3\times 10^{4}  \left( \frac{\omega}{\Omega}\right)^{1/2}\left( \frac{r}{\mathrm{AU}}\right)^{-13/28},\label{eq: viscosity}
\end{equation}
where $\omega$ here includes both the real and imaginary parts.

\item[\textbf{Dust separation}] The fluid approximation used to model the dust is valid only 
for scales larger than the separation between individual grains. This is simply $\lambda^{\mathrm{dust}}_{\mathrm{sep}}\sim n_{d}^{-1/3}$, or $k_{\mathrm{max,sep}}\eta r\sim (2\pi/\lambda^{\mathrm{dust}}_{\mathrm{sep}})\eta r \sim 10^{7}(r/\mathrm{AU})^{9/14}(\mu/0.01)^{1/3} (R_{d}/\mathrm{cm})^{-1}$. Alternatively, in terms of the 
stopping time, one finds, 
 $k_{\mathrm{max,sep}}\eta r \sim 10^{4}(r/\mathrm{AU})^{15/7}(\mu/0.01)^{1/3}\tau_{s}^{-1}$ for grains in the Epstein regime and $k_{\mathrm{max,sep}}\eta r \sim 4\times 10^{5}(\mu/0.01)^{1/3}\tau_{s}^{-1/2}$ for grains in the Stokes regime. For most regimes of interest (e.g., at small $\tau_{s}$), these scales are  smaller than the viscous 
 cutoff, Eq.~\eqref{eq: viscosity}.

\item[\textbf{Background turbulence}]If, on some scale $k$, the turnover time of an eddy is faster than the
growth rate of the RDI, we cannot treat the background as in equilibrium over this timescale. The actual effects of turbulence in this limit are unclear---while it is commonly treated as a diffusive process, numerous studies have shown that it can drive very strong grain concentration on small scales \citep{bracco:1999.keplerian.largescale.grain.density.sims,cuzzi:2001.grain.concentration.chondrules,pan:2011.grain.clustering.midstokes.sims,2016MNRAS.455...89H}. A commonly used model for turbulence in disks \citep{Shakura:1973uy} is to assume that the accretion is caused primarily by the turbulent stress, and that
the level of turbulence is $u^{2} \sim \alpha c_{s}^{2}$ (where $u$ here is the rms turbulent velocity 
at the outer scale). A very simplistic estimate of $u_{k}$, the strength of the 
turbulent velocity field on scale $k$, then comes from assuming a Kolmogorov cascade \citep{1941DoSSR..30..301K} with outer scale turnover time  $t_{\mathrm{eddy,outer}}\sim \Omega^{-1}$. This leads to $u_{k}\sim \alpha^{1/3}c_{s}^{2/3}\Omega^{1/3} k^{-1/3}$ (the outer scale of the turbulence is $k_{\mathrm{outer}}\sim \alpha^{-1/2} \Omega/c_{s}$), and suggests that the 
RDI can grow so long as $\Im[\omega(k)]\gtrsim t_{\mathrm{eddy}}^{-1}\sim \alpha^{1/3}c_{s}^{2/3}\Omega^{1/3} k^{2/3}$.
 However, this picture of turbulent accretion in protoplanetary disks has recently been called into question by both  observations \citep{2015ApJ...813...99F,2016A&A...592A..49T,2017ApJ...843..150F} and theory (e.g., \citealp{1996ApJ...457..355G,2011MNRAS.415.3291G,Lesur:2014cc,2015MNRAS.454.1117S,2017ApJ...845...75B}).  In particular, the 
 turbulence  throughout much of the disk may be much weaker than inferred from accretion rates, because the angular momentum transport necessary for accretion can instead be caused by winds. Given this,  it may in fact be more appropriate to  estimate instability properties assuming a laminar 
disk profile.

\item[\textbf{Grain charge}] We have neglected the influence of grain charging, which can be important, both to 
grain dynamics and the chemistry of the disk, especially 
for smaller grains \citep{2006A&A...445..205I,2009ApJ...698.1122O,2012A&A...538A.124I}. A simple estimate for when grain charging becomes 
important to dust dynamics is when the influence of magnetic fields on the grain is greater than that of the
neutral drag, \emph{viz.,} when $t_{L}=m_{d}c/(|q_{d}| B)\lesssim t_{s}$, where $t_{L}$ is the Larmor time, and $m_{d}$ and $q_{d}=Ze$ are the dust mass and charge. Translating this condition into $\tau_{s}$ and the MMSN model, one finds that 
grains of size $\tau_{s}\lesssim 2\times 10^{-11} (r/\mathrm{AU})^{2.2}\beta^{-1/4}|Z|^{1/2}$ (where $\beta=8\pi p_{0}/B^{2}$) will be directly 
affected by magnetic fields (this expression is for grains in the Epstein regime, charge is effectively never important for grains in the Stokes regime).
We note that although grain charge can become very large, perhaps up to $Z\sim -10^{5}$ \citep{2009ApJ...698.1122O}, this only occurs for larger grains (large $\tau_{s}$), so it is not likely that this effect will be important in disks. However in more rarefied regions (e.g.\ disk winds) the effects could be important. Similarly, charged grains experience electrostatic interactions (Coulomb drag, as well as photo-electric and photo-desorptive interactions in the presence of hard radiation sources) with the gas. However as noted in \citet{lee:dynamics.charged.dust.gmcs} (see also \citealt{1979ApJ...231...77D}), Coulomb drag only dominates Epstein/Stokes drag when the ionization fractions in the gas exceed $x_{e} \gtrsim 0.03$, vastly larger than expected in protoplanetary disks.

\item[\textbf{Grain-grain collisions}]We have neglected any influence from grain-grain collisions, which would act to 
thermalize the grains and invalidate our assumption that they behave as a pressureless fluid \citep{Jacquet:2011cy,2012A&A...537A.125J}. This is a reasonable approximation considering that we have focused 
on the low metallicity limit ($\mu\ll1$) throughout this work (except App.~\ref{app: high mu streaming}). Of course, collisions could become important in the non-linear evolution of the RDIs here, as they produce strong local dust clumping.
\end{description}

\section{Application to planetesimal formation}\label{sec: astrophysics}

In this section, we explore some possible physical consequences
of the instabilities discussed in \S\S\ref{sec: resonance epicyclic}--\ref{sec: MHD and others}. In particular, we
consider physical scenarios where the RDIs discussed above may be important, organizing the discussion around the instability type (i.e., epicyclic,  BV, or magnetic), in a similar way to \S\S\ref{sec: resonance epicyclic}--\ref{sec: MHD and others}. 

\subsection{The YG streaming instability}

The primary contribution of this work to the theory of the Youdin-Goodman (YG) streaming instability, which is applicable to dust streaming in the midplane of the disk,
 has been to give simple  analytic 
expressions for the YG streaming instability's  growth rate and fastest-growing wavenumbers (e.g., Eq.~\eqref{eq: epicyclic RDI growth small tau}). To our knowledge, these have not appeared in previous works, but may be useful for simple  estimates and/or numerical tests.

In addition, we have provided a simple interpretation for why the instability exists: it arises 
due to the resonance between the dust drift velocity $\driftvel$ and the gas epicyclic modes in the disk, which propagate 
with phase velocity $V_{\mathrm{epi}}= \pm {\bm{k}}\cdot \bm{\Omega}/k^{2}$.
In the frame of the drifting dust, the epicyclic wave is stationary. As described in 
the simple model of Fig.~\ref{fig: toy model}, the wave's pressure perturbations  attract the dust, and the 
dust feedback acts to further enhance the magnitude of the pressure perturbations---a feedback that results in instability. It is therefore the ``epicyclic RDI.'' 
We reiterate that this interpretation is not in conflict with previous 
interpretations of the mechanism for the YG streaming instability \citep{2005ApJ...620..459Y,2007ApJ...662..613Y,Jacquet:2011cy}---indeed, it is through its interaction with the epicyclic wave that the dust is attracted to pressure maxima. Rather, the interpretation provides a 
clear prediction of the fastest-growing wavenumbers ($\bm{k}\cdot \driftvel = \hat{\bm{k}}\cdot \bm{\Omega}$), the specific reason that the instability relies 
on the rotational support of the gas (this allows the gas to support epicyclic oscillations), and a clear method for extending the analysis to 
more complex physical situations.

We have also shown (App.~\ref{app: high mu streaming}) that the fastest-growing mode with horizontal dust streaming when $\mu>1$ (dust dominates gas by mass), is not technically the same mode as the YG streaming instability at $\mu<1$ (the epicyclic RDI), even though both are commonly named the streaming instability. While the epicyclic RDI  does exist at $\mu>1$ (or more precisely its continuation, which involves similar physics), it is no longer the fastest-growing mode: a different mode appears which is unstable {only} for $\mu>1$, and has a higher maximum growth rate. The maximum growth rate of this mode increases with decreasing $\tau_{s}$ at modest  $\tau_{s}$ (see Fig.~\ref{fig:app: high mu epi}),  but it operates only at very short wavelengths when $\tau_{s}\ll 1$. We again provide simple analytic expressions for the growth rate and fastest-growing wavenumbers of this instability. We also show that it is not an RDI, but arises from joint epicyclic oscillations of dust and gas that resemble a destabilized harmonic oscillator, with the dust driving the gas and destabilizing such oscillations when $\mu>1$.

\subsection{The disk ``\SI''}\label{sub: discussion epi rdi}

\subsubsection{Basic predictions: Rapid instability growth during dust settling}\label{subsub: discussion epi rdi: basics}

Perhaps the most interesting result of this work has been the discovery of a new version 
of the streaming instability---the disk ``\si''---which arises when the dust motion is dominated by its vertical settling 
towards the disk midplane, $\driftvelz \sim \tau_{s} c_{s}$. Most interestingly, unlike the YG streaming instability, 
this instability does not depend significantly on grain size. For most mode angles, the \si\ has growth rate $\Im(\omega)/\Omega \approx \mu^{1/2}\hat{k}_{x}$ around wavenumber $k\eta r\sim \eta^{1/2}/\tau_{s}$, where $\eta\sim 10^{-3}$ parameterizes the gas pressure support (see Eq.~\eqref{eq: eta definition}). (These estimates depend modestly on the stratification profile and we shall consider an approximately neutrally buoyant fluid here and below for simplicity; see \S\ref{subsub: epi BV compared to epi}). This means that, for small grains in particular, the growth rates can be orders-of-magnitude faster than the YG streaming instability---comparable, in fact, to the disk orbital time. Moreover, the characteristic wavelengths are larger than the YG streaming instability by a factor $\sim \!\eta^{-1/2}\!\sim\! 30$. Thus, a larger volume of grains can be concentrated into the structures produced by the instability, suggesting the concentrations could be more likely to be gravitationally unstable and better able to resist destruction via turbulence. 

Even more surprising,  at  a particular ``double-resonant'' mode angle $\theta_{\mathrm{res}}$, where $\hat{\bm{k}}\cdot \driftvel =0$,
$\Im(\omega)$ formally increases without bound as $\Im(\omega)/\Omega\sim (\tau_{s}\,\mu\,k_{x}\eta r)^{1/3}$, once 
$\Im(\omega)$ is larger than $\omega_{\mathrm{EBV}}= (\hat{k}_{z}^{2}\Omega^{2}+\hat{k}_{x}^{2}N_{\mathrm{BV}}^{2})^{1/2} $. We shall consider this double-resonant mode separately in estimates below, although it appears in 
exactly the same physical set up as the standard \si\ (just at smaller scales). Here $\omega_{\mathrm{EBV}}$ is the frequency of joint epicyclic-buoyancy oscillations, or inertia-gravity waves, which is the natural oscillation frequency in stratified regions of the gas ($N_{\mathrm{BV}}$ is  the \BV\ frequency; see Eq.~\eqref{eq: epi-BV gas frequency} and \S\ref{sub: discussion BV rdi} below).

Importantly, these growth rate estimates are easily faster than the time required for small grains to settle to the 
midplane of the disk, which scales as $\Omega\,t_{\mathrm{settle}}\sim (\Omega t_{s})^{-1}= \tau_{s}^{-1}$ (for 
grains starting approximately one scale height above the midplane).
Thus, although necessarily 
transient, halting once grains reach the midplane, the instability will evolve well into its nonlinear 
phase long before its driving force (the downwards drift) is removed.
This leads us to suggest a scenario where smaller grains clump significantly, due to the \si,
in the process of settling towards the disk midplane.

\subsubsection{Potential nonlinear consequences \&\ appearance}

The \si\ modes grow fastest when $\hat{k}_{x}>\hat{k}_{z}$, so would
have the appearance of concentric axisymmetric cylinders of higher dust concentration that form as the dust settles. Their fastest-growing wavelength (see Eq.~\eqref{eq: vertical epicyclic resonant omega mid angles}) is  $2\pi/k\sim \lambda\sim \lambda_{x}\sim 2\pi \,\eta^{1/2} r \tau_{s}\sim 2\pi\, h_{g}\tau_{s}$ (this estimate can increase somewhat 
depending on the temperature stratification profile; see \S\ref{subsub: epi BV compared to epi}).
This suggests that a cylinder of dust of area $A\sim 2\pi\,r\,\lambda$ (in an MMSN-type disk with with density $\Sigma \sim 2200\,{\rm g\,cm^{-2}}\,\Sigma_{\rm MMSN}\,\,(r/\mathrm{AU})^{-3/2}$; \citealp{Chiang:2010ew}) contains a mass of dust $M\sim 4\times10^{24}\,{\rm g}\,(\tau_{s}/0.001)\,(\mu/0.01)\,\Sigma_{\rm MMSN}\,(r/\mathrm{AU})^{11/14}$, or enough mass to form a planetesimal of size $R_{\mathrm{plan}} \sim 1000\,{\rm km}\,(\tau_{s}/0.001)^{1/3}\,(\mu/0.01)^{1/3}\,\Sigma_{\rm MMSN}^{1/3}\,(r/\mathrm{AU})^{-0.26}\,\bar{\rho}_{\mathrm{solid}}^{-1/3}$
(here $\bar{\rho}_{\mathrm{solid}}$ is the mass density of the planetesimal in ${\rm g\,cm^{-3}}$).\footnote{Alternatively, a  more conservative estimate of the enclosed mass (for modes further from the fastest-growing mode orientation) could consider a ring with $|k_{z}|\sim |k_{x}|$, with volume $V\sim\!2\pi r \lambda^{2}$. This  ring would contain a mass sufficient to form a planetesimal of size $R_{\mathrm{plan}}\sim 8\times 10^{3}\tau_{s}^{2/3}(r/\mathrm{AU})^{11/42}(\mu/0.01)^{1/3}\,\bar{\rho}_{\mathrm{solid}}^{-1/3}\, \mathrm{km}$.}

While of course we cannot simply extrapolate from the linear behavior of the \si\ to directly form a planetesimal (given the very large nonlinear concentration of grains that would be necessary), this estimate does 
show that the overdensities it creates contain a significant amount of mass. Thus, some possible outcomes could include: (i) direct planetesimal formation as grains sediment vertically; (ii) the creation of dust clumps that later act as high-metallicity ($\mu\gtrsim 1$) seeds for the  YG  streaming instability in the disk midplane (which operates on smaller scales, see App.~\ref{app: high mu streaming}); or (iii) the generation  of alternating over-dense and under-dense rings as the dust settles,  which could, given sufficiently strong dust concentration, act as ``pressure bumps'' in the midplane and so potentially trap even more dust (thus continuing to grow in density contrast). 

Noting that simulations have found a strong  dependence of the efficiency of planetesimal formation on  metallicity,
with a cutoff metallicity of $\mu\approx 0.02$ for $\tau_{s}\gtrsim 0.1$ grains  \citep{Johansen:2009ih,Bai:2010gm}, even if the local enhancement of the metallicity were 
only a factor of several, it might strongly influence the planetesimal formation process. Further, small grains---which  seem to require a higher local metallicity to  form planetesimals  (according to numerical simulations; \citealp{2015A&A...579A..43C,2016arXiv161107014Y})---may also be more
efficiently clumped by the \si\ because of their longer settling times, thus reaching higher local metallicities. 
Such clumping may also increase the collision rate of grains (presumably 
proportionally to the relative density increase in the clumps),  perhaps enhancing  grain coagulation rates 
to the point where a significant population of larger grains could form at relatively low metallicities ($\mu\ll1$) {before} the solids settle completely in 
the midplane of the disk \citep{2013A&A...556A..37D,2014A&A...572A..78D}. 

Interestingly, if we speculate that some constant fraction of the dust mass concentrated in an annulus by the initial \si\ eventually ends up in a planet, and that the maximum grain size  is constant in terms of $\tau_{s}$ (as models of grain-growth suggest; \citealt{birnstiel:2012.grain.size.distribution.models,drazkowska.2014:dust.growth.protoplanetary.disk}), then our estimates above imply the resulting object size would depend very weakly on location in the disk, as $R_{\mathrm{plan}} \propto r^{0.26}$ (or $R_{\mathrm{plan}}\propto P^{0.17}$, in terms of the orbital period $P$). This could conceivably provide a partial explanation for the observed tendency of planet sizes to vary weakly within the same disk \citep{weiss:planets.same.size.vs.radius,millholland:multi.planet.systems.uniform.sizes}.

\subsubsection{Range of grain sizes where growth is rapid}

Of course,  simulations will be required to assess the scenarios discussed in the previous paragraphs 
in detail. Nonetheless, let us consider the relevant timescales more quantitatively, starting with the standard (RDI) \si\ and then considering  modes at the double-resonant angle $\hat{\bm{k}}\cdot \driftvel=0$ (these modes have higher growth rates, but smaller scales).
When  $\hat{k}_{x}\gtrsim\hat{k}_{z}$, the \si\ has 
growth rate $\Im(\omega)/\Omega \approx \mu^{1/2}\approx 0.1(\mu/0.01)^{1/2}$ at $k\eta r\sim \eta^{1/2}/\tau_{s}\approx 0.03 (r/\mathrm{AU})^{2/7}\tau_{s}^{-1}\sim (h_{g}\tau_{s})^{-1}$, and thus is not affected by viscosity for $\tau_{s}\gtrsim 10^{-6}\,(r/\mathrm{AU})^{3/4}$, or $R_{d}\gtrsim 8\,(r/\mathrm{AU})^{-3/4}\,\mu\mathrm{m}$. So long as the disk is sufficiently quiescent over the growth time scales (see below), this mode  grows sufficiently fast to clump the dust before 
it reaches the disk midplane for grains of size $\tau_{s}\lesssim 0.1\,(\mu/0.01)^{1/2}$. As discussed above, the 
mass contained in a single ring  the size of this mode is sufficient to form a $\mathrm{km}$-sized planetesimal for effectively any grain size ($\tau_{s} \gtrsim 10^{-12}$ at $\sim \!1\mathrm{AU}$).

The double-resonant mode can grow much faster than the standard (RDI) \si, but is also active on smaller 
scales and so does not contain as much mass.
We can estimate its maximum growth rate from Eq.~\eqref{eq: viscosity} for the smallest scales allowed due to gas viscosity.\footnote{Note also that the scale limit due to 
interparticle  separation can be more severe than that due to viscosity for grains with $\tau_{s}\gtrsim 0.1$ at reasonable metallicities.} This shows that
$(\omega/\Omega)_{\mathrm{max}}\sim 12\, (\mu/0.01)^{2/5}\tau_{s}^{2/5}(r/\mathrm{AU})^{-0.2}$, so 
long as this value is larger than the gas epicyclic-\BV\ frequency $\omega_{\mathrm{EBV}}  \sim \Omega$ (at one scale height, depending on the temperature stratification\footnote{There is ambiguity here because the \BV\ frequency depends on the temperature stratification profile, which is uncertain in disks. However, as discussed in \S\ref{subsub: epi BV compared to epi}, changes to the temperature stratification  cause only minor (factor several) changes to what we most care about---the maximum growth rate as a function of $k\eta r$---so it is not worth considering the stratification profile in detail for these simple estimates.}).
This suggests that grains with $\tau_{s}\gtrsim 10^{-3}$ are unstable to the double-resonant mode on small scales, with a growth rate $\Im(\omega)>\Omega$; i.e., the instability grows faster than the disk dynamical time (recall from above that grains with $\tau_{s} \gtrsim 10^{-5}$ have maximum $\Im(\omega) \sim 0.1\Omega$). This is sufficiently large to clump grains with $\tau_{s}\lesssim1$ before they settle into the midplane, implying that effectively \emph{all} grain sizes are unstable to instabilities  that grow more rapidly than their vertical settling time. This rapid clumping on small scales   could significantly 
modify grain  coagulation or other properties of the dusty gas.

\subsubsection{The role of turbulence}

Of course, the caveat about the quiescence of the gas is an important one. 
The effect of turbulence in protoplanetary disks is rather difficult to estimate and quite poorly understood at the present time, but there is now significant evidence from observations \citep{2015ApJ...813...99F,2016ApJ...816...25P,2016A&A...592A..49T} and theory \citep{Bai:2013jg,2017ApJ...845...75B} that turbulence is quite weak in most regions of the disk. For example, the observations of \citet{2015ApJ...813...99F} place an upper limit on the turbulence level  around the young star HD 163296 of $u_{\mathrm{rms}}\lesssim 0.03 c_{s}$, which is significantly lower than would be inferred from the accretion rate (if accretion proceeded primarily through turbulent stresses). Instead, it has been argued recently that accretion is driven primarily by winds, with a largely laminar profile in the bulk disk (see, e.g., \citealt{2017ApJ...845...75B} for comprehensive simulations of protoplanetary disk accretion physics, which show a mostly laminar disk profile).

Despite this uncertainty, assuming some turbulence exists in the disk, the disk \si\ is likely to be  less sensitive to the presence of turbulent motions than the standard YG streaming instability. This is for two reasons: (1) the growth rates for small grains are much faster, and (2) the characteristic wavelengths are  larger.

If we parameterize the level of turbulence using $\alpha$ (which is likely to be overly pessimistic) and balance the RDI growth rate and eddy turnover time (as discussed in \S\ref{sec: other effects}), we estimate that turbulence would likely influence the \si\ when $\tau_{s} \lesssim (0.003\rightarrow0.03)\,(\alpha/10^{-4})^{1/2}\,(\mu/0.1)^{-3/4}$ (where the range depends on assumptions about the largest eddy turnover times in units of $\Omega^{-1}$, and the width of the resonance). A similar estimate argues that turbulence is important for the  YG streaming instability when $\tau_{s}\lesssim (0.1\rightarrow 1)\,(\alpha/10^{-4})^{1/5}\,(r/\mathrm{AU})^{4/35}\,(\mu/0.1)^{-3/10}$. In other words, compared to the YG streaming instability,  the \si\ is likely to be less affected by external turbulence and survive for smaller grain sizes.\footnote{A similar estimate of when eddy turnover times become faster than the growth rate, applied to the high-$\mu$ mode of the standard streaming instability discussed in App.~\ref{app: high mu streaming}, gives $\tau_{s}\lesssim 0.1 (\alpha/10^{-4})^{2/5}\,\mu^{13/10}\,(r/\mathrm{AU})^{8/35}$. This arises because the growth rates of this mode, while large, are restricted to very high $k$ (see Eqs.~\eqref{eq:app: omega expansion low tau}--\eqref{eq:app: higher k epi high mu}).}
For lower levels of turbulence, the \si\ becomes even more robust compared to the YG streaming instability (the scaling with $\alpha$ is different), while the opposite occurs at lower $\mu$.
The double-resonant mode can be influenced by turbulence for effectively all grain sizes, because its growth rate increases (in a homogeneous background) without limit with $k$ as $\sim k^{1/3}$: in a Kolmogorov-type cascade, this suggests there will be some sufficiently high $k$ where eddy turnover times ($\sim k^{-2/3}$) become shorter than the mode growth timescales.

We also note that dust-induced turbulence caused by the extra mass loading in the midplane is likely only relevant  in a thin layer near the midplane \citep{Garaud:2004,2012ApJ...744..101T}, and thus presumably more important for the YG streaming instability than the \si. In contrast, some other turbulence-generation mechanisms, such as shear or  buoyancy induced instability \citep{2013MNRAS.435.2610N,0004-637X-788-1-21,2017arXiv171006007F} would presumably more strongly affect  regions away from the disk midplane.
Ultimately, any useful turbulence-related constraints  on grain concentration due to the disk \si\ will require nonlinear simulations and better theoretical understanding of disk accretion mechanisms and instabilities. Given 
the potential importance of global effects, disk thermodynamics, and other complicated nonideal effects, this is a rather
difficult computational problem to tackle in detail.

\subsubsection{Robustness of the \SI}

The fundamental character of the \si\ (vertical-epicyclic RDI) is robust to a wide range of assumptions or details of our derivation, including: (1) vertical or radial stratification of the disk (assuming $k h_{g} \gg 1$); (2) gas compressibility; (3) the form of the drag law (Epstein or Stokes drag); (4) the gas equation-of-state; (5) including or ignoring radial or azimuthal streaming velocities (in the dust or gas); (6) including or ignoring gas streaming motion; (7)  external magnetic forces (ideal or non-ideal MHD in the gas, and Lorentz forces on the dust, so long as the Lorentz forces are sub-dominant to drag); (8) changing the gravitational potential (from Keplerian), which simply modifies the epicyclic frequency by an order-unity constant; and (9) self-gravity (at least for linear perturbations assuming the disk initially has $Q\gg1$). 
Note that we have explicitly verified all of these properties, but did not show several in detail precisely because they have no significant effect in the relevant limits. As discussed above, the \si\ is also unstable for {\em any} finite dust-to-gas ratio $\mu$ and grain size/stopping time $\tau_{s}$. The relevant question, as discussed above, is whether or not the instability grows fast enough to generate interesting nonlinear behavior before grains settle into the disk midplane (or before structures are disrupted by external turbulence).

\subsubsection{The \SI\ in simulations}\label{subsub: vertical RDI in simulations}

A question that naturally arises is whether the disk \si\ has been observed
in previous simulations. So far as we are aware, it has not, for two likely reasons: first, it has been common 
(e.g., \citealt{Bai:2010gm}) to simulate only a small portion of the disk plane, so as to capture 
more accurately the concentration of dust at the disk midplane; second, most works have focused on larger 
grain sizes. Both of these choices decrease the settling time of dust (or eliminate settling entirely), and it seems that this has been too short  to see the instability develop in previous simulation work. Let us consider 
\citet{2016arXiv161107014Y} in more detail, which, to our knowledge, has been the closest to resolving the 
vertical streaming instability, due their small particles and high-resolution two-dimensional domains. 
Their highest resolution simulation, with $5120$ grid cells per $h_{g}$ and $\tau_{s}=0.01$, can  
resolve scales up to $k\eta r\sim 160$, suggesting the growth rate of the double-resonant mode\footnote{Note that \citet{2016arXiv161107014Y} do not include gas stratification, which implies that we should use the estimates of \S\ref{subsub: epicyclic vertical streaming}, rather than those of \S\ref{sub: BV epi RDI}, although they are very similar anyway.} at the smallest 
scales is $\Im(\omega)/\Omega\approx 0.3$. However, their simulation initializes grains at a height of $h_{d,\mathrm{init}}\sim 0.02 h_{g}$ (barely above the midplane), which suggests that they settle to the midplane within $t\sim 2\Omega^{-1}$, and there is simply not enough time for grains to clump significantly due to either the \si\ or the double-resonant mode. Similarly, while the study of \citet{2014ApJ...792...86Y} uses domains of large vertical extent and an initially uniform dust density distribution in $z$, their rather large particles ($\tau_{s}\approx 0.3$) again cause the particles to settle quickly, and the resolution is too low ($160$ cells per $h_{g}$ for the largest domains) to see the fast-growing double-resonant mode.
Nonetheless, it seems quite feasible to study this mode in simulation, at least in two dimensions, potentially using a setup similar to \citet{2016arXiv161107014Y}, or
to \citet{Lambrechts:2016fg} but with rotation included (see \S\ref{subsub: BV RDI in simulations}).

\subsection{\BV\ RDI}\label{sub: discussion BV rdi}

As a subsidiary result of this work, we analyzed the \BV\ (BV) RDI, which arises 
as a result of the resonance between streaming dust and BV oscillations in  a stratified fluid. 
In essence, the instability is relatively simple: the streaming dust sees a stationary BV oscillation, which 
involves  motions  of the gas that advect the background density/entropy gradient, and a small pressure perturbation. The dust can interact 
with gas density perturbations through the dependence of the stopping time $t_{s}$ on the gas density; this enables the alternate RDI mechanism discussed in \S\ref{sub: toy model} and is parameterized through the $\zeta_{\rho}\Lambda_{S}$ term in Eq.~\eqref{eq: final compressible BV mode}. Dust  is also drawn into 
  pressure maxima, which, as shown in Fig.~\ref{fig: toy model},  helps to destabilize the mode (note that there remain theoretical uncertainties regarding the exact form of this term; see \S\ref{subsub: local linear BV caution}). When the dust-gas drift is dominated by the dust settling in the direction of gravity, 
  the mode's growth rate is \begin{equation}
\Im(\omega) \approx \left(\frac{\mu}{2}\right)^{1/2} \hat{k}_{x} \left[ L_{0}^{-1}\frac{\driftvelz}{t_{s}}\left(1+\frac{\Lambda_{S}}{2}\right)\right]^{1/2},\end{equation}
 for grains in the Epstein drag regime. It grows at the scale where $\bm{k}\cdot \driftvel$ matches the \BV\ frequency, or at $(kL_{0})^{-1}\sim \lambda/L_{0}\sim\driftvelmag/c_{s}$. Here $L_{0}^{-1}=\gamma_{\mathrm{gas}}^{-1}\nabla\ln P$, $\Lambda_{S}$ parameterizes the entropy/temperature stratification (see Eq.~\eqref{eq: stratified equilibrium defs}), and 
 $(L_{0}^{-1}\driftvelz/t_{s})^{1/2}\approx (\driftvelz/c_{s})\,t_{s}^{-1} \sim c_{s}L_{0}^{-1}$ for grains that are passively settling through the gas (i.e., when there
 are no  external forces on grains other than gravity). The instability is likely only weakly modified  by dust stratification,
 so long as the dust stratification is similar to, or less than, the gas stratification (see App.~\ref{app: dust stratification}).
 
  As can be seen through analysis of the RDI including both 
  rotation and stratification (\S\ref{sub: BV epi RDI}), in disks,  this BV RDI  is of minor relevance compared to the epicyclic RDI, usually increasing growth rates by $\mathcal{O}(1)$ factors without changing its qualitative behavior (some of these effects were mentioned briefly above in the qualitative discussion of \S\ref{subsub: discussion epi rdi: basics}).
  Nonetheless,  the BV RDI  can occur whenever grains settle through a stratified atmosphere (with or without rotation).
  We can estimate when the instability could be important for dust clumping by asserting that  the growth timescale  
  timescale be shorter than the time for the particles to move through the stratified atmosphere, which is of length $\sim\! L_{0}$. From the estimates above, this requires $\Im(\omega)\sim \mu^{1/2} t_{s}^{-1}\driftvelmag/c_{s}\gtrsim \driftvelmag/L_{0}$, which may be rearranged to $\mu^{1/2}\gtrsim \driftvelmag/c_{s} \sim t_{s} g/c_{s}\sim \lambda/L_{0}$. Thus we see that the instability becomes faster growing,  compared to the settling time, for smaller grain sizes (small $t_{s}$), although it also grows on smaller scales.\footnote{Note that this condition, $\mu^{1/2}\gtrsim \lambda/L_{0}$, is also required for the validity of our analysis; see \S\ref{subsub: local linear BV caution}.}
There 
  may, therefore, be a  variety of non-disk systems where the instability plays a critical role---for example, in the atmospheres of forming giant planets \citep{2000ApJ...537.1013I,Lambrechts:2016fg}.  The BV RDI  may also be observable in terrestrial experiments \citep{2011AnRFM..43...97G}

\subsubsection{The \BV\ RDI in simulations}\label{subsub: BV RDI in simulations}

In  \citet{Lambrechts:2016fg}, the authors numerically set up 
 a stratified pressure-supported  gaseous atmosphere and allowed grains 
 to settle in the direction of gravity, observing significant clumping of the grains as they settled. 
 Using the parameters of their fiducial setup (their \texttt{run2} or \texttt{run3}), we estimate that
 the BV RDI in their setup should have a growth rate of approximately 
 $\Im(\omega)\, t_{s}\approx 0.08\,\mu^{1/2} \hat{k}_{x}$ at wavenumber $g t_{s}^{2} k\approx 0.08\, \hat{k}_{x}/\hat{k}_{z}$ (we use the 
 ``friction units'' of \citealt{Lambrechts:2016fg} here, with $z$ the direction of stratification). This suggests 
 that the BV RDI is unstable for in their simulations for oblique wavenumbers ($\hat{k}_{x}>\hat{k}_{z}$).\footnote{The simplified analytic expressions we derived in the text are not valid when $\mu\sim 1$, as was adopted in most of the \citet{Lambrechts:2016fg} simulations. However, numerical solutions of the dispersion relation (not shown) show that the approximate analytic expression is reasonable accurate (within a factor of a few of the growth rate), although the resonances become broader than shown in Fig.~\ref{fig: BV RDI} when $\mu\gtrsim 1$.} Indeed,  their measured growth rates in their lower-$\mu$ runs (\texttt{runs1.01}--\texttt{runs4.01}), where our analytic expressions are most accurate, agree within a factor of several with our BV-RDI predictions and scale as $\Im(\omega)\sim \mu^{1/2}$ as we predict (see  \citealt{Lambrechts:2016fg}, Figure~10). This interpretation is also commensurate with the toy model 
put forth by \citet{Lambrechts:2016fg}, which invoked the importance of buoyancy for the instability. However, 
the instability was not found in their 
linear stability analysis (their Appendix~A) due to the neglect of stratification  of the gas in that analysis, which is crucial to the BV RDI (including compressibility could also have  led to different instabilities in their analysis, see \papertwo).

\subsection{Magnetic RDIs}\label{sub: discussion magnetic rdi}

Our final, albeit brief, exploration of RDIs in protoplanetary disks in this work 
concerned RDIs that result from resonances with magnetic waves. 
 In principle, such instabilities could grow rapidly in a well-ionized gas on small scales, for 
dust-gas drift velocities comparable to those  in protoplanetary disks. However,
we found that the  significant non-ideal effects, which arise due to the low ionization fraction in the bulk regions
of stellar disks,
will likely tend to damp out any such RDIs at scales well below where their growth rates become astrophysically 
interesting (see Fig.~\ref{fig: standard MHD}). While MHD with one such non-ideal effect, the Hall effect, can support
a number of RDIs  because it has undamped waves (the Hall-whistler and Hall-Alfv\'en RDIs, see Fig.~\ref{fig: Hall MHD}), both Ohmic and Ambipolar diffusion act to damp such  instabilities, and thus magnetic instabilities are unlikely to be of interest under standard midplane disk conditions (i.e., those in the bulk of the disk in the  MMSN model).
Magnetic RDIs may, however, be more important in the inner regions of the disk, or in the  outer layers, where 
the proximity to the protostar suggests the gas/plasma is likely well approximated by ideal MHD (see, e.g., \citealt{2017ApJ...835..230F}). Depending on the levels of turbulence in such regions, the slow magnetosonic RDI could act to clump grains on small scales, including those grains with dynamically important charge \citep{Hopkins:2018rdi}.

Another place where the magnetosonic and acoustic RDIs may be relevant is in disk winds, which are now thought to be a key accretion mechanism in protoplanetary disks
(see, e.g., \citealt{Bai:2013jg,2015MNRAS.454.1117S,2017ApJ...845...75B}).
Such winds can reach supersonic velocities and are likely well ionized \citep{2016ApJ...818..152B}. Thus, if a wind contained dust, it could be unstable to both the slow and the fast magnetosonic RDIs (or the related acoustic RDI). This  
would cause dust clumping over quite short timescales (see \papertwo), potentially  
 modifying important properties of the dust-laden gas, e.g., its opacity.
One can estimate a critical grain size that is swept up by a wind (or by gas evaporation) by
balancing the downwards force due to gravity against the drag from the wind 
 (see \citealt{2015ApJ...804...29G}). For a wind  launched with a velocity $\sim c_{s}$
 from a scale height $z\sim 4h_{g}$, this suggests particles with $R_{d}\lesssim 6 (r/\mathrm{AU})^{-3/2} \mathrm{cm}$ can be swept up by the wind, if they exist at this scale height. 
 As they are swept up and accelerated by the gas, the particles reach  the 
 gas velocity after $t\sim t_{s}$, implying that the RDI must have $\Im(\omega)\gtrsim t_{s}^{-1}$
 for the instability to have time to develop. Noting that the  growth rate of both the hydrodynamic and magnetic RDIs scale as $\Im(\omega)t_{s}\sim \mu^{1/2}(k\,c_{s}\,t_{s})^{1/2}$  when $\driftvelmag \gtrsim c_{s}$, we see that modes with $k\eta r\gtrsim 3 (R_{d}/\mathrm{cm})^{-1}(T_{\mathrm{wind}}/1000\mathrm{K})^{-1}(\mu/0.01)^{-1}$ grow sufficiently 
fast to clump the dust (where $T_{\mathrm{wind}}$ is the temperature of the wind). Of course, because 
all of the larger grains and most of the smaller grains  will have sedimented towards the
midplane,  the dust-to-gas ratio $\mu$ at the wind launch point may be very low  \citep{2015ApJ...804...29G}.

\section{Conclusion}\label{sec: conclusion}

In this work, we have introduced and studied a variety of  well-known and new instabilities of streaming dust, exploring their relevance to planetesimal formation in protoplanetary disks. 
Each of these instabilities is related to the well-studied ``streaming instability'' \citep{2005ApJ...620..459Y} through
the recognition that they are all---including the streaming instability---members of the broad class of \emph{Resonant Drag Instabilities} (RDIs; see \paperone\ and \papertwo). In a dust-gas mixture
where there is a nonzero relative velocity $\driftvel$ between the two phases, an RDI 
occurs at wavenumber $\bm{k}$ whenever the dust streaming frequency, $\bm{k}\cdot \driftvel$,
resonates with (equals) the frequency $\omega_{\mathcal{F}}$ of an undamped wave in the gas. In the 
frame of the drifting dust grains,  a resonant wave is stationary and attracts dust towards its pressure maxima. The
backreaction  of the dust on the gas then acts as a force towards these same pressure maxima, enhancing them
further and promoting exponential growth of the perturbation (see Fig.~\ref{fig: toy model}).
At low metallicities,   RDI modes, with $\bm{k}\cdot \driftvel=\omega_{\mathcal{F}}$, are always the fastest-growing
drag-induced instabilities in the system (\paperone). Further, they always act to concentrate grains as the instability grows.

This RDI  theory described above  suggests a general algorithm for discovering {new} dust-gas streaming instabilities: (i) chose an undamped
gas wave, which oscillates with some frequency $\omega_{\mathcal{F}}$; (ii) compute 
the resonant wavenumbers $\bm{k}_{\mathrm{res}}$ for which $\bm{k}\cdot \driftvel =\omega_{\mathcal{F}}$;   (iii) 
use the RDI formula, Eq.~\eqref{eq: dust eval pert}, to compute the growth rate of the fastest-growing modes in the system, which occur at wavenumber  $\bm{k}=\bm{k}_{\mathrm{res}}$.
The results of this paper have simply been an application of this algorithm to some different
oscillation modes of disks. Remarkably, we have shown this leads to several new instabilities, which (to our knowledge) have not been previously recognized. For smaller dust grains, these can have growth rates that are orders-of-magnitude faster than the  \citet{2005ApJ...620..459Y} streaming instability.
An aspect of this result that deserves emphasis is that {\em even the smallest grains}
are subject to fast-growing RDIs, suggesting that a separate treatment of dust and gas dynamics 
may be important for many applications.

\subsection{Relation to known instabilities}

One  purpose of this work has been to interpret the standard (YG) streaming instability within 
the RDI framework, give simple analytic expressions for its growth rates 
and fastest-growing wavenumbers, and put forward a heuristic toy model for its operation (\S\ref{sub: toy model}). The high-metallicity $\mu>1$ case, which we show is actually a different instability, is analyzed separately in App.~\ref{app: high mu streaming}. The expressions we derive compare well 
against numerical solutions of the dispersion relation (e.g., Figs.~\ref{fig: 2D epicyclic from YG}--\ref{fig: epicyclic 1D}), while the interpretation of the streaming instability as an RDI is helpful for gaining a general physical 
picture for its mechanism and extensions.

\subsection{New instabilities with rapid growth rates}

Going beyond the well-studied streaming instability, we have explored several different RDIs, all of which should be present in disks. Most interestingly, we have shown that when grains  settle towards the midplane of the disk, a new instability---the disk \si---appears. Its growth rate is approximately  {\em independent} of grain size ($\tau_{s}$), and for small grains, is orders of magnitude larger than the standard YG streaming instability. Moreover, the \si\  grows  on  larger  wavelengths (by a factor $\eta^{-1/2}\sim 30$) than the YG streaming instability, potentially allowing it to concentrate a larger mass of grains and likely making it more robust against external turbulence in the disk. We show that for a wide range of grain sizes, the growth timescales are significantly shorter than the dust's vertical settling time, and, even at low dust-to-gas ratios, comparable to or shorter than the disk orbital period.

This suggests a picture where small grains could clump significantly in the process of settling towards the  midplane of the disk, with potentially interesting consequences for grain growth and other properties (e.g., opacity). We expect the instabilities to aggregate dust into narrow radial annuli or bands as it sediments, potentially building pressure bumps and dust traps into the initial distribution of dust in the midplane, without any external processes required. If the clumping goes further during sedimentation, it could possibly nonlinearly reach sufficient densities to trigger planetesimal formation via gravitational collapse, or (more conservatively) could generate high-metallicity seeds for the standard YG  streaming instability once the dust clumps reach the  midplane. This suggests a mechanism that could allow planetesimal formation at lower metallicities, for smaller grain sizes, than inferred from simulations up to now \citep{Johansen:2009ih,Bai:2010gm,2015A&A...579A..43C}.

We have also examined a variety of other RDIs, including those caused by a resonance with buoyancy (\BV) oscillations, sound waves, and various MHD waves. 
While we have mostly found that these are less important than the  \si,  each has potential relevance in some regimes. 
For the reader interested in a quick overview of these results, we have outlined the behavior of each RDI and given 
astrophysically relevant estimates their properties in \S\ref{sec: astrophysics}, while \S\ref{sec:overview.of.modes} lists each of the instabilities covered in this work.

\subsection{Future work}

While we have analyzed the main types of oscillations that could be expected 
in cooler disks around young stars, there remain a variety of
interesting avenues for exploration. On the linear side, it will be interesting to 
explore the interaction of dust with non-axisymmetric waves, which 
have been completely ignored in this work (because a time-dependent or global analysis 
is necessary to study such modes). One interesting possibility is spiral density waves \citep{2004MNRAS.350..849N,Heinemann:2009jv,2016Sci...353.1519P}, which, due to their 
large gas  perturbations, may interact relatively strongly with dust. We have also largely focused our analysis on MMSN-type disks, at modest distances ($\sim$\,AU). Under more extreme conditions---e.g., around massive stars, or very close/far from the star---it is possible that the different gas conditions could change the relative importance of  different RDIs  or even produce new RDIs. 

On the computational front, the path forward is  clear:  the role of the \si\ in planetesimal formation can only be studied in real detail using  simulations of its nonlinear evolution. These
are quite feasible with present-day computational  resources, at least in local two-dimensional domains,  
and require simulating the settling of dust through a rotating stratified disk atmosphere (see \S\ref{subsub: vertical RDI in simulations} and \S\ref{subsub: BV RDI in simulations} for further discussion of simulations).
 Whether the \si\ can, ultimately, have a significant effect on the planetesimal formation process will depend on the relative clumping 
of dust that occurs during its nonlinear evolution.

\vspace{-0.2cm}
\acknowledgments 
It is a pleasure to thank E.~Chiang, E.~Quataert, and A.~Youdin for helpful comments and discussion. JS was funded in part by the Gordon and Betty Moore Foundation through Grant GBMF5076 to Lars Bildsten, Eliot Quataert and E. Sterl Phinney. Support for PFH was provided by an Alfred P. Sloan Research Fellowship, NSF Collaborative Research Grant \#1715847 and CAREER grant \#1455342.   
\vspace{0.7cm}


\begin{thebibliography}{}

\bibitem[\protect\citeauthoryear{Alfv{\'e}n}{Alfv{\'e}n}{1942}]{Alfven:1942hl}
Alfv{\'e}n H.,  1942, Nature, 150, 405

\bibitem[\protect\citeauthoryear{Armitage, Eisner \& Simon}{Armitage
  et~al.}{2016}]{2041-8205-828-1-L2}
Armitage P.~J.,  Eisner J.~A.,    Simon J.~B.,  2016, \apjl, 828, L2

\bibitem[\protect\citeauthoryear{{Bai}}{{Bai}}{2017}]{2017ApJ...845...75B}
{Bai} X.-N.,  2017, \apj, 845, 75

\bibitem[\protect\citeauthoryear{{Bai} \& {Goodman}}{{Bai} \&
  {Goodman}}{2009}]{2009ApJ...701..737B}
{Bai} X.-N.,  {Goodman} J.,  2009, \apj, 701, 737

\bibitem[\protect\citeauthoryear{Bai \& Stone}{Bai \& Stone}{2010}]{Bai:2010gm}
Bai X.-N.,  Stone J.~M.,  2010, \apj, 722, 1437

\bibitem[\protect\citeauthoryear{{Bai} \& {Stone}}{{Bai} \&
  {Stone}}{2010}]{2010ApJ...722L.220B}
{Bai} X.-N.,  {Stone} J.~M.,  2010, \apjl, 722, L220

\bibitem[\protect\citeauthoryear{Bai \& Stone}{Bai \& Stone}{2013}]{Bai:2013jg}
Bai X.-N.,  Stone J.~M.,  2013, \apj, 769, 76

\bibitem[\protect\citeauthoryear{{Bai}, {Ye}, {Goodman} \& {Yuan}}{{Bai}
  et~al.}{2016}]{2016ApJ...818..152B}
{Bai} X.-N.,  {Ye} J.,  {Goodman} J.,    {Yuan} F.,  2016, \apj, 818, 152

\bibitem[\protect\citeauthoryear{{Baines}, {Williams} \& {Asebiomo}}{{Baines}
  et~al.}{1965}]{1965MNRAS.130...63B}
{Baines} M.~J.,  {Williams} I.~P.,    {Asebiomo} A.~S.,  1965, \mnras, 130, 63

\bibitem[\protect\citeauthoryear{{Baines} \& {Mitsudera}}{{Baines} \&
  {Mitsudera}}{1994}]{1994JFM...276..327B}
{Baines} P.~G.,  {Mitsudera} H.,  1994, \jfm, 276, 327

\bibitem[\protect\citeauthoryear{Balbus \& Terquem}{Balbus \&
  Terquem}{2001}]{2001ApJ...552..235B}
Balbus S.~A.,  Terquem C.,  2001, \apj, 552, 235

\bibitem[\protect\citeauthoryear{{Barge} \& {Sommeria}}{{Barge} \&
  {Sommeria}}{1995}]{barge:1995.vortex.trap.idea}
{Barge} P.,  {Sommeria} J.,  1995, \aap, 295, L1

\bibitem[\protect\citeauthoryear{{Barker} \& {Latter}}{{Barker} \&
  {Latter}}{2015}]{2015MNRAS.450...21B}
{Barker} A.~J.,  {Latter} H.~N.,  2015, \mnras, 450, 21

\bibitem[\protect\citeauthoryear{Bender \& Orszag}{Bender \&
  Orszag}{1978}]{bender1978advanced}
Bender C.~M.,  Orszag S.~A.,  1978, {Advanced Mathematical Methods for
  Scientists and Engineers I: Asymptotic Methods and Perturbation Theory}.
Springer

\bibitem[\protect\citeauthoryear{{Birnstiel}, {Klahr} \&
  {Ercolano}}{{Birnstiel}
  et~al.}{2012}]{birnstiel:2012.grain.size.distribution.models}
{Birnstiel} T.,  {Klahr} H.,    {Ercolano} B.,  2012, \aap, 539, A148

\bibitem[\protect\citeauthoryear{{Blaes} \& {Balbus}}{{Blaes} \&
  {Balbus}}{1994}]{1994ApJ...421..163B}
{Blaes} O.~M.,  {Balbus} S.~A.,  1994, \apj, 421, 163

\bibitem[\protect\citeauthoryear{Blum \& Wurm}{Blum \&
  Wurm}{2008}]{2008ARA&A..46...21B}
Blum J.,  Wurm G.,  2008, \araa, 46, 21

\bibitem[\protect\citeauthoryear{{Bowler}}{{Bowler}}{2016}]{2016PASP..128j2001B}
{Bowler} B.~P.,  2016, \pasp, 128, 102001

\bibitem[\protect\citeauthoryear{Bracco, Chavanis, Provenzale \&
  Spiegel}{Bracco
  et~al.}{1999}]{bracco:1999.keplerian.largescale.grain.density.sims}
Bracco A.,  Chavanis P.~H.,  Provenzale A.,    Spiegel E.~A.,  1999, \pfluid,
  11, 2280

\bibitem[\protect\citeauthoryear{{Brauer}, {Dullemond} \& {Henning}}{{Brauer}
  et~al.}{2008}]{2008A&A...480..859B}
{Brauer} F.,  {Dullemond} C.~P.,    {Henning} T.,  2008, \astronastro, 480, 859

\bibitem[\protect\citeauthoryear{Carrera, Gorti, Johansen \& Davies}{Carrera
  et~al.}{2017}]{0004-637X-839-1-16}
Carrera D.,  Gorti U.,  Johansen A.,    Davies M.~B.,  2017, \apj, 839, 16

\bibitem[\protect\citeauthoryear{{Carrera}, {Johansen} \& {Davies}}{{Carrera}
  et~al.}{2015}]{2015A&A...579A..43C}
{Carrera} D.,  {Johansen} A.,    {Davies} M.~B.,  2015, \astronastro, 579, A43

\bibitem[\protect\citeauthoryear{{Cassan}, {Kubas}, {Beaulieu}
  et~al.,}{{Cassan} et~al.}{2012}]{2012Natur.481..167C}
{Cassan} A.,  {Kubas} D.,  {Beaulieu} J.-P.,    et~al., 2012, \nature, 481, 167

\bibitem[\protect\citeauthoryear{Chiang \& Youdin}{Chiang \&
  Youdin}{2010}]{Chiang:2010ew}
Chiang E.,  Youdin A.~N.,  2010, \arpes, 38, 493

\bibitem[\protect\citeauthoryear{{Cuzzi}, {Hartlep} \& {Estrada}}{{Cuzzi}
  et~al.}{2016}]{2016LPI....47.2661C}
{Cuzzi} J.~N.,  {Hartlep} T.,    {Estrada} P.~R.,  2016, in Lunar and Planetary
  Science Conference Vol.~47 of Lunar and Planetary Inst.~Technical Report,
  {Planetesimal Initial Mass Functions and Creation Rates Under Turbulent
  Concentration Using Scale-Dependent Cascades}.
p.~2661

\bibitem[\protect\citeauthoryear{{Cuzzi}, {Hogan}, {Paque} \&
  {Dobrovolskis}}{{Cuzzi}
  et~al.}{2001}]{cuzzi:2001.grain.concentration.chondrules}
{Cuzzi} J.~N.,  {Hogan} R.~C.,  {Paque} J.~M.,    {Dobrovolskis} A.~R.,  2001,
  \apj, 546, 496

\bibitem[\protect\citeauthoryear{{Dittrich}, {Klahr} \& {Johansen}}{{Dittrich}
  et~al.}{2013}]{dittrich:2013.grain.clustering.mri.disk.sims}
{Dittrich} K.,  {Klahr} H.,    {Johansen} A.,  2013, \apj, 763, 117

\bibitem[\protect\citeauthoryear{Dobson, Zhang, Greene, Engdahl \&
  Sauer}{Dobson et~al.}{2001}]{dobson2001strong}
Dobson I.,  Zhang J.,  Greene S.,  Engdahl H.,    Sauer P.~W.,  2001, IEEE
  Transactions on Circuits and Systems I: Fundamental Theory and Applications,
  48, 340

\bibitem[\protect\citeauthoryear{Draine \& Salpeter}{Draine \&
  Salpeter}{1979}]{1979ApJ...231...77D}
Draine B.~T.,  Salpeter E.~E.,  1979, \apj, 231, 77

\bibitem[\protect\citeauthoryear{{Dr{\c a}{\.z}kowska} \& {Dullemond}}{{Dr{\c
  a}{\.z}kowska} \& {Dullemond}}{2014}]{2014A&A...572A..78D}
{Dr{\c a}{\.z}kowska} J.,  {Dullemond} C.~P.,  2014, \astronastro, 572, A78

\bibitem[\protect\citeauthoryear{{Dr{\c a}{\.z}kowska}, {Windmark} \&
  {Dullemond}}{{Dr{\c a}{\.z}kowska} et~al.}{2013}]{2013A&A...556A..37D}
{Dr{\c a}{\.z}kowska} J.,  {Windmark} F.,    {Dullemond} C.~P.,  2013,
  \astronastro, 556, A37

\bibitem[\protect\citeauthoryear{{Dr{\c a}{\.z}kowska}, {Windmark} \&
  {Dullemond}}{{Dr{\c a}{\.z}kowska}
  et~al.}{2014}]{drazkowska.2014:dust.growth.protoplanetary.disk}
{Dr{\c a}{\.z}kowska} J.,  {Windmark} F.,    {Dullemond} C.~P.,  2014, \aap,
  567, A38

\bibitem[\protect\citeauthoryear{Drew}{Drew}{1983}]{doi:10.1146/annurev.fl.15.010183.001401}
Drew D.~A.,  1983, \arfm, 15, 261

\bibitem[\protect\citeauthoryear{Epstein}{Epstein}{1923}]{Epstein}
Epstein P.~S.,  1923, \pr, 22, 1

\bibitem[\protect\citeauthoryear{{Flaherty}, {Hughes}, {Rose}, {Simon}, {Qi},
  {Andrews}, {K{\'o}sp{\'a}l}, {Wilner}, {Chiang}, {Armitage} \&
  {Bai}}{{Flaherty} et~al.}{2017}]{2017ApJ...843..150F}
{Flaherty} K.~M.,  {Hughes} A.~M.,  {Rose} S.~C.,  {Simon} J.~B.,  {Qi} C.,
  {Andrews} S.~M.,  {K{\'o}sp{\'a}l} {\'A}.,  {Wilner} D.~J.,  {Chiang} E.,
  {Armitage} P.~J.,    {Bai} X.-n.,  2017, \apj, 843, 150

\bibitem[\protect\citeauthoryear{{Flaherty}, {Hughes}, {Rosenfeld}, {Andrews},
  {Chiang}, {Simon}, {Kerzner} \& {Wilner}}{{Flaherty}
  et~al.}{2015}]{2015ApJ...813...99F}
{Flaherty} K.~M.,  {Hughes} A.~M.,  {Rosenfeld} K.~A.,  {Andrews} S.~M.,
  {Chiang} E.,  {Simon} J.~B.,  {Kerzner} S.,    {Wilner} D.~J.,  2015, \apj,
  813, 99

\bibitem[\protect\citeauthoryear{{Flock}, {Fromang}, {Turner} \&
  {Benisty}}{{Flock} et~al.}{2017}]{2017ApJ...835..230F}
{Flock} M.,  {Fromang} S.,  {Turner} N.~J.,    {Benisty} M.,  2017, \apj, 835,
  230

\bibitem[\protect\citeauthoryear{{Flock}, {Nelson}, {Turner}, {Bertrang},
  {Carrasco-Gonzalez}, {Henning}, {Lyra} \& {Teague}}{{Flock}
  et~al.}{2017}]{2017arXiv171006007F}
{Flock} M.,  {Nelson} R.~P.,  {Turner} N.~J.,  {Bertrang} G.~H.-M.,
  {Carrasco-Gonzalez} C.,  {Henning} T.,  {Lyra} W.,    {Teague} R.,  2017,
  ArXiv e-prints

\bibitem[\protect\citeauthoryear{{Gammie}}{{Gammie}}{1996}]{1996ApJ...457..355G}
{Gammie} C.~F.,  1996, \apj, 457, 355

\bibitem[\protect\citeauthoryear{Garaud \& Lin}{Garaud \&
  Lin}{2004}]{Garaud:2004}
Garaud P.,  Lin D. N.~C.,  2004, \apj, 608, 1050

\bibitem[\protect\citeauthoryear{Goldreich \& Lynden-Bell}{Goldreich \&
  Lynden-Bell}{1965}]{Goldreich:1965tg}
Goldreich P.,  Lynden-Bell D.,  1965, \mnras, 130, 125

\bibitem[\protect\citeauthoryear{{Goldreich} \& {Ward}}{{Goldreich} \&
  {Ward}}{1973}]{1973ApJ...183.1051G}
{Goldreich} P.,  {Ward} W.~R.,  1973, \apj, 183, 1051

\bibitem[\protect\citeauthoryear{{Goodman} \& {Pindor}}{{Goodman} \&
  {Pindor}}{2000}]{2000Icar..148..537G}
{Goodman} J.,  {Pindor} B.,  2000, \icarus, 148, 537

\bibitem[\protect\citeauthoryear{{Gorti}, {Hollenbach} \& {Dullemond}}{{Gorti}
  et~al.}{2015}]{2015ApJ...804...29G}
{Gorti} U.,  {Hollenbach} D.,    {Dullemond} C.~P.,  2015, \apj, 804, 29

\bibitem[\protect\citeauthoryear{{Gressel}, {Nelson} \& {Turner}}{{Gressel}
  et~al.}{2011}]{2011MNRAS.415.3291G}
{Gressel} O.,  {Nelson} R.~P.,    {Turner} N.~J.,  2011, \mnras, 415, 3291

\bibitem[\protect\citeauthoryear{{Guazzelli} \& {Hinch}}{{Guazzelli} \&
  {Hinch}}{2011}]{2011AnRFM..43...97G}
{Guazzelli} {\'E}.,  {Hinch} J.,  2011, \arfm, 43, 97

\bibitem[\protect\citeauthoryear{Heinemann \& Papaloizou}{Heinemann \&
  Papaloizou}{2009}]{Heinemann:2009jv}
Heinemann T.,  Papaloizou J. C.~B.,  2009, \mnras, 397, 52

\bibitem[\protect\citeauthoryear{{Hopkins}}{{Hopkins}}{2016a}]{2016MNRAS.455...89H}
{Hopkins} P.~F.,  2016a, \mnras, 455, 89

\bibitem[\protect\citeauthoryear{{Hopkins}}{{Hopkins}}{2016b}]{hopkins:2014.pebble.pile.formation}
{Hopkins} P.~F.,  2016b, \mnras, 456, 2383

\bibitem[\protect\citeauthoryear{{Hopkins} \& {Christiansen}}{{Hopkins} \&
  {Christiansen}}{2013}]{2013ApJ...776...48H}
{Hopkins} P.~F.,  {Christiansen} J.~L.,  2013, \apj, 776, 48

\bibitem[\protect\citeauthoryear{{Hopkins} \& {Squire}}{{Hopkins} \&
  {Squire}}{2017}]{Hopkins:2017rdi}
{Hopkins} P.~F.,  {Squire} J.,  2017, ArXiv e-prints

\bibitem[\protect\citeauthoryear{{Hopkins} \& {Squire}}{{Hopkins} \&
  {Squire}}{2018}]{Hopkins:2018rdi}
{Hopkins} P.~F.,  {Squire} J.,  2018, ArXiv e-prints

\bibitem[\protect\citeauthoryear{{Hubbard}}{{Hubbard}}{2016}]{2016MNRAS.456.3079H}
{Hubbard} A.,  2016, \mnras, 456, 3079

\bibitem[\protect\citeauthoryear{{Ikoma}, {Nakazawa} \& {Emori}}{{Ikoma}
  et~al.}{2000}]{2000ApJ...537.1013I}
{Ikoma} M.,  {Nakazawa} K.,    {Emori} H.,  2000, \apj, 537, 1013

\bibitem[\protect\citeauthoryear{{Ilgner}}{{Ilgner}}{2012}]{2012A&A...538A.124I}
{Ilgner} M.,  2012, \astronastro, 538, A124

\bibitem[\protect\citeauthoryear{{Ilgner} \& {Nelson}}{{Ilgner} \&
  {Nelson}}{2006}]{2006A&A...445..205I}
{Ilgner} M.,  {Nelson} R.~P.,  2006, \astronastro, 445, 205

\bibitem[\protect\citeauthoryear{Jacquet, Balbus \& Latter}{Jacquet
  et~al.}{2011}]{Jacquet:2011cy}
Jacquet E.,  Balbus S.,    Latter H.,  2011, \mnras, 415, 3591

\bibitem[\protect\citeauthoryear{{Johansen}, {Blum}, {Tanaka}, {Ormel},
  {Bizzarro} \& {Rickman}}{{Johansen} et~al.}{2014}]{2014prpl.conf..547J}
{Johansen} A.,  {Blum} J.,  {Tanaka} H.,  {Ormel} C.,  {Bizzarro} M.,
  {Rickman} H.,  2014, Protostars and Planets VI, pp 547--570

\bibitem[\protect\citeauthoryear{Johansen, Low, Lacerda \& Bizzarro}{Johansen
  et~al.}{2015}]{Johansen:2015he}
Johansen A.,  Low M. M.~M.,  Lacerda P.,    Bizzarro M.,  2015, \sciencea, 1,
  e1500109

\bibitem[\protect\citeauthoryear{{Johansen}, {Oishi}, {Mac Low}, {Klahr},
  {Henning} \& {Youdin}}{{Johansen} et~al.}{2007}]{2007Natur.448.1022J}
{Johansen} A.,  {Oishi} J.~S.,  {Mac Low} M.-M.,  {Klahr} H.,  {Henning} T.,
  {Youdin} A.,  2007, \nature, 448, 1022

\bibitem[\protect\citeauthoryear{{Johansen} \& {Youdin}}{{Johansen} \&
  {Youdin}}{2007}]{2007ApJ...662..627J}
{Johansen} A.,  {Youdin} A.,  2007, \apj, 662, 627

\bibitem[\protect\citeauthoryear{Johansen, Youdin \& Klahr}{Johansen
  et~al.}{2009}]{Johansen:2009jf}
Johansen A.,  Youdin A.,    Klahr H.,  2009, \apj, 697, 1269

\bibitem[\protect\citeauthoryear{Johansen, Youdin \& Mac~Low}{Johansen
  et~al.}{2009}]{Johansen:2009ih}
Johansen A.,  Youdin A.,    Mac~Low M.-M.,  2009, \apj, 704, L75

\bibitem[\protect\citeauthoryear{{Johansen}, {Youdin} \& {Lithwick}}{{Johansen}
  et~al.}{2012}]{2012A&A...537A.125J}
{Johansen} A.,  {Youdin} A.~N.,    {Lithwick} Y.,  2012, \astronastro, 537,
  A125

\bibitem[\protect\citeauthoryear{{Kennel} \& {Wong}}{{Kennel} \&
  {Wong}}{1967}]{1967JPlPh...1...75K}
{Kennel} C.~F.,  {Wong} H.~V.,  1967, \jplp, 1, 75

\bibitem[\protect\citeauthoryear{Klahr \& Hubbard}{Klahr \&
  Hubbard}{2014}]{0004-637X-788-1-21}
Klahr H.,  Hubbard A.,  2014, \apj, 788, 21

\bibitem[\protect\citeauthoryear{Kolmogorov}{Kolmogorov}{1941}]{1941DoSSR..30..301K}
Kolmogorov A.,  1941, \dansssr, 30, 301

\bibitem[\protect\citeauthoryear{{Kowalik}, {Hanasz}, {W{\'o}lta{\'n}ski} \&
  {Gawryszczak}}{{Kowalik} et~al.}{2013}]{2013MNRAS.434.1460K}
{Kowalik} K.,  {Hanasz} M.,  {W{\'o}lta{\'n}ski} D.,    {Gawryszczak} A.,
  2013, \mnras, 434, 1460

\bibitem[\protect\citeauthoryear{{Krijt}, {Ormel}, {Dominik} \&
  {Tielens}}{{Krijt} et~al.}{2015}]{2015A&A...574A..83K}
{Krijt} S.,  {Ormel} C.~W.,  {Dominik} C.,    {Tielens} A.~G.~G.~M.,  2015,
  \astronastro, 574, A83

\bibitem[\protect\citeauthoryear{Kunz \& Balbus}{Kunz \&
  Balbus}{2004}]{Kunz:2004ib}
Kunz M.~W.,  Balbus S.~A.,  2004, \mnras, 348, 355

\bibitem[\protect\citeauthoryear{{Laibe} \& {Price}}{{Laibe} \&
  {Price}}{2014}]{2014MNRAS.440.2136L}
{Laibe} G.,  {Price} D.~J.,  2014, \mnras, 440, 2136

\bibitem[\protect\citeauthoryear{Lambrechts, Johansen, Capelo, Blum \&
  Bodenschatz}{Lambrechts et~al.}{2016}]{Lambrechts:2016fg}
Lambrechts M.,  Johansen A.,  Capelo H.~L.,  Blum J.,    Bodenschatz E.,  2016,
  \astronastro, 591, A133

\bibitem[\protect\citeauthoryear{{Lee}, {Hopkins} \& {Squire}}{{Lee}
  et~al.}{2017}]{lee:dynamics.charged.dust.gmcs}
{Lee} H.,  {Hopkins} P.~F.,    {Squire} J.,  2017, \mnras, 469, 3532

\bibitem[\protect\citeauthoryear{Lesur, Kunz \& Fromang}{Lesur
  et~al.}{2014}]{Lesur:2014cc}
Lesur G.,  Kunz M.~W.,    Fromang S.,  2014, \astronastro, 566, A56

\bibitem[\protect\citeauthoryear{{Lin} \& {Youdin}}{{Lin} \&
  {Youdin}}{2017}]{2017arXiv170802945L}
{Lin} M.-K.,  {Youdin} A.~N.,  2017, ArXiv e-prints

\bibitem[\protect\citeauthoryear{{Lor{\'e}n-Aguilar} \&
  {Bate}}{{Lor{\'e}n-Aguilar} \& {Bate}}{2016}]{2016MNRAS.457L..54L}
{Lor{\'e}n-Aguilar} P.,  {Bate} M.~R.,  2016, \mnras, 457, L54

\bibitem[\protect\citeauthoryear{Marble}{Marble}{1970}]{doi:10.1146/annurev.fl.02.010170.002145}
Marble F.~E.,  1970, \arfm, 2, 397

\bibitem[\protect\citeauthoryear{{Marcus}, {Pei}, {Jiang} \&
  {Hassanzadeh}}{{Marcus} et~al.}{2013}]{2013PhRvL.111h4501M}
{Marcus} P.~S.,  {Pei} S.,  {Jiang} C.-H.,    {Hassanzadeh} P.,  2013, \prl,
  111, 084501

\bibitem[\protect\citeauthoryear{{Millholland}, {Wang} \&
  {Laughlin}}{{Millholland}
  et~al.}{2017}]{millholland:multi.planet.systems.uniform.sizes}
{Millholland} S.,  {Wang} S.,    {Laughlin} G.,  2017, \apj, in press,
  arXiv:1710.11152

\bibitem[\protect\citeauthoryear{{Nakagawa}, {Sekiya} \& {Hayashi}}{{Nakagawa}
  et~al.}{1986}]{1986Icar...67..375N}
{Nakagawa} Y.,  {Sekiya} M.,    {Hayashi} C.,  1986, \icarus, 67, 375

\bibitem[\protect\citeauthoryear{{Nelson}, {Gressel} \& {Umurhan}}{{Nelson}
  et~al.}{2013}]{2013MNRAS.435.2610N}
{Nelson} R.~P.,  {Gressel} O.,    {Umurhan} O.~M.,  2013, \mnras, 435, 2610

\bibitem[\protect\citeauthoryear{{Nelson} \& {Papaloizou}}{{Nelson} \&
  {Papaloizou}}{2004}]{2004MNRAS.350..849N}
{Nelson} R.~P.,  {Papaloizou} J.~C.~B.,  2004, \mnras, 350, 849

\bibitem[\protect\citeauthoryear{{Okuzumi}}{{Okuzumi}}{2009}]{2009ApJ...698.1122O}
{Okuzumi} S.,  2009, \apj, 698, 1122

\bibitem[\protect\citeauthoryear{Pan \& Padoan}{Pan \&
  Padoan}{2013}]{Pan:2013bm}
Pan L.,  Padoan P.,  2013, \apj, 776, 12

\bibitem[\protect\citeauthoryear{{Pan}, {Padoan}, {Scalo}, {Kritsuk} \&
  {Norman}}{{Pan} et~al.}{2011}]{pan:2011.grain.clustering.midstokes.sims}
{Pan} L.,  {Padoan} P.,  {Scalo} J.,  {Kritsuk} A.~G.,    {Norman} M.~L.,
  2011, \apj, 740, 6

\bibitem[\protect\citeauthoryear{Papaloizou \& Pringle}{Papaloizou \&
  Pringle}{1985}]{Papaloizou:1985vf}
Papaloizou J. C.~B.,  Pringle J.~E.,  1985, \mnras, 213, 799

\bibitem[\protect\citeauthoryear{{P{\'e}rez}, {Carpenter}, {Andrews}
  et~al.,}{{P{\'e}rez} et~al.}{2016}]{2016Sci...353.1519P}
{P{\'e}rez} L.~M.,  {Carpenter} J.~M.,  {Andrews} S.~M.,    et~al., 2016,
  \science, 353, 1519

\bibitem[\protect\citeauthoryear{{Pinte}, {Dent}, {M{\'e}nard}, {Hales},
  {Hill}, {Cortes} \& {de Gregorio-Monsalvo}}{{Pinte}
  et~al.}{2016}]{2016ApJ...816...25P}
{Pinte} C.,  {Dent} W.~R.~F.,  {M{\'e}nard} F.,  {Hales} A.,  {Hill} T.,
  {Cortes} P.,    {de Gregorio-Monsalvo} I.,  2016, \apj, 816, 25

\bibitem[\protect\citeauthoryear{{Ruden}, {Papaloizou} \& {Lin}}{{Ruden}
  et~al.}{1988}]{1988ApJ...329..739R}
{Ruden} S.~P.,  {Papaloizou} J.~C.~B.,    {Lin} D.~N.~C.,  1988, \apj, 329, 739

\bibitem[\protect\citeauthoryear{{Sch{\"a}fer}, {Yang} \&
  {Johansen}}{{Sch{\"a}fer} et~al.}{2017}]{2017A&A...597A..69S}
{Sch{\"a}fer} U.,  {Yang} C.-C.,    {Johansen} A.,  2017, \astronastro, 597,
  A69

\bibitem[\protect\citeauthoryear{Sch{\"a}fer, Yang \& Johansen}{Sch{\"a}fer
  et~al.}{2017}]{Schafer:2017hu}
Sch{\"a}fer U.,  Yang C.-C.,    Johansen A.,  2017, \astronastro, 597, A69

\bibitem[\protect\citeauthoryear{Shadmehri}{Shadmehri}{2016}]{0004-637X-817-2-140}
Shadmehri M.,  2016, \apj, 817, 140

\bibitem[\protect\citeauthoryear{Shakura \& Sunyaev}{Shakura \&
  Sunyaev}{1973}]{Shakura:1973uy}
Shakura N.~I.,  Sunyaev R.~A.,  1973, \astronastro, 24, 337

\bibitem[\protect\citeauthoryear{{Simon}, {Armitage}, {Li} \& {Youdin}}{{Simon}
  et~al.}{2016}]{2016ApJ...822...55S}
{Simon} J.~B.,  {Armitage} P.~J.,  {Li} R.,    {Youdin} A.~N.,  2016, \apj,
  822, 55

\bibitem[\protect\citeauthoryear{Simon, Armitage, Youdin \& Li}{Simon
  et~al.}{2017}]{2041-8205-847-2-L12}
Simon J.~B.,  Armitage P.~J.,  Youdin A.~N.,    Li R.,  2017, \apjl, 847, L12

\bibitem[\protect\citeauthoryear{{Simon}, {Lesur}, {Kunz} \&
  {Armitage}}{{Simon} et~al.}{2015}]{2015MNRAS.454.1117S}
{Simon} J.~B.,  {Lesur} G.,  {Kunz} M.~W.,    {Armitage} P.~J.,  2015, \mnras,
  454, 1117

\bibitem[\protect\citeauthoryear{{Spitzer}}{{Spitzer}}{1965}]{1965pfig.book.....S}
{Spitzer} L.,  1965, Physics of fully ionized gases.
Interscience Tracts on Physics and Astronomy, New York

\bibitem[\protect\citeauthoryear{Squire \& Bhattacharjee}{Squire \&
  Bhattacharjee}{2014}]{Squire:2014es}
Squire J.,  Bhattacharjee A.,  2014, \apj, 797, 67

\bibitem[\protect\citeauthoryear{{Squire} \& {Hopkins}}{{Squire} \&
  {Hopkins}}{2017}]{Squire:2017rdi}
{Squire} J.,  {Hopkins} P.~F.,  2017, ArXiv e-prints

\bibitem[\protect\citeauthoryear{Sundaresan}{Sundaresan}{2003}]{doi:10.1146/annurev.fluid.35.101101.161151}
Sundaresan S.,  2003, Annual Review of Fluid Mechanics, 35, 63

\bibitem[\protect\citeauthoryear{{Takeuchi}, {Muto}, {Okuzumi}, {Ishitsu} \&
  {Ida}}{{Takeuchi} et~al.}{2012}]{2012ApJ...744..101T}
{Takeuchi} T.,  {Muto} T.,  {Okuzumi} S.,  {Ishitsu} N.,    {Ida} S.,  2012,
  \apj, 744, 101

\bibitem[\protect\citeauthoryear{{Teague}, {Guilloteau}, {Semenov}, {Henning},
  {Dutrey}, {Pi{\'e}tu}, {Birnstiel}, {Chapillon}, {Hollenbach} \&
  {Gorti}}{{Teague} et~al.}{2016}]{2016A&A...592A..49T}
{Teague} R.,  {Guilloteau} S.,  {Semenov} D.,  {Henning} T.,  {Dutrey} A.,
  {Pi{\'e}tu} V.,  {Birnstiel} T.,  {Chapillon} E.,  {Hollenbach} D.,
  {Gorti} U.,  2016, \aap, 592, A49

\bibitem[\protect\citeauthoryear{Trefethen, Trefethen, Reddy \&
  Driscoll}{Trefethen et~al.}{1993}]{Trefethen:1993bb}
Trefethen L.~N.,  Trefethen A.~E.,  Reddy S.~C.,    Driscoll T.~A.,  1993,
  \science, 261, 578

\bibitem[\protect\citeauthoryear{{Umurhan}, {Shariff} \& {Cuzzi}}{{Umurhan}
  et~al.}{2016}]{2016ApJ...830...95U}
{Umurhan} O.~M.,  {Shariff} K.,    {Cuzzi} J.~N.,  2016, \apj, 830, 95

\bibitem[\protect\citeauthoryear{{Wardle}}{{Wardle}}{1999}]{1999MNRAS.307..849W}
{Wardle} M.,  1999, \mnras, 307, 849

\bibitem[\protect\citeauthoryear{{Wardle}}{{Wardle}}{2007}]{2007Ap&SS.311...35W}
{Wardle} M.,  2007, \astross, 311, 35

\bibitem[\protect\citeauthoryear{{Weidenschilling}}{{Weidenschilling}}{1977}]{1977Ap&SS..51..153W}
{Weidenschilling} S.~J.,  1977, \astross, 51, 153

\bibitem[\protect\citeauthoryear{{Weiss}, {Marcy}, {Petigura}, {Fulton},
  {Howard}, {Winn}, {Isaacson}, {Morton}, {Hirsch}, {Sinukoff}, {Cumming},
  {Hebb} \& {Cargile}}{{Weiss}
  et~al.}{2017}]{weiss:planets.same.size.vs.radius}
{Weiss} L.~M.,  {Marcy} G.~W.,  {Petigura} E.~A.,  {Fulton} B.~J.,  {Howard}
  A.~W.,  {Winn} J.~N.,  {Isaacson} H.~T.,  {Morton} T.~D.,  {Hirsch} L.~A.,
  {Sinukoff} E.~J.,  {Cumming} A.,  {Hebb} L.,    {Cargile} P.~A.,  2017, ArXiv
  e-prints

\bibitem[\protect\citeauthoryear{White}{White}{2010}]{white2010asymptotic}
White R.~B.,  2010, {Asymptotic Analysis of Differential Equations}.
Imperial College Press

\bibitem[\protect\citeauthoryear{{Yang} \& {Johansen}}{{Yang} \&
  {Johansen}}{2014}]{2014ApJ...792...86Y}
{Yang} C.-C.,  {Johansen} A.,  2014, \apj, 792, 86

\bibitem[\protect\citeauthoryear{{Yang}, {Johansen} \& {Carrera}}{{Yang}
  et~al.}{2016}]{2016arXiv161107014Y}
{Yang} C.-C.,  {Johansen} A.,    {Carrera} D.,  2016, ArXiv e-prints

\bibitem[\protect\citeauthoryear{Youdin \& Johansen}{Youdin \&
  Johansen}{2007}]{2007ApJ...662..613Y}
Youdin A.,  Johansen A.,  2007, \apj, 662, 613

\bibitem[\protect\citeauthoryear{Youdin \& Goodman}{Youdin \&
  Goodman}{2005}]{2005ApJ...620..459Y}
Youdin A.~N.,  Goodman J.,  2005, \apj, 620, 459

\bibitem[\protect\citeauthoryear{{Zhang}, {Qin}, {Davidson}, {Liu} \&
  {Xiao}}{{Zhang} et~al.}{2016}]{2016PhPl...23g2111Z}
{Zhang} R.,  {Qin} H.,  {Davidson} R.~C.,  {Liu} J.,    {Xiao} J.,  2016, \pop,
  23, 072111

\bibitem[\protect\citeauthoryear{{Zhu} \& {Stone}}{{Zhu} \&
  {Stone}}{2014}]{zhu:2014.non.ideal.mhd.vortex.traps}
{Zhu} Z.,  {Stone} J.~M.,  2014, \apj, 795, 53

\bibitem[\protect\citeauthoryear{{Zsom}, {Ormel}, {G{\"u}ttler}, {Blum} \&
  {Dullemond}}{{Zsom} et~al.}{2010}]{2010A&A...513A..57Z}
{Zsom} A.,  {Ormel} C.~W.,  {G{\"u}ttler} C.,  {Blum} J.,    {Dullemond} C.~P.,
   2010, \astronastro, 513, A57

\end{thebibliography}

\appendix
\label{APPENDIX STARTS HERE}

\section{The high-$\mu$ limit of the streaming instability}\label{app: high mu streaming}

In this appendix, we study the standard YG streaming instability, with horizontal drift in the disk midplane, in the high-$\mu$ limit. Simulations 
show that the YG streaming instability concentrates grains to sufficient densities to form planetesimals only 
after they have reached high local metallicities (e.g., \citealp{Johansen:2009ih,Bai:2010gm}), so this limit is particularly important physically. However, as we show 
below, the fastest growing mode in this regime has a  different character to the RDIs studied in the
main text\footnote{In fact, one can analyze the $\mu\gg1$, $\tau_{s}\gg1$ limit as an 
RDI by carrying out a similar matrix analysis to that described in \S\ref{sec: resonance}, but using  $\mu^{-1}\ll1$ as
the perturbation parameter. However, because this only gives simple expressions for the less physically relevant case with $\tau_{s}\gg 1$, we do not present this calculation here. } (see also, e.g., Fig.~4 of \citealt{2005ApJ...620..459Y}), and is in fact a separate mode that becomes unstable only once $\mu>1$. Our purpose 
here is to illustrate this different character of the high-$\mu$ instability, and  give analytic expressions 
for its fastest growing wavenumber and growth rate. We do so via expansions of the the dispersion relation 
in $\tau_{s}\ll1$, quantifying 
how the instability is confined to very short wavelengths when $\tau_{s}\ll1$ and $\mu\gg1$ (see also  \citealp{2005ApJ...620..459Y,2007ApJ...662..613Y}).

\subsection{Instability criterion and growth rates}

Following \citet{2005ApJ...620..459Y},
we shall treat the instability assuming incompressibility of the gas. We shall proceed by expanding the dispersion
relation, which is derived from the characteristic polynomial of the linearized (unstratified) epicyclic system (Eqs.~\eqref{eq: gas NL rho}--\eqref{eq: dust NL v local kep}), after
inserting the NSH drift velocities (Eqs.~\eqref{eq: full NSH drift U}--\eqref{eq: full NSH drift V}). Note that we do not
transform into the frame of the gas here---i.e., we insert $\bm{u}=\bm{u}_{0}+\delta \bm{u}$ and $\bm{v}=\bm{v}_{0}+\delta \bm{v}$---because it makes the character of the instability more clear. 

Our first step is to note that the instability 
grows fastest when $k_{z}\gg k_{x}$ \citep{2005ApJ...620..459Y,Jacquet:2011cy}, and expand 
the dispersion relation in $(k_{z}\eta r)^{-1}$. Noting that $k_{x}$ of the fastest-growing mode
scales as $k_{x}\eta r\sim \tau_{s}^{-1}$, we then insert $k_{x}\eta r\sim\tau_{s}^{-1}$ and expand the 
resulting expression in $\tau_{s}$. The result (at order $\tau_{s}^{-3}$) is effectively the terminal velocity approximation 
of \citet{2005ApJ...620..459Y} (see their appendix), and yields the simple polynomial 
\begin{equation}
(\mu+1)^{2} \left(\frac{\omega}{\Omega}\right)^3+2  \tau_{s}k_x\eta r\,\left(\frac{\omega}{\Omega}\right)^2-(\mu+1)^{2} \left(\frac{\omega}{\Omega}\right)+2\, (\mu-1) \, \tau_{s}\, k_{x}\eta r=0.\label{eq:app: low k polynomial for high mu}
   \end{equation}
  While the solutions to this equation (which is identical to Eq.~(39) of \citet{2005ApJ...620..459Y} with $k_{z}\approx k$), can  be written in closed form if desired, they are
not very enlightening. 
However, expanding the solutions of \eqref{eq:app: low k polynomial for high mu} to second order in $\tau_{s}k_{x}\eta r\gg1$, two of the roots are
\begin{equation}
\frac{\omega}{\Omega} \approx \pm i \sqrt{\mu
   -1}+\frac{\mu  (\mu +1)^2}{4\, k_x\eta r\, \tau _s} \mp\frac{i \mu  (5 \mu -4) (\mu +1)^4}{32\,(k_x \eta r \,\tau_s )^2\sqrt{\mu -1} }+\ldots,\label{eq:app: omega expansion low tau}\end{equation}
which shows that there is a separate root, {\em unrelated} to the low-$\mu$ RDI discussed in \S\ref{sec: resonance epicyclic}, which has growth rate $\Im(\omega)/\Omega=\sqrt{\mu-1}$ and so becomes unstable only for $\mu>1$ (when $\tau_{s}\ll 1$).\footnote{Note that, because  \citet{2005ApJ...620..459Y}
expand their Eq.~(39) in $\tau_{s}\ll 1$, without also taking $k_{x}\eta r\,\tau_{s}\gg 1$, they do not discuss this root in 
their \S5.3.} 

\subsection{Characteristic wavenumbers}

Equating the first and third terms in Eq.~\eqref{eq:app: omega expansion low tau}, we can estimate that the  $\Im(\omega)/\Omega\approx \sqrt{\mu-1}$ solution is valid once $k_{x} \eta r\,\tau_{s}\gtrsim \mu^{1/2}(5\mu/2-2)^{1/2}(\mu-1)^{-1/2}(\mu+1)^{2}/4$ (for lower $k_{x}$ one must compute the full solutions to Eq.~\eqref{eq:app: low k polynomial for high mu}).
We can also evaluate the lowest unstable wavenumber from Eq.~\eqref{eq:app: low k polynomial for high mu} by 
evaluating the point at which its discriminant crosses zero, which shows that this mode is stable  
for 
\begin{equation}
k_{x}\eta r\, \tau_{s}<(k_{x}\eta r\,\tau_{s})_{\mathrm{cutoff}} \approx \frac{\mu}{3\sqrt{3}}+\frac{8}{9\sqrt{3}}+\mathcal{O}(\mu^{-1}).\label{eq:app: kxmax for high mu epi}
\end{equation}
This condition illustrates the  very short wavelength nature of this mode---it requires $k\eta r \gg \mu\,\tau_{s}^{-1}$---and, given that high-$k$ modes are  damped by viscosity or dust interparticle spacing (see \S\ref{sec: other effects}), this can be rather restrictive when $\tau_{s}\ll 1$ and $\mu\gg 1$.\footnote{We have assumed $k_{z}\gg k_{x}$ in deriving Eq.~\eqref{eq:app: kxmax for high mu epi}, and this limit is approximately valid until $k_{x}\sim k_{z}$. In the
opposite limit, $k_{z}\ll k_{x}$, one can carry out an expansion in $\tau_{s}\ll1$ and $k_{z}/k_{x}\ll1$ to obtain
the polynomial,  $(\mu+1)^{2}({\omega}/\Omega)^3+2  \tau_{s}k_x \eta r({\omega}/\Omega)^2-(\mu+1)^{2} ({\omega}/\Omega)(k_{z}/k_{x})^{2}+2 (\mu-1)  \tau_{s} k_{x} \eta r (k_{z}/k_{x})^{2}=0$. This polynomial is nearly identical to Eq.~\eqref{eq:app: low k polynomial for high mu} aside from the modification of the last two terms. 
Using the same methodology as in the text leading up to Eq.~\eqref{eq:app: kxmax for high mu epi}, one 
finds solutions with maximum growth rate $\Im(\omega)/\Omega = (k_{z}/k_{x})\sqrt{\mu-1}$, which 
are unstable for 
$k_{x}\eta r \,\tau_{s}<(k_{x}\eta r\,\tau_{s})_{\mathrm{cutoff}} \approx [{\mu}/({3\sqrt{3}})+{8}/{9\sqrt{3}}](k_{z}/k_{x})+\mathcal{O}(\mu^{-1})$. Thus, for $k_{z}/k_{x}\ll 1$, the high-$\mu$ streaming instability is, in principle, unstable for wavenumbers below the cutoff wavenumber Eq.~\eqref{eq:app: kxmax for high mu epi}, by a factor  $\sim \!k_{z}/k_{x}$; however, such modes also have  growth rates that are $k_{z}/k_{x}$ times smaller than modes with $k_{z}\gg k_{x}$, so are less astrophysically interesting. 
This implies that, to be able to strongly clump grains (which likely requires $\Im(\omega)\gtrsim \Omega$), the  high-$\mu$ streaming instability is confined to very short wavelength modes when $\tau_{s}\ll1$ and $\mu\gg 1$.    }

\begin{figure}
\begin{center}
\includegraphics[width=1.0\columnwidth]{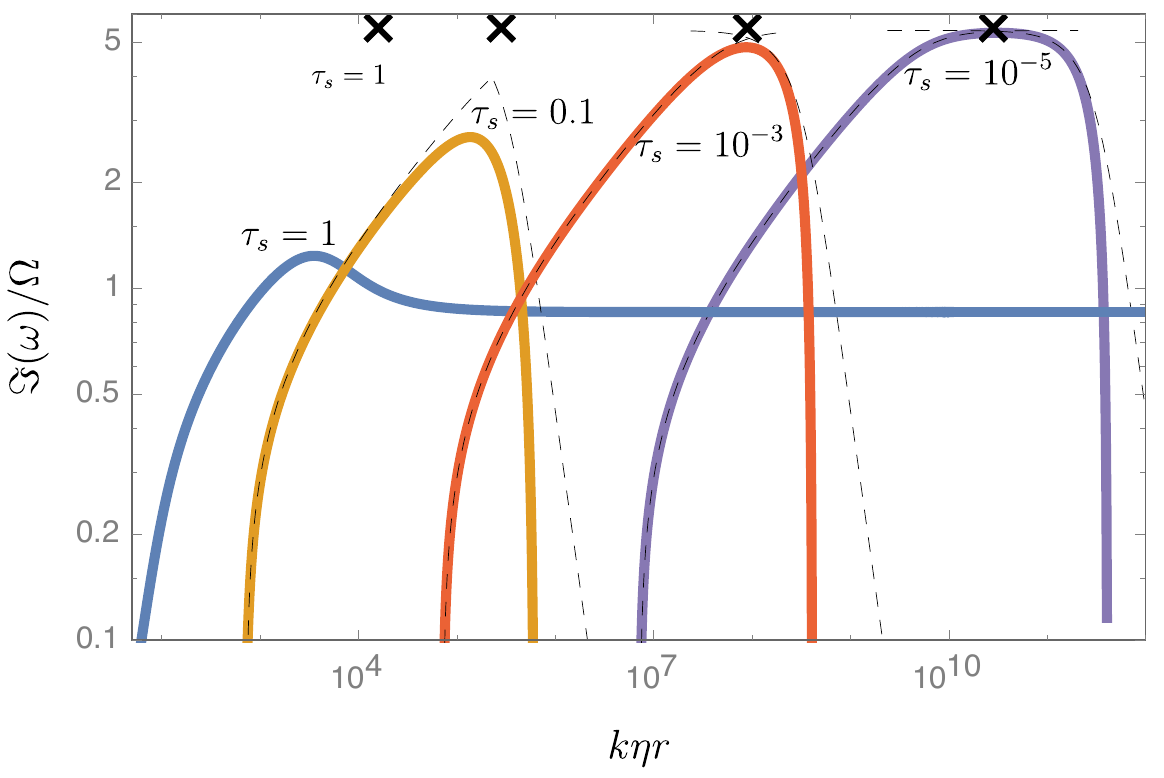}
\end{center}
\caption{The growth rate of the ``high-$\mu$ streaming instability''---the fastest growing unstable mode when $\mu>1$ and there is no vertical dust drift. We derive its properties in App.~\ref{app: high mu streaming}, and show that it is a {\em fundamentally different instability} from the low-metallicity ($\mu<1$) YG streaming instability (the epicyclic RDI). The epicyclic RDI is the fastest-growing mode at $\mu<1$ (where the high-$\mu$ mode becomes stable), but (without vertical drift) it is slower-growing than this ``high-$\mu$'' mode (which  is technically not an RDI) once $\mu>1$. We compare numerical solutions of the full dispersion 
relation  (thick colored lines; only fastest-growing mode shown) with analytic solutions from App.~\ref{app: high mu streaming} derived from asymptotic expansion, for different $\tau_{s}$ as labeled. We take $\mu=30$ and $k_{x}\approx 0.09\,k_{z}$ ($\theta_{k}=85^{\circ}$; this mode angle is close enough to $90^{\circ}$ such that the $k_{x}\gg k_{z}$ limit studied in the text applies well). Black crosses show the maximum growth rate ($\omega=i\sqrt{\mu-1}\,\Omega$) and fastest-growing wavenumber ($k_{x,\mathrm{max}}$; Eq.~\eqref{eq:app: kxmax for high mu epi}) for small $\tau_{s}$, while dashed lines show the analytic approximations (Eq.~\eqref{eq:app: low k polynomial for high mu} and Eq.~\eqref{eq:app: higher k epi high mu}, below/above $k_{x,\,\mathrm{max}}$). 
\label{fig:app: high mu epi}}
\end{figure}

While Eq.~\eqref{eq:app: omega expansion low tau} (or the full solutions of Eq.~\eqref{eq:app: low k polynomial for high mu}) predict that $\Im(\omega)$ asymptotes to a constant value at high $k_{x}$, this behavior 
is spurious due to our neglect of terms at next order in $\tau_{s}$. Indeed, carrying out the same expansion
as led to Eq.~\eqref{eq:app: low k polynomial for high mu} but now keeping the  next-order terms in 
$\tau_{s}$, one finds terms of order $(k_{x}\eta r \,\tau_{s})^{3}/\tau_{s}^{2}$ (compared to terms of order $(k_{x}\eta r\,\tau_{s})/\tau_{s}^{3}$ in Eq.~\eqref{eq:app: low k polynomial for high mu}), which become important when $k_{x}\eta r \gtrsim \tau_{s}^{-3/2}$. We thus expand again in large $k_{x}$, but now with $k_{x}\eta r\, \tau_{s}^{3/2}\sim 1$, finding
the roots,
\begin{equation}
\frac{\omega}{\Omega}\approx \pm i \left[ \left(\frac{4 \mu (k_{x} \eta r)^{2}\tau_{s}^{3}}{(1+\mu)^{5}}\right)^{2}+ \mu-1  \right]^{1/2} \mp i\frac{4 \mu (k_{x} \eta r)^{2}\tau_{s}^{3}}{(1+\mu)^{5}}.\label{eq:app: higher k epi high mu}
\end{equation}
This solution approaches $\omega/\Omega = \pm i\sqrt{\mu-1}$ for small $k_{x}\eta r$---i.e., the opposite limit 
to Eq.~\eqref{eq:app: omega expansion low tau}---and there is a sharp decrease towards 
$\omega=0$ once $(k_{x} \eta r)^{2}\tau_{s}^{3}\gtrsim \mu^{-1}(\mu-1)^{1/2}(\mu+1)^{5}/4$.
The fastest growing wavenumber of the full solution, which we term $k_{x,\mathrm{max}}$, is then well approximated as 
the geometric mean of the $k_{x} \eta r\,\tau_{s}\sim 1$ and $k_{x}\eta r \,\tau_{s}^{3/2}\sim 1$ solutions (Eqs.~\eqref{eq:app: omega expansion low tau} and \eqref{eq:app: higher k epi high mu}, respectively), 
giving
\begin{equation}
k_{x,\mathrm{max}}\eta r \sim \frac{(\mu +1)^{9/4} ({10\, \mu -8})^{1/4}}{4 (\mu -1)^{1/8}
   \tau_s^{5/4}}\approx 0.4 \mu^{19/8}\tau_{s}^{-5/4}+\mathcal{O}(\mu^{11/8}).
\end{equation}
As shown in Fig.~\ref{fig:app: high mu epi} with the black crosses,  this approximation of $k_{x,\mathrm{max}}$ and the solution $\omega/\Omega=i\sqrt{1+\mu}$ compare well against solutions  of the full characteristic polynomial 
for $\tau_{s}\lesssim1$, as do the solutions of Eq.~\eqref{eq:app: low k polynomial for high mu} and Eq.~\eqref{eq:app: higher k epi high mu} (shown with thin dashed lines) in their respective regimes of validity. 
At higher $\tau_{s}\gtrsim 0.1$, there are relatively significant errors as expected (given the expansion in $\tau_{s}$),
although the prediction for the fastest growing wavenumber remains reasonable up to $\tau_{s}\approx 1$ (blue curve in Fig.~\ref{fig:app: high mu epi}). Finally, we note parenthetically that the qualitatively different behavior of the $\tau_{s}=1$ dispersion 
relation  at high $k$ (it remains unstable as $k\rightarrow \infty$) can also occur at lower $\tau_{s}$ so long as $\mu$ is sufficiently large; however, this does not seem to be of any profound physical importance, as it occurs only at very high $k$ and  will be damped by viscosity.

\subsection{Mode structure \&\ the mechanism for the high-$\mu$ streaming instability}

Using the same expansions as above at wavenumbers around $k_{x,\,{\rm max}}$ (Eq.~\eqref{eq:app: kxmax for high mu epi}; this is where the growth rate peaks), in a frame moving with the gas drift, the leading-order expression for the eigenfunctions of the high-$\mu$ mode can be written in  particularly simple form:
\begin{align}
&\left(\delta u_{x},\,\delta u_{y},\,\delta u_{z},\,\frac{\delta \rho}{\rho_{0}}\right) \approx \left(1,\,-i\frac{\Omega}{2\omega},\,-\frac{k_{x}}{k_{z}},\,0\right), \label{eq: gas efunction high mu} \\ &
\left(\delta v_{x},\,\delta v_{y},\,\delta v_{z},\,\frac{\delta\rho_{d}}{\rho_{d0}}\right) \approx \left(1-i\,\tau_{s}\frac{\Omega}{\omega},\, -i\frac{\Omega}{2\,\omega}, -\frac{k_{x}}{k_{z}},\, i\frac{\Omega}{\omega}\frac{\tau_{s}}{\driftvelx}\right). \label{eq: dust efunction high mu}\end{align}
Here $\omega$ is the mode frequency and the normalization  is arbitrary (we set $\delta u_{x}=1$ for convenience).

At this order (for a given $\omega$), the gas eigenfunction is exactly that of a normal, incompressible epicyclic oscillation (which would exist without dust, but with different $\omega$). Further, despite high $\mu$, the low-$\tau_{s}$ assumption means that the dust is tightly coupled to the gas and the two (approximately) move  together as a single fluid with a heavier mean molecular weight. Thus, the gas  and dust eigenfunctions (Eqs.~\eqref{eq: gas efunction high mu} and  \eqref{eq: dust efunction high mu}) are almost identical: the azimuthal and vertical components trace the gas to leading order, but the drift in the radial direction generates a small phase offset in the dust velocity in that direction ($\delta v_{x}-\delta u_{x} = -i\,\tau_{s}\,\Omega/\omega$), which (as we would expect) vanishes as $\tau_{s} \rightarrow 0$. 
For finite $\tau_{s}$, this offset means the dust mode is not exactly incompressible, and generates a density fluctuation $\delta \rho_{d}$, which will in turn increase or decrease the strength of the drag acceleration on $\delta u_{x}$ (the gas acceleration in the drift direction). 

Inserting these eigenfunctions into the equations of motion, the $\delta u_{x}$ equation at this order (recall, this is valid for  $k$ around the peak growth rate) is: 
\begin{equation}
i\,\omega\,\delta u_{x} = -2\,\Omega\,\delta u_{y} - \mu\,\frac{\driftvelx}{t_{s,0}}\frac{\delta\rho_{d}}{\rho_{d0}} \approx i\,\frac{\Omega^{2}}{\omega}\,\delta u_{x} - i\,\mu\,\frac{\Omega^{2}}{\omega}\,\delta u_{x},
\end{equation}
which leads to the same dispersion relation as above $\omega^{2} = \Omega^{2}\,(1-\mu)$. We see  that the instability occurs when the second term on the right-hand side---the forcing of $\delta u_{x}$ by the drag from dust on gas---becomes larger than the first term, which is the restoring force from the normal epicyclic acceleration. 

In other words, around this mode, the gas acts as a harmonic oscillator, with the natural oscillations being the epicyclic, incompressible mode, and the dust drag acting to decrease the normal frequency until it passes through zero and becomes imaginary (i.e., the dust-drag generically destabilizes the oscillator when the drag ``driving'' becomes larger in magnitude than the restoring force). 

We also immediately see that at low $\mu \ll 1$, this mode does persist, but it becomes uninteresting: it is simply dust and gas executing stable epicyclic motion in concert.

\subsubsection{Required physics \&\  the relationship to Resonant Drag Instabilities}

By straightforward, albeit tedious, extension of the above arguments, one can confirm that the fundamental character of the high-$\mu$ mode is not altered if we: (1) allow gas compressibility (at the high wavenumbers of the mode, the resulting eigenfunction is incompressible in the gas to leading order); (2) change the drag law; (3) add external magnetic forces (MHD in the gas or Lorentz forces on the dust, provided the Lorentz forces on dust are sub-dominant to drag); (4) add radial stratification (assuming $k\,L_{0}\gg1$ for the local approximation to be valid); (5) include/exclude azimuthal streaming, and/or ignore the gas streaming velocity; (6) change the potential shape (from Keplerian), so the epicyclic frequency is modified (this only systematically changes the growth rate by an order-unity constant). 
We also see that the key ingredients for such an instability to exist are: (1) dust-gas coupling with $\mu > 1$; (2) non-vanishing radial dust drift; and (3) an appropriate  restoring force on the gas and dust
that scales with the velocity (e.g.,\ $d{\bf v}/d t= \mathbb{F}_{d}\cdot{\bf v}$ and $d{\bf u}/dt = \mathbb{F}_{d}\cdot{\bf u}$, where $\mathbb{F}_{d}$ is a tensor) in order to generate harmonic motion (in this case, the epicycles).\footnote{In order for the ``natural'' response to a forcing $\mathbb{F}_{d}\cdot {\bf v}$ to resemble otherwise stable incompressible oscillations, it requires $\mathbb{F}_{d}$ have appropriately paired off-diagonal terms (with arbitrary absolute value but opposite signs (in e.g.\ the $x,y$ and $y,x$ terms). If the terms have the same sign, the system is unstable even without dust. } 
While the other, non-rotating, coupled dust-gas systems studied in the main text---e.g., the stratified atmosphere of \S\ref{sub: BV RDI} (\BV\ RDI)---share properties (1) and (2) with the rotating system, they do not 
share property (3), because they each have no dust restoring force, only a gas restoring force. Thus these other systems do not exhibit this instability, even at $\mu>1$.

Importantly, the high-$\mu$ streaming instability is not an RDI, at least in the sense discussed in the main text, which must be: (1) unstable for {\em all} $\mu$, (2) have  growth rate is maximized when the ``resonant condition'' $\driftvel\cdot\bm{k} = \omega_{0} = \bm{\Omega}\cdot{\bf k}$ is satisfied (hence, resonance with epicyclic oscillations), and (3) (more formally) arise because the linear equation matrix is defective and thus have a growth rate that scales as $\Im(\omega)\sim \mu^{1/2}$ (or $\Im(\omega)\sim \mu^{1/3}$).
This fast-growing, high-$k$ mode, although traditionally also called the ``streaming instability,'' is a fundamentally different mode, which  shares none of these features and appears only when $\mu>1$ (even though it is a de-stabilized epicycle). 
Instead, it is more akin to to a destabilized harmonic oscillator. These observations explain the sudden increase in the growth 
rate of the YG streaming instability at $\mu=1$ (see, for example, figure 8 of \citealt{2005ApJ...620..459Y}).

\section{Dust stratification}\label{app: dust stratification}

In this appendix, we recompute the \BV\ (BV) RDI, allowing for stratification of the dust density and the 
streaming velocity $\driftvel$ (recall, in \S\ref{sec: adding stratification} we considered stratified gas, but ignored possible stratification of the dust for simplicity). As will become clear below, in the presence of a $\driftvel$
stratification, our block-matrix formalism for analyzing the RDI cannot be trivially applied in its standard form, and we therefore use an expansion 
of the (polynomial) dispersion relation to carry out our analysis.
As as outlined in \S\ref{subsub: local linear BV caution}, there are caveats regarding the validity of the local dispersion relation treatment, which also apply here (more formally a true WKBJ analysis should be used, which is beyond the scope of this work).
Nonetheless, it is likely that the analysis here gives a basic picture for the effect of dust stratification.

\begin{figure}
\begin{center}
\includegraphics[width=1.0\columnwidth]{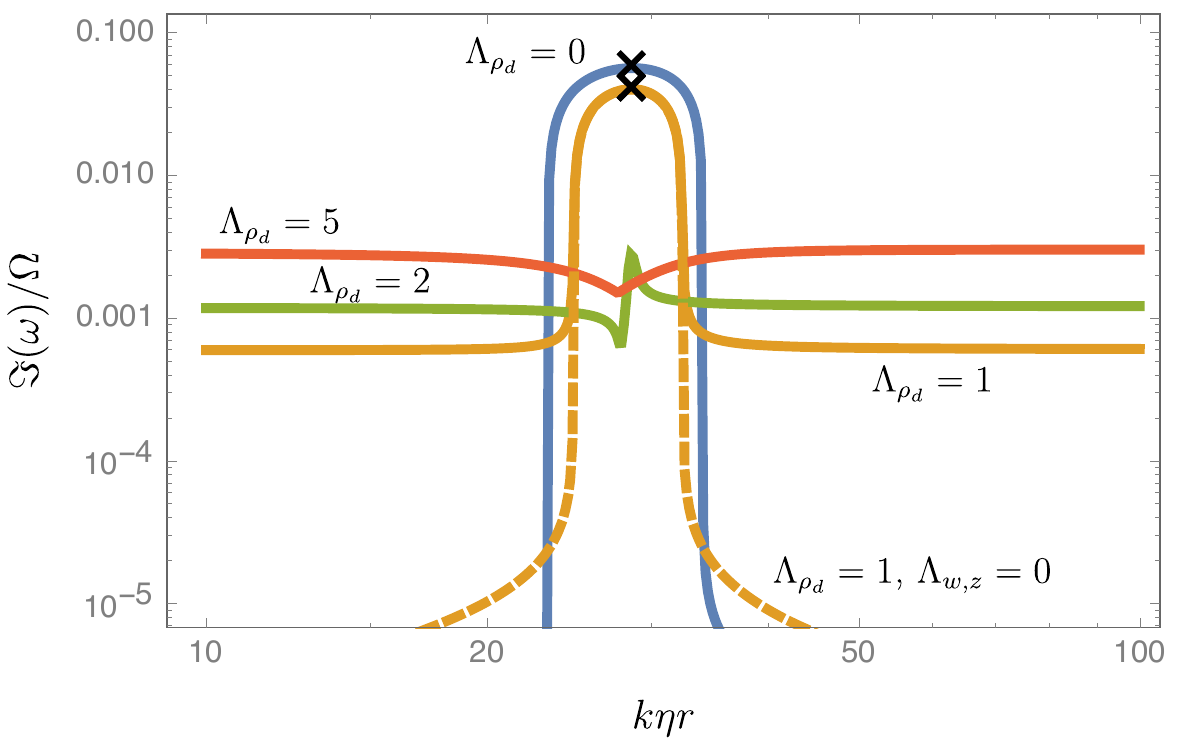}
\end{center}
\caption{The \BV\ RDI including strong stratification in the {dust density and drift velocity} along the direction of the dust drift (dust density and velocity stratification perpendicular to the drift direction cause much smaller effects; see \papertwo). The dust drift is vertical, in the direction of gravity. 
Each line shows the exact solution for a mode with $\hat{\bm{k}}$ at $\theta_{k}=45^{\circ}$ to the stratification direction, with $\tau_{s}=10^{-3}$, $\mu=0.01$, $\eta=0.001$,  $\zeta_{\rho}=1/2$ (Epstein drag), and stratification parameter $\Lambda_{S}=2$, such that $1+\zeta_{\rho}\Lambda_{S}=2$ (thus the system is stably stratified in the absence of dust). 
We compare different strengths of dust-density stratification $\Lambda_{\rho_{d}}$ (ratio of pressure scale length to dust density gradient scale-length): unstratified dust ($\Lambda_{\rho_{d}}=0$; blue), and dust stratification $\Lambda_{\rho_{d}}=1$ (orange), $\Lambda_{\rho_{d}}=2$ (green), and $\Lambda_{\rho_{d}}=5$ (red). For each of the solid curves, we 
choose the drift-velocity stratification ($\Lambda_{w,z}$) such that the background state does not vary in time ($\Lambda_{w,z}=-\Lambda_{\rho_{d}}$). For comparison, the dashed orange curve shows $\Lambda_{\rho_{d}}=1$ with no 
velocity stratification, $\Lambda_{w,z}=0$ (in this case the background state would slowly vary in time).
Our leading-order analytic expectation (Eq.~\eqref{eq: final compressible BV mode app}) is that BV RDI becomes stable when $1+\zeta_{\rho}\Lambda_{S}-\Lambda_{\rho_{d}}<0$ (i.e.,\ when $\Lambda_{\rho_{d}} \ge 2$ for the parameters used here). This is born out to lowest order---the growth rate at the resonant $k\eta r$ decreases significantly for $\Lambda_{\rho_{d}}\ge 2$---although the mode is still unstable (a more formal, global treatment is needed to treat such higher-order effects correctly).
By comparing the dashed and solid $\Lambda_{\rho_{d}}=1$ cases (orange curves), we also see that drift-velocity stratification ($\Lambda_{w,z}$) modifies the dispersion relation significantly away from resonance, but does not change the fastest-growing
resonant modes, as  predicted (see Eq.~\eqref{eq: final compressible BV mode app}). 
\label{fig:app: dust stratification}}
\end{figure}

For simplicity, we assume all quantities are stratified in the
same direction, because misaligned stratifications can introduce new instabilities related to 
baroclinicity, which are not our concern here (see  App.~3 of \papertwo\ for a
complete treatment of stratification for the acoustic RDI). 
We thus use the same style of notation as in \S\ref{sub: BV RDI}, defining,
\begin{equation}
\Lambda_{\rho_{d}}\equiv L_{0} \frac{d\ln\rho_{d}}{dz},\quad \Lambda_{w,j}\equiv L_{0} \frac{d\ln\driftvelj}{dz},\end{equation}
such that the linearized dust equations for local perturbations ($kL_{0}\gg1$) become
\begin{align}
(-i\omega+&i\bm{k}\cdot \driftvel+\driftvelz \Lambda_{w,z}L_{0}^{-1})  \frac{{\delta\rho_{d}}}{\rho_{d0}} + i \bm{k}\cdot \delta \bm{v} + \delta {v}_{z} \Lambda_{\rho_{d}} L_{0}^{-1} =0,\label{eq: BV rho d app}\\[1ex]
(-i\omega+&i\bm{k}\cdot \driftvel )\,  \delta {v}_{j}  + \delta v_{z}\Lambda_{w,j}\driftvelj L_{0}^{-1}= -2\Omega (\hat{\bm{z}}\times \delta \bm{v})_{j}\nonumber \\ & \qquad\qquad\qquad+ \frac{3}{2}\Omega\delta v_{x}\hat{\bm{y}} - \frac{\delta{v}_{j}-\delta {u}_{j}}{t_{s0}}- \driftvelj\frac{\delta t_{s}}{t_{s0}}.\label{eq: BV v d app}\end{align}
To be a true equilibrium with $\partial_{t} \rho_{d0}=0$, the stratification must satisfy $\Lambda_{\rho_{d}}=-\Lambda_{w,z}$, 
and there is also a (small) inertial stress $\driftvelz \driftvelj \Lambda_{w,j} L_{0}^{-1}$, which adds to the drag and any 
external acceleration in the background state.  However, so long as $|\partial_{t}\ln \rho_{d0}|=|\driftvelz L_{0}^{-1}(\Lambda_{\rho_{d}}+\Lambda_{w,z})|\ll \Im(\omega)$, we may consider the system to be a local expansion in time, and we do not explicitly enforce $\Lambda_{\rho_{d}}=-\Lambda_{w,z}$.  In future work, a more formal global WKBJ analysis could be used to apply 
this approximation more formally.

A key step in the RDI analysis outlined in \S\ref{subsub: RDIs} was the identification of the dust-density-perturbation
eigenmode with $\omega=\bm{k}\cdot \driftvel$. This occurred  due to 
the lack of a dust pressure response, because dust density perturbations are simply advected without modification by $\driftvel$. However, with a background gradient in $\driftvel$, this eigenmode becomes 
$\omega = \bm{k}\cdot \driftvel - i\driftvelz \Lambda_{w,z}L_{0}^{-1}$ (or more generally $\omega = \bm{k}\cdot \driftvel - i\nabla\cdot \driftvel$). Physically, this is just the statement that an advected dust perturbation is stretched or compressed along with the background (equilibrium) dust flow. However, this means that there is no longer an exact resonance (at least formally) between the dust 
mode and the gas mode (unless the gas mode is also weakly damped or growing at the same rate).\footnote{Although, it transpires that the RDI result is correct in this case anyway, because the drift-velocity stratification
does not affect the growth rates at resonance; see Eq.~\eqref{eq: final compressible BV mode app}.}

For this reason, we instead carry out the analysis in this section from the dispersion 
relation, obtained from Eqs.~\eqref{eq: BV rho}--\eqref{eq: BV p} and \eqref{eq: BV rho d app}--\eqref{eq: BV v d app}. We insert known scalings obtained from the RDI analysis of \S\ref{sub: BV RDI}, specifically 
 $\omega = \omega_{\mathcal{F}}+ \mu^{1/2}\omega^{(1)}=\hat{k}_{x}N_{BV} + \mu^{1/2}\omega^{(1)}$ and 
 $k = \omega_{\mathcal{F}}/\hat{\bm{k}}\cdot \driftvel$, and 
expand in $\mu\ll1$ and $\tau_{s}\ll1$  (with $\driftvelx\sim \driftvelz\sim \tau_{s}$ and $\driftvely\sim \tau_{s}^{2}$).
To lowest order in $\tau_{s}$ and first order in $\mu$, this yields a second-order 
polynomial for $\omega^{(1)}$,
with solutions 
\begin{equation}
\omega^{(1)} = \pm \mu^{1/2}\left[\frac{\hat{k}_{x}(\hat{\bm{k}}\times\driftvel)_{y}}{2t_{s0}L_{0}}(1+\zeta_{\rho} \Lambda_{S}-\Lambda_{\rho_{d}})\right]^{1/2}.\label{eq: final compressible BV mode app}\end{equation}
We see that the dust density stratification adds a simple correction to the BV RDI growth rate (c.f., Eq.~\eqref{eq: final compressible BV mode}), and that there is no contribution at this order from $\driftvel$ stratification.
This behavior is confirmed using numerical solutions of the local dispersion relation in Fig.~\ref{fig:app: dust stratification}, which illustrates the precipitous drop in the growth rate when $1+\zeta_{\rho} \Lambda_{S}-\Lambda_{\rho_{d}}<0$ and the lack of dependence on $\Lambda_{w,z}$. We reiterate from  \S\ref{sub:sub: BV RDI properties} that when 
$\Theta_{S}=1+\zeta_{\rho} \Lambda_{S}-\Lambda_{\rho_{d}}>0$, the BV RDI is most unstable for dust streaming (vertically) in the 
direction of gravity, while if $\Theta_{S}<0$, it is most unstable 
for dust streaming against the direction of gravity (e.g., streaming upwards when above the disk midplane). When 
$\driftvel$ is perpendicular to $\hat{\bm{g}}$, the system is unstable for either sign of $\Theta_{S}$ depending on the sign of $\hat{k}_{x}\hat{k}_{z}$.
Finally, we note that the correction for the joint epicyclic-BV RDI (\S\ref{sub: BV epi RDI}) is identical---i.e., $1+\zeta_{\rho}\Lambda_{S}$ in Eq.~\eqref{eq: epi BV RDI} becomes $1+\zeta_{\rho}\Lambda_{S}-\Lambda_{\rho_{d}}$, as expected---and there is no 
modification to the double-resonant mode from dust stratification.


\end{document}